\RequirePackage{fix-cm}
\documentclass[12pt,a4paper]{article}
\usepackage[utf8]{inputenc}
\usepackage[T1]{fontenc}
\usepackage[english]{babel}
\usepackage{scrextend}
\usepackage{amsmath}
\usepackage{amsfonts}
\usepackage{amssymb}
\usepackage{mathtools}
\usepackage{titlesec}
\usepackage{xfrac}
\usepackage[left=2.5cm,right=2.5cm,top=2.5cm,bottom=2.5cm]{geometry}
\usepackage{bm}
\usepackage{setspace}
\usepackage{graphicx}
\usepackage[usenames,dvipsnames,table]{xcolor}
\usepackage[position=top]{floatrow}
\usepackage[position=top,skip=5pt]{subfig}
\usepackage[para,flushleft]{threeparttable}
\usepackage{booktabs}
\usepackage[detect-all]{siunitx}
\usepackage{multirow}
\usepackage{makecell}
\usepackage{color}
\usepackage{colortbl}
\usepackage{array}
\usepackage[multiple]{footmisc}
\usepackage{natbib}
\usepackage{pdflscape}
\usepackage{lmodern}
\usepackage{tikz}
\usepackage{tikzscale}
\usepackage{bbm}
\usepackage{totcount}
\usepackage{microtype}

\bibpunct{(}{)}{,}{a}{}{,}
\def\citeapos#1{\citeauthor{#1}'s (\citeyear{#1})}

\usepackage{hyperref}
\hypersetup{%
	breaklinks=true,
	colorlinks=false,
	citecolor=black,
	pdfborder={0 0 0}
}

\usepackage{url}
\usepackage[textsize=scriptsize,backgroundcolor=none,bordercolor=none,linecolor=black]{todonotes}
\newcommand{\nc}{\newcommand}
\nc{\lra}{\Leftrightarrow}
\newcolumntype{s}{D{3}{3}{-1}}

\titleformat*{\subsection}{\normalsize\bfseries}

\makeatletter
\renewcommand\paragraph{\@startsection{paragraph}{4}{\z@}%
	{-3.25ex\@plus -1ex \@minus -.2ex}%
	{0.5ex \@plus .2ex}%
	{\normalfont\normalsize\bfseries}}
\makeatother

\let\oldfootnote\footnote
\renewcommand\footnote[1]{%
	\oldfootnote{\hspace{1.5mm}#1}}

\usepackage{titling}
\usepackage{microtype}

\title{\LARGE{Does external medical review reduce disability insurance inflow?}}

\author{Helge Liebert%
	\thanks{Center for Disability and Integration, Department of Economics,
		University of St.\ Gallen, Rosenbergstr.\ 51, 9000 St.\ Gallen, Switzerland.
		Email: helge.liebert@unisg.ch.}
	\textsuperscript{ ,\,}%
	\thanks{I thank the editor and three anonymous referees for their valuable
		comments. The paper benefited from discussions with Simone Balestra, Eva
		Deuchert, Beatrix Eugster, Per Johansson, Rafael Lalive, Michael Lechner, Nicole
		Maestas, Beatrice Mäder and seminar participants at the University of St.\
		Gallen, the University of Uppsala/IFAU, the 2015 SOLE/EALE meeting in Montreal
		and the 2018 European Workshop on Health Economics and Econometrics in
		Groningen. All remaining errors are my own. This work was funded by the Swiss
		National Science Foundation under grant no. 100018\_143317/1.}}

\date{\href{https://doi.org/10.1016/j.jhealeco.2018.12.005}%
  {\textit{Journal of Health Economics}, 2019, Vol.~64, 108--128
	\texttt{\small https://doi.org/10.1016/j.jhealeco.2018.12.005}}}

\begin{document}
\maketitle
\thispagestyle{empty}
\begin{abstract}

	\noindent This paper investigates the effects of introducing external medical
	review for disability insurance (DI) in a system relying on treating physician
	testimony for eligibility determination. Using a unique policy change and
	administrative data from Switzerland, I show that medical review reduces DI
	incidence by 23\%. Incidence reductions are closely tied to
	difficult-to-diagnose conditions, suggesting inaccurate assessments by
	treating physicians. Due to a partial benefit system, reductions in full
	benefit awards are partly offset by increases in partial benefits. More
	intense screening also increases labor market participation. Existing benefit
	recipients are downgraded and lose part of their benefit income when scheduled
	medical reviews occur. Back-of-the-envelope calculations indicate that
	external medical review is highly cost-effective. Under additional
	assumptions, the results provide a lower bound of the effect on the false
	positive award error rate.

\end{abstract}
\clearpage

%%%%%%%%%%%%%%%%%%%% Introduction %%%%%%%%%%%%%%%%%%%%%%%%%%%%%%%%%%%%%%%%%%%%%

\section{Introduction}
\label{sec:Introduction}

Targeted programs constitute the most common form of social protection
worldwide. Benefit payments are disbursed to groups identified by a common
characteristic -- families, the unemployed or persons with a work-limiting
disability. Among the different social programs, disability insurance (DI) is by
far the most costly. The average OECD country spends about 2.3\% of GDP on
disability-related benefits \citep{OECD2010}. In both the United States and
Europe, the number of DI beneficiaries has been rising throughout the late 20th
and early 21st century and recently stabilized on a high level---on average
about 6\% of the working age population in OECD countries receive disability
benefits \citep{OECD2010}. Increases in DI beneficiaries have often been
associated with imperfect screening of DI applicants
\citep[e.g.][]{Autor2003rise}. One indication for this is that the relative
prevalence of difficult-to-diagnose health conditions like musculoskeletal or
mental health problems on the DI rolls has increased at a higher rate than
prevalence in the general population \citep{Campolieti2002moral,OECD2010}.
Across OECD countries, 60\% of DI inflow can be attributed to muscoloskeletal
conditions or mental health claims \citep{OECD2009sickness}.

Disability benefit decisions are made based on medical assessments of
individuals' residual functional capacity, i.e., their remaining ability to
work. However, the medical assessment process required for eligibility
determination differs across countries. In 40\% of the OECD countries surveyed
in \citet{OECD2003transforming}, the first gatekeeper to the DI system is the
treating physician. In Norway, Switzerland and the United States---countries
which are characterized by high rates of DI prevalence---treating physician
testimony has historically often been decisive for claims decisions. Treating
physicians also hold an influential role in the DI determination process in
Australia, Denmark, Germany, Sweden, and the United Kingdom. In these DI
systems, the treating physician submits the medical documentation of applicants'
diagnosis and treatment history to the DI administration. After submission, the
documentation is reviewed by caseworkers---and potentially also by DI
physicians.

Whether treating physicians or DI-appointed physicians alone should assess
residual functional capacity of DI applicants remains an open question. Treating
physicians are considered to have an informational advantage, hence their
recommendation is often influential in award decisions. The United States Social
Security Administration (SSA) even adopted a `treating physician rule' in 1991,
giving `controlling weight' to the treating physician's opinion. At the same
time, treating physicians are known to diagnose clients favorably in the context
of sick-listing, possibly to prevent harming a long-standing physician-patient
relationship
\citep[e.g.][]{Zinn1996physician,Englund2000variations,Kankaanpaeae2012variations}.
Moreover, treating physicians are often general practitioners and not clinical
specialists, and it is unclear whether complex disabling conditions can be
accurately diagnosed by treating physicians. For these reasons, treating
physicians' assessments are commonly subjected to \textit{medical review} by DI
physicians, who are often clinical specialists.

This paper evaluates the effectiveness of external medical review and its
implications. Identification relies on quasi-experimental policy variation
generated by an extensive pilot program that preceded the nationwide
introduction of mandatory medical review in Switzerland. For the analysis, I
develop a combined difference-in-differences and spatial matching approach,
embedded in an age-based duration analysis framework for estimation. The results
indicate that introducing medical review reduces DI admissions by 23\%.
Reductions are closely tied to psychological and musculoskeletal conditions,
diseases which are more prone to inaccurate diagnoses. Medical review also
increases labor market participation. In an extension to the main analysis, I
provide explicit identifying conditions under which the inflow reduction can be
interpreted as a bound on the reduction in DI award errors. Looking at the
stock, I find that existing benefit recipients are downgraded and lose part of
their benefit income when scheduled medical reviews occur. Finally, I demonstrate that
medical review is highly cost effective.

In 2005, external medical review became mandatory for all DI applications in
Switzerland. This reform was preceded by a pilot, which introduced mandatory
medical review in several Swiss cantons already in 2002. Medical review in this
context means file-based review, exchange with treating physicians and personal
examinations by official DI and other third-party physicians. The reform had
three major components. First, it substantially increased the medical staff and
funding directed towards reviewing DI applicants' cases, more than doubling the
number of full-time equivalent staff positions. Screening quality was improved
by substantially reducing the individual DI physicians' caseload and by
directing cases to physicians' specialized in the relevant field. Second, the
physicians are mandated to review all DI applications, to conduct medical checks
if required and to provide the responsible DI caseworker with better information
about applicants' health. Before the policy change, caseworkers relied on
information provided by applicants' treating physicians for their decision, as
the DI offices had insufficient resources to screen individuals. Third, the
policy also abolished legal obstacles that prevented DI physicians from
examining applicants in person or requesting further documentation. Meanwhile,
the decision structure remains unchanged, the final eligibility decision remains
with the responsible DI caseworker.

This paper contributes to the literature on screening in DI by investigating
medical review, a form of screening which has so far been largely neglected. DI
screening involves two distinct aspects: \textit{stringency} and
\textit{quality}. Interestingly, while screening has received considerable
attention in the literature on DI, studies on screening in DI almost exclusively
focus on variations in screening stringency and use them to obtain a
control group to identify the disincentive effect of DI on labor supply
\citep[e.g.][]{Karlstroem2008employment,Mitra2009disability,Jong2011screening,Staubli2011impact,Maestas2013does,French2014}.
These studies rely on either explicit or implicit changes to eligibility
criteria and the admittance threshold for identification and generally find
positive labor supply effects of screening. For example,
\citet{Jong2011screening}, \citet{Maestas2013does} and \citet{French2014} rely
on variations in adjudicator stringency, while \citet{Karlstroem2008employment}
and \citet{Staubli2011impact} rely on explicit policy reforms that limited
eligibility for certain groups. Looking at DI in Austria,
\citet{Staubli2011impact} shows that stricter eligibility requirements both
reduce insurance prevalence and increase labor supply. Naturally, these studies
also often find lower take-up rates of DI because individuals become
mechanically ineligible for DI due to changes in the admittance criteria.

In this paper, I focus on the implications of medical review, an intervention
that influences screening quality by providing more information on individuals'
underlying capacity to work. Looking at medical review allows abstracting from
mechanical inflow effects which arise due to implicit eligibility requirement
changes. Unlike stringency changes, medical review does not inherently involve a
trade-off between false positive and false negative decision errors
\citep[e.g.][]{Kleven2011transfer,low2015}. Since medical review is primarily
targeting new DI applicants, I focus explicitly on insurance \textit{incidence}
(inflow) in the analysis, since \textit{prevalence} (stock) is likely to be more
inert. In addition, research has shown that inducing work take-up among
long-term beneficiaries can be difficult and results regarding the employment
capabilities of this group are mixed
\citep[e.g.][]{Kornfeld2000,Adam2010reforming,Borghans2014,Buetler2015financial,Moore2015,GarciaMandico2018}.

Moreover, the results in this paper also relate to the findings of health
condition-dependent effect heterogeneity in the literature on disincentive
effects of DI and the literature on misreporting of health status. In a seminal
paper, \citet{Bound1989health} finds that up to half of DI recipients in the US
would be working in the absence of DI.\@ Newer studies have confirmed
\citeapos{Bound1989health} main result, but also show that there is considerable
effect heterogeneity
\citep[e.g.][]{Chen2008work,Wachter2011trends,Maestas2013does,French2014}.
Results by \citet{Wachter2011trends} indicate that especially employment of
younger individuals and those who applied based on mental health and
muscoloskeletal conditions would be non-negligible in the absence of DI.\@
Related to this, \citet{Campolieti2006disability} notes that stricter DI entry
requirements cause fewer reports of these difficult-to-diagnose conditions among
older males.%
\footnote{Other studies have observed that self-reports of
	disability differ from objective measures of functional limitations and that
	individuals out of the labor market tend to overstate health limitations
	\citep{Butler1987measurement,Kreider1999latent,Kreider2007disabilitya,Kreider2008inferring}.
	Exaggeration and malingering of health limitations by patients in anticipation
	of insurance benefits has also been documented in medical studies
	\citep[e.g.][]{Frueh2003disability} and the literature on worker compensation
	schemes \citep[e.g.][]{Staten1982information,Bolduc2002workers}.}
Using administrative records, I show that medical screening reduces insurance
inflow of difficult-to-diagnose conditions and increases labor market
participation. This effectively ties excess inflow of individuals capable of
working to certain conditions and suggests that medical review is a
cost-effective policy to reduce it.

Finally, an extensive theoretical literature investigates the implications of
imperfect tagging in social insurances. Since disability status is private
information, it is inferred by the insurance with error. The seminal work by
\citet{Akerlof1978economics} has been extended to include two-sided
classification errors and applied to the DI context by
\citet{Sheshinski1978model}, \citet{Parsons1996imperfect} and
\citet{Kleven2011transfer}, among others. Few empirical studies have attempted
to estimate the size of classification errors directly. Given auxiliary
assumptions, the results in this paper provide a tentative lower bound estimate
of the effect of medical review on the false positive classification error rate
in these models. In addition, the results suggest that award errors most likely
exceed rejection errors, a finding that diverges from the results for the US.
Although not an exact quantification, these results, unlike earlier studies, do
not rely on small sample expert reviews and the assumption of subsample perfect
classification \citep{Nagi1969disability,Smith1971social} or a comparison with
self-reported disability status \citep{Benitez-Silva2004how}.

Taken together, many results in the paper are also closely related to the
findings by \citet{low2015}, who analyze the trade-off between incentives and
insurance in DI using a life-cycle model. Among other results, they find that
false acceptances exist especially among individuals with moderate limitations,
which can be related to the result that medical review is especially effective
for soft, difficult-to-diagnose health conditions, which are only partially work
limiting. Since welfare effects in their model are dominated by coverage for the
severely work-limited, they pose whether allowing for partial
disability and partial benefits may be a way to reduce incentive costs. While I
cannot make a statement about costs relative to a binary DI system, my results
are obtained within a partial benefit system, indicating that incentive costs
still matter with partial classifications and that misclassification is a
question of degree.

In sum, the paper provides three distinct contributions. First, I show that
medical review is cost-effective in reducing and downgrading inflow of DI
recipients. Second, I demonstrate that reductions are exclusively tied to
difficult-to-diagnose conditions. Together with the fact that screening
increases labor supply, this suggests a combination of inaccurate diagnoses by
treating physicians and possible moral hazard on the side of applicants. Third,
I provide explicit conditions under which the inflow reduction implied by the
reduced-form estimate can be interpreted as a net reduction in DI award errors.

The paper proceeds as follows: The \hyperref[sec:Background]{next section}
discusses the institutional setting and the role of medical screening in DI,
\autoref{sec:Data} introduces the data, \autoref{sec:Strategy} covers
identification and estimation methods, \autoref{sec:Results} discusses the
results and \autoref{sec:Discussion} concludes.

%%%%%%%%%%%%%%%%%%%% INSTITUTIONAL BACKGROUND %%%%%%%%%%%%%%%%%%%%%%%%%%%%%%%%%

\section{Institutional background}
\label{sec:Background}

The Swiss DI system is characterized by generous benefits. Individuals can
receive benefits from three main benefit schemes: mandatory public DI, mandatory
employer-provided occupational pensions and optional private DI.\@ Eligibility
for benefits is determined by the local public DI office responsible for the
main mandatory public scheme and binding for all other benefit providers.
Replacement rates are based on an individual's previous income, contribution
history, whether the individual receives full or partial benefits and the
family situation. The full benefit amount from the mandatory public
DI scheme is capped between 1,175 CHF and 2,350 CHF per month before taxes,
depending on prior income, marriage and contribution history. Individuals with
children receive an additional 40\% of this amount for each dependent child. In
addition, there are income-contingent benefits for spouses and means-tested
supplementary benefits for recipients who fall below the subsistence earnings
threshold. The additional payouts from the mandatory occupational pension scheme
vary based on the contribution length and the employers contract terms. Focusing
only on the two mandatory schemes, a 40 year old adult with full contribution
history and average wage can expect a replacement rate of 70\% if single, 80\%
if married, and 100\% if married with two children. At earnings below the
average wage, the replacement rate increases sharply up to 120\%, exceeding the
prior earnings level \citep{OECD2006,OECD2010}.

Eligibility status and the benefit amount from the main public DI scheme are
determined based on an individual's \textit{disability degree}, a measure of
work incapacity calculated as one minus the ratio of potential labor market
income with disability to the potential income without disability (typically
prior earnings). The determination of potential income is directly tied to a
medical assessment of individuals' \textit{residual work capacity}. If granted,
benefits are paid indefinitely, and are only revised if applicants' health or
earnings change substantially, or they become eligible for retirement pay.
Unlike unemployment insurance (UI), DI benefits are not attached to return-to-work
measures. The Swiss system allows for partial disability benefits in quarterly
increments.

\begin{figure}[t]
    \centering
	\caption{Cantons with medical review during the pilot}
    \label{fig:map}
	\begin{threeparttable}
            \includegraphics[width=\textwidth]{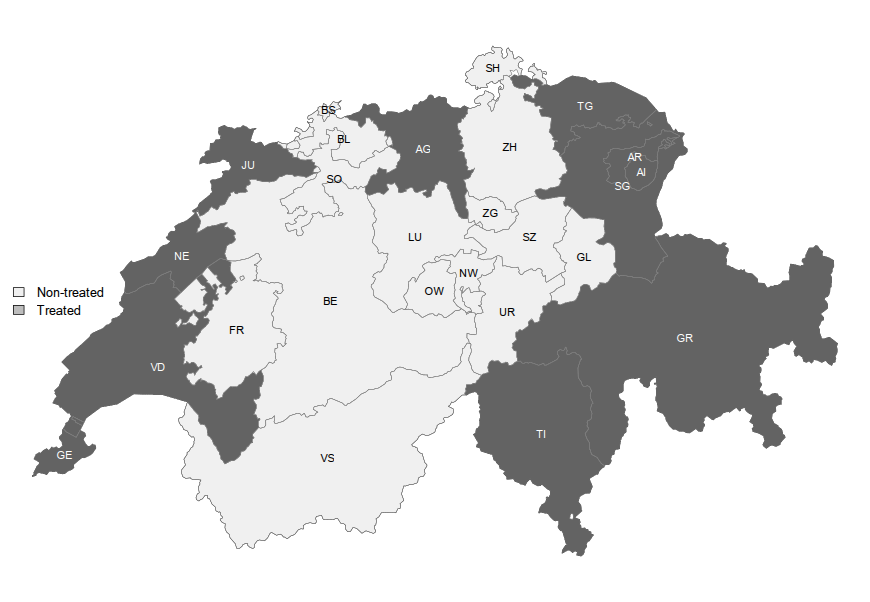}
        \begin{tablenotes}[paragraph,flushleft]
			\scriptsize
			\item Note: Pilot cantons shaded gray. Legend: ZH:\ Z\"urich,
			BE:\ Bern, LU:\ Lucerne, UR:\ Uri, SZ:\ Schwyz, OW:\ Obwalden, NW:\
			Nidwalden, GL:\ Glarus, ZG:\ Zug, FR:\ Fribourg, SO:\ Solothurn, BS:\
			Basel-Stadt, BL:\ Basel-Landschaft, SH:\ Schaffhausen, AR:\ Appenzell
			A.-Rh., AI:\ Appenzell I.-Rh., SG:\ St.\ Gallen, GR:\ Graub\"unden, AG:\
			Aargau, TG:\ Thurgau, TI:\ Ticino, VD:\ Vaud, VS:\ Valais, NE:\
			Neuch\^atel, GE:\ Geneva, JU:\ Jura.
		\end{tablenotes}
	\end{threeparttable}
\end{figure}

%%% Local Variables:
%%% mode: latex
%%% TeX-master: "../rad"
%%% End:

The Swiss parliament passed a reform of the DI system in 2003 (\textit{4.\
	Revision des Bundesgesetzes über die Invalidenversicherung}). Prior to this,
medical review occured infrequently and DI caseworkers made their decisions
based on medical assessments submitted by the applicants' treating physician.
The treating physician-based screening procedure had been in place unrevised
since 1973. The reform resulted in a large expansion of the medical staff
available for review of insurance applications and substantially extended their
legal competences. Physicians were tasked to conduct (re-) appraisals of benefit
claims and authorized to carry out medical examinations.

To assess the effect of the institutional changes, the Federal Ministry of
Social Insurances devised a pilot scheme. Beginning in 2002, 11 out of 26
cantons could already hire new staff and conduct medical review. In the
remaining cantons, operation began in 2005 as scheduled by the reform proposal.
Following the nationwide implementation in 2005, staff funding was expanded
further. The cantons that introduced medical review in 2002 are shown in
\autoref{fig:map}. The cantonal DI offices operate autonomously, but hold a
yearly joint conference, during which participation in the early adopter program
was decided (endogenous self-selection is addressed in more detail in
\autoref{sec:Strategy}). The program was fully funded by the federal ministry.

To become eligible for DI, individuals have to register with their local DI
office. Applicants must register with the DI office corresponding to their place
of residence and cannot file for benefits elsewhere. When filing a benefit
claim, applicants have their treating physician submit the medical documentation
of their condition and their previous earnings records. The earnings loss
induced by the condition must span at least twelve months to qualify for
benefits. The disability insurance office then assesses the individual earnings
loss based on the severity of the condition and its impact on work capability.
Based on the assessment, the caseworker makes a decision whether the person
qualifies for benefits.

Prior to 2002, the insurance office could only assess eligibility from the
medical certificates issued by the applicant's chosen treating physician,
typically the applicant's general practitioner. DI offices were legally not
allowed to examine the applicant, even when in doubt about the credibility or
severity of the impediment. The DI caseworkers deciding on the application have
no medical training themselves, but could consult with physicians working at the
DI offices if they deemed it necessary. However, the DI offices were notoriously
understaffed with physicians. In 2006, the average DI physician reviewed about
612 dossiers per year. Considering the changes in manpower, this figure would
have to be 2.25 times as high prior to the reform to ensure the same coverage
given that application numbers remained constant (Appendix \autoref{fig:diapp}).
For this reason, only a subset of selected dossiers were passed to the DI
physicians for inspection. Caseworkers were reliant on the medical assessment
provided by the treating physician when awarding benefits.

This situation changed with the reform, which essentially strengthened the role
of independent DI physicians in the application process. There are three major
changes attached to the policy. First, the reform substantially increased the
medical staff working for the DI offices. Aggregate figures indicate that the
number of full-time equivalent positions increased by 125\%. Nationwide, the
number of staff positions increased from 105 to 235 due to the reform. Positions
are distributed among cantons proportional to the insured population, implying
that the relative increase is the same for every region. Pilot cantons
experienced this increase three years earlier (see Appendix
\autoref{fig:diphys}).%
\footnote{Since the reform more than doubled the number of physicians working at
	the DI offices, there is concern about delays in hiring staff and filling
	positions. However, comparing the average share of vacancies filled in 2006
	between offices in pilot and late adopter regions does not indicate that such
	delays did occur.}
New physicians are selected to have specialized in fields relevant to diagnose
difficult cases (e.g.\ rheumatology, orthopedics or psychiatry) and are trained
in actuarial regulation. Second, medical review became mandatory for DI claims.
Every applicants' medical history is reviewed and summarized in a non-technical
report for the DI caseworker. Third, physicians were given the authority to
screen people in person, to consult with treating physicians and order further
examinations with other specialists. Before, reviews were legally restricted to
file-based review. The staff is instructed to focus on new DI applicants and aid
with scheduled revisions of existing beneficiaries claim status.

\begin{figure}
    \centering
	\caption{The DI application and decision process}
    \label{fig:info}
  \makebox[\textwidth][c]{%
      \includegraphics[width=1.1\textwidth]{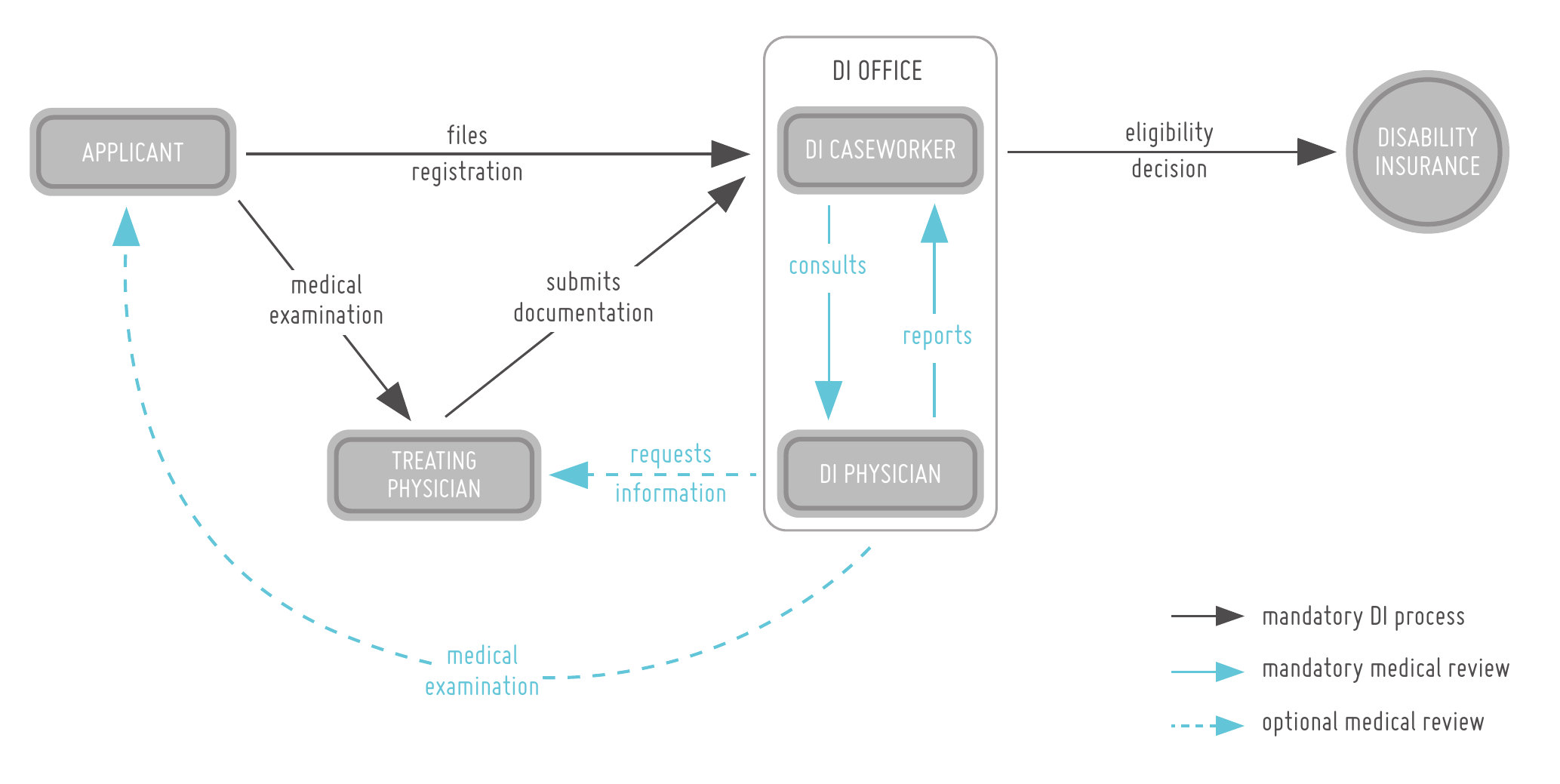}
      }
\end{figure}

%%% Local Variables:
%%% mode: latex
%%% TeX-master: "../rad"
%%% End:

A schematic overview of the application process and the additional processes is
depicted in \autoref{fig:info}. Under the new system, the responsible DI
physician always receives a complete copy of an individual's insurance
application, including the medical documentation of potential limitations. The
DI physician then provides an evaluation of the applicant's eligibility for the
DI caseworker. If the documentation is considered insufficient, additional
information can be requested from treating physicians. Furthermore, if the
physicians notice inconsistencies in the application or deem it to be invalid,
they have the authority to consult with the treating physician, to conduct
further examinations or request visits to other clinical specialist.%
\footnote{Examples for inconsistencies are an applicant claiming benefits on
	grounds of depression without a sufficiently documented history of therapy or
	medication, or an individual with moderate chronic pain claiming full work
	incapacity.}
The DI frequently uses the available channels to gather additional information:
Aggregate figures suggest that in-house examinations occur in up to 10\% of
cases, specialist consultations are decreed in up to 12\% of cases and special
multidisciplinary reports when multiple conditions are present are requested in
up to 6\% \citep{Wapf2007evaluation}.

The DI physicians' eligibility evaluation is not binding. The final decision on
whether benefits are granted remains with the responsible insurance caseworker
and the actuarial requirements are the same. This implies that the regulatory
framework remains unchanged, only the provision of information about the
subjects' eligibility regarding health limitations is affected by the reform.

%%%%%%%%%%%%%%%%%%%% DATA %%%%%%%%%%%%%%%%%%%%%%%%%%%%%%%%%%%%%%%%%%%%%%%%%%%%%

\section{Data}
\label{sec:Data}

The main analysis regarding insurance inflow and the analysis of the labor
market response are both based on the SESAM (\textit{Syntheserhebung soziale
	Sicherheit und Arbeitsmarkt}) data set provided by the Swiss Federal Statistical
Office. The SESAM data link the official Swiss labor force survey (SAKE,
\textit{Schweizerische Arbeitskräfteerhebung}) to administrative records. The
sample period ranges from 1999--2011. I rely on the SESAM data to analyze the DI
hazard because they are the largest representative administrative data source
available which combines different social security and labor market registers
and has sufficient coverage over time. Given the survey weights, the data is
representative of the Swiss population.

SESAM is a rotating panel which tracks individuals for five years until they
drop out and each year 20\% of individuals are resampled. Due to the small
incidence of disability insurance in the population (at most 0.5\% per year) and
the limited number of individuals that can be tracked over several years, the
longitudinal sample dimension cannot be used for the analysis. Instead, the most
recent observation for each individual is used (the choice of observation does
not influence the results in the paper). This sample restriction results in a
large dataset of repeated cross-sections of individual spells. The subsequent
analysis relies on the longitudinal information contained within the dates of
each spell.

DI receipt is measured in the data with the year in which the individual became
eligible for benefits. Since benefits are paid retrospectively, this date
usually coincides with the date the claim was filed. DI receipt is only observed
for those who receive benefits at the time of sampling. The main outcome in the
analysis is DI inflow, measured using the age of disability benefit receipt. The
treatment region is defined as the cantons participating in the pilot project
and the treatment period comprises the years 2002--2004. The data also contains
information about the specific health limitations that ultimately lead to the DI
award. In addition, the data provides a rich set of information about income,
labor market history, current welfare receipt, education, family background and
a wealth of other socio-economic characteristics.

Although the panel dimension in the data does not provide a sufficiently large
enough sample to analyze inflow, when pooled across the whole observation period
it does provide some insight into the dynamics surrounding the time of DI
receipt (see Appendix \autoref{tab:beforeafter}). The statistics show that even
though individuals reduce their labor supply before filing for benefits, a
non-negligible share of beneficiaries continues to work. This reflects the fact
that DI insures earnings losses and that the benefit system is graduated.
Individuals might still work part-time or be absent from work with a sickness
note from a physician. Still, the share of individuals engaging in work drops
from 62\% to 36\% before filing for benefits. About 30\% of recipients continue
to work two years after filing for benefits. Starting from a lower baseline
rate, there is a similarly sharp drop in the share of individuals actively
looking for work before filing for benefits. Very few persons look for a job two
years after filing for benefits. In addition, almost half of all individuals
report having been absent at work despite having a valid contract one year prior
to filing for benefits. This figure decreases after filing, suggesting that
these individuals either leave employment or find a more stable work
arrangement. The income loss that is a requirement for DI eligibility can also
be seen in the data. Income one year after filing for benefits is only about one
third of the income one year prior to filing. The share of people reporting
dismissal from the employment office due to exhausting UI benefits increases one
year prior to filing but is generally low, indicating that individuals
transition to DI either smoothly from UI or directly from work. Finally,
individuals self-reported health declines sharply when filing for benefits. Two
years prior to filing, 23\% of future beneficiaries report having physical or
psychological problems (about twice the unconditional population rate). This
share increases to 84\% after filing for benefits.

The empirical strategy outlined in \autoref{sec:Strategy} partly relies on a
local estimation approach and requires geospatial information to identify
municipalities in the vicinity of administrative borders. The SESAM data
contain information about individuals' municipality of residence. I augment the
data with information about distances between municipal centroids obtained from
www.search.ch. For each municipality, I compute the distance to the nearest
treated/non-treated counterpart sampled in the same year.

Based on this, I construct two estimation samples from the SESAM data, a
\textit{global} sample (containing all individuals in all regions) and a
\textit{local} sample (containing only individuals in municipalities near the
border between treated and control regions). Distance information is available
as both actual travel distance and travel time by car. I choose a travel
distance of 20 kilometers between municipalities as the threshold for the local
sample.%
\footnote{Microcensus data on mobility show that 80\% of commuters stay within
	this distance limit, and it corresponds approximately to the average commuting
	distance and time in Switzerland \citep{BSV2012statistiken,eugsterparchet2018}.}
I then compute nearest-neighbor estimation weights for this sample. The
unrestricted global sample comprises 259,323 individuals, the local sample is
restricted to 133,549 individuals. (descriptive statistics are given in Appendix
\autoref{tab:desc}, the sample composition is mapped in Appendix
\autoref{fig:map2}). In the estimations, I use the survey weights for the global
sample and nearest-neighbor weights for the local sample. All results in the
paper are robust to the choice of distance measure, variations in the threshold
level and whether weights are applied.

As discussed in \autoref{sec:Background}, medical review also applies to
scheduled reassessments of existing beneficiaries' claim status. In the second
part of the analysis, I investigate the effects of medical review on existing
beneficiaries. For this analysis, I use a second administrative dataset provided
by the Swiss Federal Ministry of Social Insurances. I use the data to estimate
the effects of medical review on the disability degree classification and
benefit payment in the beneficiary stock. Moreover, I rely on this data to
investigate potential outflow effects in the beneficiary stock which could
confound the main results (see \autoref{sec:Strategy}).

The stock data tracks the stock of all existing DI recipients from 2001 onwards.
For each individual, I observe the age of entry and the time spent on the DI
rolls. In addition, the data register the actual disability degree, the benefit
amount paid out by the state insurance and the health limitations the person
suffers from, among other socio-economic variables. However, the stock data only
register the region of residence, rendering localized analyses impossible. All
stock analyses condition on individuals with benefit receipt prior to treatment
in 2001, such that results are unconfounded by new entries to the DI payroll.

%%%%%%%%%%%%%%%%%%%% EMPIRICAL STRATEGY %%%%%%%%%%%%%%%%%%%%%%%%%%%%%%%%%%%%%%%

\section{Empirical strategy}
\label{sec:Strategy}

In this section, I develop the empirical approach used in the remainder of the
paper. Section~\ref{sec:ident} discusses identification and introduces the
duration model used in the main analysis. Section~\ref{sec:did} provides
explicit identifying conditions for difference-in-differences in a
\citet{Cox1972regression} proportional hazards model. Section~\ref{sec:threats}
discusses potential mechanisms that could violate these conditions and provides
evidence to support their validity. Finally, section~\ref{sec:idtype} explores
and discusses additional identifying conditions which tighten the interpretation
of the reduced-form estimate, bounding the effect of medical review on the false
positive award error rate.

\subsection{Identification approach and estimation method}%
\label{sec:ident}

The main quantity of interest is the change in the population DI hazard induced
by external medical review, i.e., the change in the rate of newly awarded
benefits among previously non-receiving working-age individuals. However, due
to an opaque political decision process and self-selection into the early
adopter scheme, treatment assignment cannot be assumed to be fully random. The
cantons participating in the pilot program are a mixture of high and low
prevalence regions, and regional cooperation considerations were relevant in
the assignment process.

A difference-in-differences identification approach is used to evaluate the
impact of the medical review institutions. Differencing removes
time-invariant influences on potential outcomes. This removes bias due to
selection into the program based on fixed or inert aggregate regional
differences. However, identification still requires a common development of DI
incidence in the absence of the expansion of medical review. This assumption
raises concerns related to regional heterogeneity and selection. The
remainder of this section introduces the modeling approach, the following
sections present the identifying assumptions and discuss potential threats to
their validity.

As \citet{Autor2003rise} illustrate, people rarely transition directly from
employment into DI, but typically apply conditional on job loss. One concern in
the present context is that labor markets may be less resilient in some regions,
or that regions with strong industrial and commercial hubs are more affected by
common economic shocks. If screening is imperfect and disability insurance is
used as an extension to unemployment insurance or an early retirement vehicle in
case of job loss, differential labor market trends can confound the results.
Since Switzerland is a country with historically tight labor markets, such
concerns are alleviated to some degree. Nevertheless, there may also be other
underlying differences between regions based on the self-selection into the
pilot program that cause time-variant divergence. Remaining time-variant
heterogeneity among Swiss regions may raise concerns about biased treatment
effect estimates.

To address this issue, I follow a twofold approach. A first set of results is
based on the full sample of individuals across all regions. A more narrow
identification approach focuses on individuals in border regions within
commuting distance between treated and control areas. Focusing on these
regions generates samples that are balanced in observable characteristics ex
ante and increases the credibility of the common trend assumption. Similar
strategies are used by \citet{Froelich2010exploiting} and
\citet{Campolieti2012disability}.

However, local estimation approaches relying on sampling based on the distance
to a border can suffer from problems due to spatial clustering on different
sides along the border \citep[cf.][]{keeletitiunik2016}. To alleviate these
concerns, I compute weights corresponding to nearest-neighbor pairwise
differences and use them in the estimations. This weighting approach is
equivalent to spatial matching. The main advantage of weighting is that it
creates a sample that is well-balanced in observables and increases the
credibility of the identifying assumptions introduced in the next section.
Weighting reduces the bias of the estimator by restricting comparisons to a more
similar control group. The bias reduction potentially comes at the cost of an
increase in variance, since the estimator may not use all available data. In the
context of matching, this bias-variance trade-off is often favorable, as the
gain from finding good matches dominates the loss due to higher variance.

For estimation, I exploit the spell format of the data and model insurance
take-up as a duration problem. The main specification uses a stratified
\citet{Cox1972regression} proportional hazard model to estimate the impact of
the reform on DI incidence. The hazard rate is modeled as
\begin{gather}\label{eq:main}
	h(t,P,D|X<\bar{x})
	= h_{0g}(t)\exp{\left(\beta_{0}P+\beta_{1}D+\beta_{2}PD\right)} \ ,
\end{gather}
where $h_{0g}(t)$ is the non-parametric baseline hazard within birth cohort
stratum $g$, $t$ denotes time in years, $D \in \left\{0,1\right\}$ is a binary
treatment group indicator and $P \in \left\{0,1\right\}$ is a binary
time-varying indicator for the pilot period during $t \in \left\{
	2002,2003,2004 \right\}$. Samples are restricted to individuals in
border municipalities between treated and control regions within an absolute
distance threshold $\bar{x}$ (20 km in the main specification), where
individuals are similar in observables and remaining differences can credibly
be assumed to be time-constant.%
\footnote{All estimates are robust across a large set of bandwidths and whether
	travel distance or travel time is chosen as the distance metric. Moreover, the
	results are also robust to replacing \eqref{eq:main} with a more flexible
	specification containing cantonal fixed effects.}

The model is specified using age as the time scale. This is preferable to using
time-on-study as analysis time due to the age-dependent nature of the disability
hazard, the rich cohort data available and the interest in the effect of a
time-varying covariate \citep{Kom1997time,Thiebaut2004choice}. All models are
stratified by five-year birth cohorts to account for cohort-specific differences
in health environments. Individuals become at risk when they are eligible for
insurance at age 18. Censoring occurs at the sampling date or when individuals
reach the retirement age, whichever occurs first. Disability benefit receipt
constitutes failure. Due to data limitations, the analysis is restricted to
single spells and disability insurance is assumed to be an absorbing state.
However, this is not much of an abstraction. Actual outflow rates due to reasons
other than death or moving to the old-age pension system amount to less than 1\%
of the stock per year \citep{BSV2012statistiken}. Previous research for
Switzerland has shown that DI recipients are loath to give up safe benefits even
when faced with strong financial incentives to do so
\citep[][]{Buetler2015financial}.

A duration approach has a number of advantages compared to a linear
difference-in-differences framework in this setting. It corresponds naturally to
the spell format of the available cross-sectional data and the fact that DI
entry is essentially a survival outcome. Data issues also limit the feasibility
of the standard difference-in-differences approach. DI receipt is observed
retrospectively as year of entry and only repeated cross-sections of a
representative sample of the population are available. Since total DI incidence
in the population is low, actual DI entry observed in each sampling year is low
and insufficient for the analysis. Note that DI entry year and sampling year can
be distinct. As the DI entry year is observed for each recipient, irrespective
of the sampling date, pooling all data increases power substantially. This is
due to the fact that all information on DI entry in any given year which is
available from subsequent years in which data was sampled can be utilized.

Pooling all cross-sectional data and conducting the analysis by age instead of
sampling year (time-on-study) also limits the possibility of implicit sampling
bias. With inflow observed retrospectively, relying on absolute sampling time as
the time measure for the analysis would require creating a pseudo-panel
structure by inferring past incidence figures from a post-treatment
cross-section and adjusting for past eligibility. Since the disability risk is
concentrated at older ages near the official retirement age, extrapolating past
incidence causes bias due to intermittent entry into the retirement scheme. A
non-negligible share of those in the old-age pension system at the sampling date
may have received DI previously, but are not observed to do so any more when
they are sampled. This share will increase the further past incidence figures
are inferred retrospectively. Incidence figures inferred this way will be
artificially low and the cross-sectional data ceases to be representative.%
\footnote{Comparisons with aggregate data indicate that the reported aggregate
	rates are underestimated by about 20\% going back five years. Inferring
	incidence further retrospectively, inferred inflow continues to decrease as
	attrition caused by moving to the old age pension system and mortality
	increase. Going back 30 years, inferred incidence converges to zero and is
	almost exclusively driven by small-sample variation of individuals who were
	awarded DI when they were very young.}

Finally, estimation of effects on incidence rates in a standard
difference-in-differences framework would require modifying the standard
common trend assumption in a way which prohibits a more detailed analysis. Since
incidence is defined as new benefit awards among previously non-receiving
working-age individuals, it is necessary to condition on the absence of benefit
receipt in the previous period when calculating the incidence rate for each
period. Since the pilot program spans three years, only incidence rates within
this time frame can effectively be compared without biasing results by
conditioning on an outcome. In contrast, a model built around the hazard as the
parameter of interest lends itself naturally for this purpose.

In follow-up analyses, I investigate possible labor market responses and how
existing beneficiaries react to the medical review process. Unlike the inflow
setting above, these measures can be analyzed in a linear model framework. In
the analysis, I estimate a linear difference-in-differences specification with
canton and year fixed effects and the interaction of the treated cantons with
the pilot period.

\subsection{Identification: Difference-in-differences for duration analysis}%
\label{sec:did}

The standard assumptions for difference-in-differences estimation have to be
restated for proportional hazard models. The exponentiated coefficient on the
interaction between treatment time and region represents a ratio of hazard
ratios
\begin{gather}
	\exp{\left(\beta_{2}\right)} =
	\frac{\sfrac{h(t|D=1,P=1)}{h(t|D=1,P=0)}}{\sfrac{h(t|D=0,P=1)}{h(t|D=0,P=0)}} \ .
\end{gather}
The distance condition has been dropped to ease notation. The effect of
interest is the relative change in the hazard for the treated, a relative
average treatment effect on the treated (rATT),
\begin{gather}\label{eq:att}
	\text{rATT} = \dfrac{h^{1}(t|D=1,P=1)}{h^{0}(t|D=1,P=1)} \ ,
\end{gather}
where $h^{D}$ denotes potential hazard rates. I assume SUTVA
\citep{Rubin1977assignment} holds, i.e., either of the two potential treatment
states is observed. As disability insurance applicants are a small fraction of
the population, it is credible that general equilibrium effects are absent.
Identification then requires the two usual conditions in restated form
\begin{gather}
	h^{1}(t|D=1,P=0) = h^{0}(t|D=1,P=0) \ ,
	\stepcounter{equation}\tag{\textit{no anticipation}, \theequation}
\end{gather}
and
\begin{gather}
	\frac{h^{0}(t|D=1,P=1)}{h^{0}(t|D=1,P=0)} = \dfrac{h^{0}(t|D=0,P=1)}{h^{0}(t|D=0,P=0)} \ .
	\stepcounter{equation}\tag{\textit{common trend}, \theequation}
\end{gather}
The main identifying assumption (5) is that in the absence of mandatory medical
review, incidence for individuals in both pilot and non-pilot (border) regions
would have changed proportionally. The common trend assumption is not invariant
to the scaling of the dependent variable \citep[e.g.][]{Lechner2010estimation}
and is modified accordingly. Instead of assuming a common trend between regions
over time in differences, I am assuming a constant hazard ratio, i.e., a common
relative change or a common absolute change in logs. In addition, I assume that
anticipation effects are absent. Given these assumptions, the coefficient of the
interaction identifies the hazard ratio of interest, the relative ATT.\@

\subsection{Potential threats}%
\label{sec:threats}

The two main threats to identification are a violation of the no anticipation
condition and the common trend assumption. Prospective or ongoing reform changes
may induce some individuals to change their behavior in anticipation of future
loss or gain. The main confounding mechanisms are mobility (individuals move to
untreated regions to apply for DI) and the timing of applications (early application
in anticipation of medical review).

The implementation and chronology of the reform alleviate these concerns. The
first draft of the reform which included the institutional changes introducing
medical review was proposed in parliament in February 2001, and underwent some
revisions until being approved by popular vote in March 2003. The pilot project
began already in January 2002, before the changes were approved. The early
adopter scheme was scheduled immediately after the reform proposal was
publicised and began only ten months afterwards.

Importantly, the pilot scheme was never publicly announced. Communication only
occurred internally between the Federal Ministry of Social Insurances and the DI
offices and was never publicised. The person responsible for the yearly
committee meeting confirmed that the medical review pilot was never publicly
communicated to outsiders. Pilots are published only since 2007, and the medical
review pilot was one of the first pilots launched by the ministry. To be
certain, I conducted a systematic news search on newspaper databases Factiva,
LexisNexis, Pressreader and Swissdox. These do not list a single record
mentioning the early adopter program. Overall, the medical review changes
implied by the reform proposal received little public attention and were only
scheduled to be implemented in 2005.\footnote{Other reform measures scheduled to
	come into effect at a later time included the introduction of a three-quarter
	benefit and the abolishment of additional benefits for spouses. These measures
	received the bulk of public attention. The changes were adopted nationwide and
	only became effective in late 2004. There were no further reforms to DI or
	other social insurances during the introduction period.}

Moreover, considering the one-year earnings loss restriction required for DI
eligibility, the time frame until implementation leaves limited scope for the
strategic timing of applications in both treated and control regions, even if
public knowledge of the program were available. In the treated regions, the
project started ten months after the first reform proposal, effectively leaving
too little time for the strategic timing of applications in treated regions.
Similarly, there is only a relatively short time period between the reforms
definite approval in March 2003 and its nationwide implementation in January
2005 that allows for strategic behavior given the one-year restriction.

Anticipation effects can also manifest in increased mobility. Individuals
considering to apply for disability benefits may anticipate the reform and move
to regions where external medical review is not implemented, generating higher
inflow in control regions and biasing results. However, as described above, the
medical review changes were not announced publicly at the time and the time
frames are relatively narrow. In addition, the amount of people moving to
another region who can be identified by tracking panel cases in the data is
negligible. Between 1999 and 2011 about 3.1\% of the people for whom time series
information is available move to another canton, and less than 0.8\% percent
move from a non-treated to a treated region. About 0.5\% of those sampled during
the pilot period do so. Mobility in Switzerland is generally low compared to
other countries.

\begin{figure}
    \centering
    \caption{Trends for aggregate disability insurance inflow}
    \footnotesize
    \label{fig:agginflowcombined}
            \includegraphics[width=0.75\textwidth]{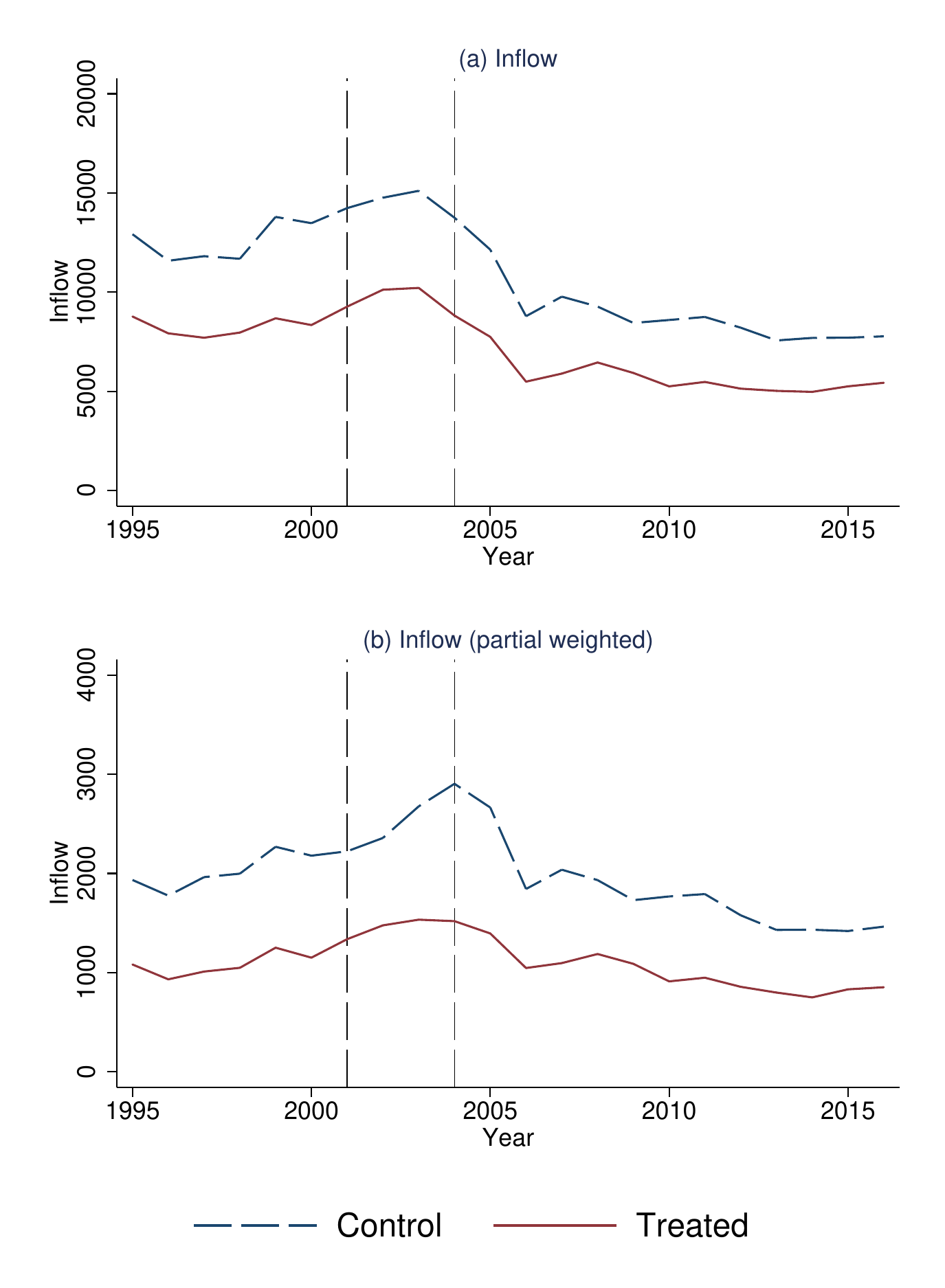}
            \floatfoot*{
            Note: Disability insurance inflow for treated and control regions in
panel (a). Insurance inflow partial weighted by pension amount shown in panel
(b). Data were provided by the Federal Ministry of Social Insurances.}
\end{figure}

%%% Local Variables:
%%% mode: latex
%%% TeX-master: "../rad"
%%% End:

Regarding the common trend assumption, I first perform a balancing test to
ensure that regions are comparable ex ante (Appendix
\autoref{tab:desc_ttest_combined}). Although balance in observables is not
strictly required for identification, the common trend assumption is more
credible if the comparison regions are similar. This exercise reduces concerns
about remaining regional heterogeneity (e.g.\ due to self-selection) that may
induce common trend violations. In the full sample there are significant
differences with regard to age, the share of foreigners, education, marriage
status and family size, characteristics which influence the propensity to
receive DI.\@ Among DI beneficiaries, musculoskeletal conditions are more
prevalent in treated regions. In the weighted local sample, balance improves
considerably. Differences are small in magnitude and mostly insignificant.
People in treated regions are on average more likely to be from a foreign
country; there are about 2\% more people with primary education in treated
regions, correspondingly less with secondary and university-level education; and
a small difference in the unemployment rate. These remaining differences in
observables are small in economic terms and will not affect the estimates unless
trends between treatment and control regions differ.

The typical diagnostic graph to inspect the validity of the common trend
assumption in difference-in-differences designs are trend plots that show how
treated and control units evolve prior to the treatment period.
\autoref{fig:agginflowcombined}, Panel (a) shows trends in the nationwide
aggregate DI inflow for all cantons based on statistics published by the Federal
Ministry for Social Insurances. Prior to 2002, the number of new DI recipients
evolves similarly in treated and control regions. Before the reform, the trends
in inflow are comparable across treated and control cantons. During the pilot,
the trend breaks and inflow declines in treated regions. This is even more
apparent when weighting partial inflow by DI coverage level, indicating that new
beneficiaries are downgraded and admitted at lower partial coverage levels
(\autoref{fig:agginflowcombined}, Panel b). After 2005, inflow rates evolve
similarly again. These developments are also apparent when looking at trends in
the aggregate beneficiary stock (Appendix \autoref{fig:aggstock}). Throughout
the whole time period, applications are largely stable and follow the same trend
in treated and control regions (Appendix \autoref{fig:diapp}). Regarding the
treatment, the number of full-time equivalent positions for DI physicians
exhibits the expected increase due to the pilot and the nationwide
implementation (Appendix \autoref{fig:diphys}, Panel a). Correspondingly, the
caseload per physician drops substantially (Appendix \autoref{fig:diphys}, Panel
b). Note that all descriptive graphs are based on data from national statistics
or federal social insurance reports and not conditioned on the local sample.

\begin{figure}%[t]
    \centering
    \caption{Log cumulative hazard by age and treatment region}
    \footnotesize
    \label{fig:lncumhaz}
            \includegraphics[width=0.75\textwidth]{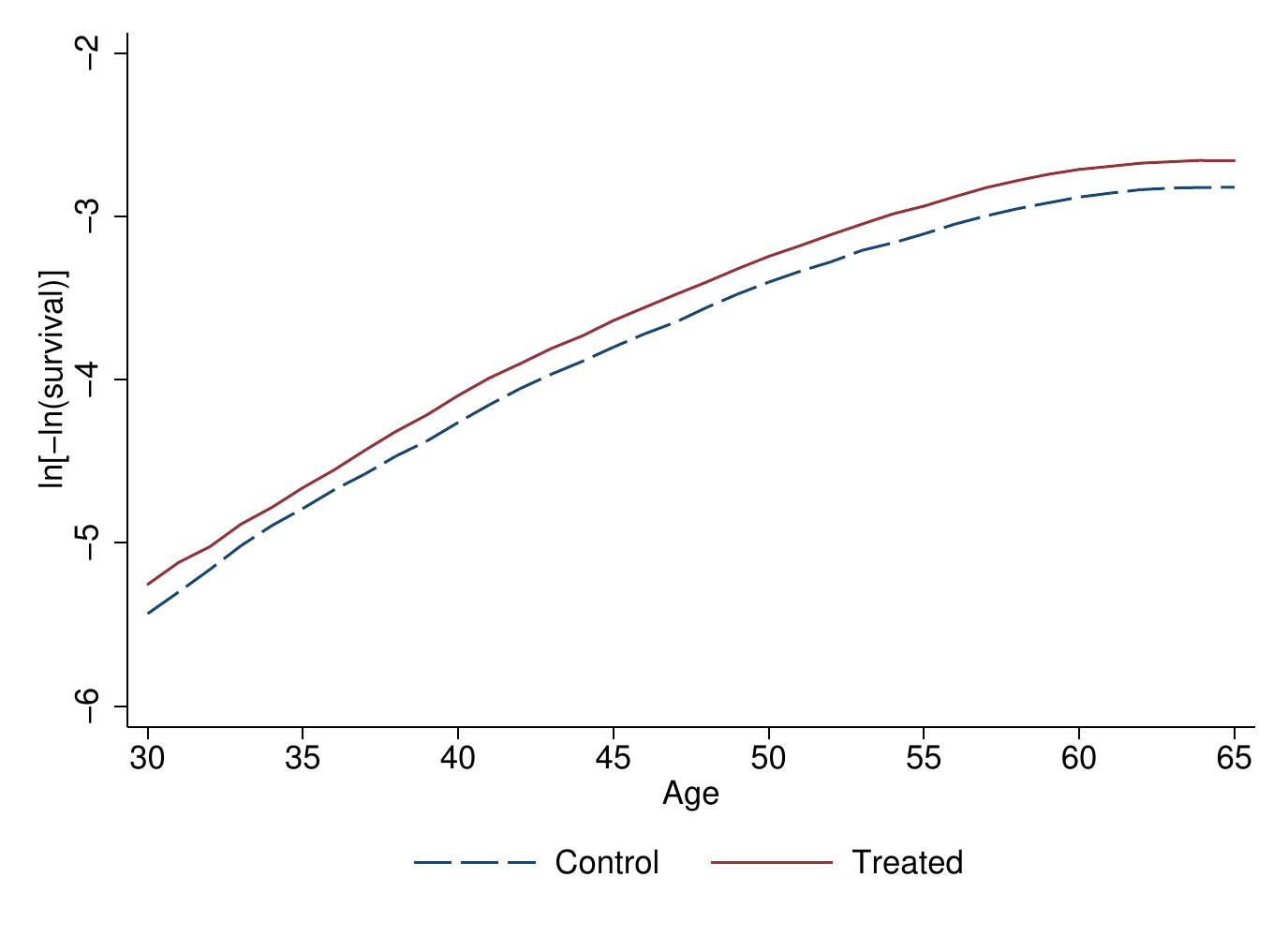}
            \floatfoot*{\centering
                Note: Log-log plot showing log cumulative hazard
            estimates by age for individuals in treated and control regions.}
\end{figure}

While an indication of comparability between regions, strictly seen, the trend
plots in \autoref{fig:agginflowcombined} do
not correspond to the dependent variable used in the estimations. Generating an
equivalent plot in an age-based duration framework is hindered by the fact that
the treatment occurs for every individual at a different time in life, i.e., the
age at which they experience the reform being implemented. An alternative test
for the common trend assumption are the typical placebo specifications for
pre-reform effects (discussed among the robustness checks in
section~\ref{sec:robustness}). These do not indicate that the common trend is
violated. Another possibility to investigate the assumption is to look at the
log cumulative hazard by age as shown in \autoref{fig:lncumhaz} (referred to as
a `log-log plot' in biostatistics). Since individuals are randomly sampled
across regions and have the same age distributions, the log cumulative hazard
estimates for both groups should be parallel.

In addition, the log-log plot is a common diagnostic to assess the validity of
the proportional hazards assumption in the Cox model, with non-parallel or
crossing lines seen as an indication that the proportionality assumption is
violated \citep[e.g.][]{vittinghoff2011regression}. Visually assessing the
validity of the assumption from the log-minus-log transformation is preferable
to comparing survival curves directly, as it is easier to determine whether two
curves are apart by a constant difference than to judge whether they are an
exponential transformation.%
\footnote{Since $h(t|x)=h_0(t)\text{exp}(x\beta)$, the equivalent relation for
	the survival curve is $S(t|x)=S_0(t)^{\text{exp}(x\beta)}$. Visual inspection
	requires identifying this exponential relationship. The log-minus-log
	transformation of the last equation gives
	$\log(-\log(S(t|x)))=\log(-\log(S_0(t)))+x\beta$, i.e., if the proportional
	hazards assumption holds, the curves of the treatment groups should be a
	constant distance apart.}
The curves in \autoref{fig:lncumhaz} appear parallel and provide no indication
that the proportional hazards assumption is violated. The same applies when the
cumulative hazard estimate is stratified by five-year birth cohorts as in the
analysis (\autoref{fig:lncumhazstrata}, Online Appendix).

Finally, another potential issue pertains to bias incurred by selective sampling
due to DI outflow. Previous benefit receipt is not registered in the data---only
current benefit receipt at the time of sampling is observed. If the reform
affected DI outflow as well, sampling may be biased, as those who were barred
from receiving insurance due to treatment are not observed in later years. This
may result in a selected sample with artificially lower inflow in treatment
regions. An actual outflow effect would be mistaken for an inflow effect due to
unobserved dropout. I test for such outflow effects using the stock data and the
results do not indicate that outflow is affected. The results for outflow are
discussed together with the robustness checks in section~\ref{sec:robustness}.

\subsection{Identification: Bounding the effect of medical review on award errors}%
\label{sec:idtype}

The assumptions outlined in section~\ref{sec:did} are sufficient to identify the
reduced form effect of introducing mandatory external medical review on the DI
inflow rate. This section explores additional conditions under which the
interpretation of the reduced form effect can be extended. The main estimation
results in the paper indicate a reduction in DI inflow. By layering two
additional assumption, this reduced form effect can be interpreted as a lower
bound of the effect on false positive award errors. Conditioning on individuals'
latent eligibility status, the total effect can be decomposed into a mixture of
effects on the false positive and false negative DI misclassification rates.

It is the main duty of the insurance office to separate meritorious from
non-meritorious claims \citep[`tag' the eligible,][]{Akerlof1978economics}. Given
the null hypothesis of `no disability', two types of classification errors can
occur in this situation: (1) Award errors (type-I, false positive) and (2)
Rejection errors (type-II, false negative). If medical review is imperfect,
benefits may be awarded to persons who are ineligible, and deserving applicants
may be denied benefits.

Hence, medical review may not reduce insurance inflow
unambiguously. Suppose that introducing mandatory medical review increases the
probability to detect applicants’ true type. This implies that medical review
can reduce \textit{both} type-I and type-II misclassification, resulting in
opposing effects on the incidence rate. The net effect on inflow is undetermined
and depends on the relative prevalence and likelihood of benefit receipt for
eligible and ineligible applicants.

To illustrate the relative prevalence of type-I and type-II errors,
decompose the relative average treatment effect in~\eqref{eq:att} by latent eligibility status
$E=\left\{0,1\right\}$,
\begin{gather}\label{eq:decomp}
	\text{rATT} = \dfrac{\splitdfrac{h^{1}(t|D=1,P=1,E=1)\cdot p^{1}(D=1,P=1,E=1)}{+h^{1}(t|D=1,P=1,E=0)\cdot\left[1-p^{1}(D=1,P=1,E=1)\right]}}
	{\splitdfrac{h^{0}(t|D=1,P=1,E=1)\cdot p^{0}(D=1,P=1,E=1)}{+h^{0}(t|D=1,P=1,E=0)\cdot\left[1-p^{0}(D=1,P=1,E=1)\right]}} \ .
\end{gather}
This underscores that the identified effect is a mixture of changes in the
hazard for both eligible and ineligible types. Using this expression, it is
possible to explore the conditions for a negative treatment effect---an
inflow reduction, corresponding to a hazard ratio smaller than one---depending on
the effect for each type separately. In the following, I simplify notation by omitting the parameters common to all objects in the conditioning set.

Unlike in other treatment effect settings, population shares in~\eqref{eq:decomp}
are superscripted by the corresponding counterfactual states. In this setting,
distinguishing them is sensible as they can be thought of as shares of
applications by eligibility types which might be influenced by the treatment.
Ruling this out to ease interpretation, assume that
\begin{gather}\label{eq:sscr}
	p^{0}(E=1) = p^{1}(E=1) \ .
	\stepcounter{equation}\tag{\textit{no self-screening}, \theequation}
\end{gather}
This assumption implies continuity in the composition of applications,
effectively ruling out that the propensity to apply for DI is influenced by the
pilot. The most likely mechanism to confound this assumption is self-screening,
i.e., individuals are selectively discouraged from applying for benefits
\citep{Parsons1991self}. For the reasons outlined in the previous section, this
behavior is unlikely since information about the pilot program did not transpire
to the public. \citeapos{Parsons1991self} original paper on the self-screening
mechanism is about how changes in screening stringency and administrative hassle
which are perfectly observed by applicants influence the application decision.
The medical review process is largely hidden to the applicants and there is no
information available to them detailing it. This view is also supported by the
data. Looking at the limited aggregate data available, application rates evolve
similarly across both groups of cantons, are very stable over time and do not
diverge during the pilot (Appendix \autoref{fig:diapp}). Due to the non-public
introduction of medical review and the common trend in applications,
differential variations in application behavior are likely to be negligible.

The estimation results in the next section indicate a reduction in inflow. This
in mind and simplifying notation due to (7), the hazard ratio must be smaller
than one,
\begin{gather}\label{eq:decomp2}
	\text{rATT} = \dfrac{h^{1}(t|E=1)\cdot p(E=1)+h^{1}(t|E=0)\cdot\left[1-p(E=1)\right]}
	{h^{0}(t|E=1)\cdot p(E=1)+h^{0}(t|E=0)\cdot\left[1-p(E=1)\right]} \leq 1 \ .
\end{gather}
Rearranging gives
\begin{gather}\label{eq:decomp3}
	\resizebox{.91\hsize}{!}{$\left[h^{1}(t|E=1)-h^{0}(t|E=1)\right] p(E=1)
		\leq -\left[h^{1}(t|E=0)-h^{0}(t|E=0)\right] \left[1-p(E=1)\right]$,}
\end{gather}
i.e., the absolute value of the population-weighted treatment effect for the
ineligible must exceed the population-weighted treatment effect for the
eligible to observe an aggregate reduction in inflow. This implies the
reduction in award errors (type-I, RHS) must exceed the reduction
in rejection errors (type-II, LHS) for the effect to be negative. This is
consistent with the interpretation of the effect in~\eqref{eq:att} as a net
effect.

Finally, assuming the treatment does not decrease inflow of eligible types,
\begin{gather}\label{eq:mtr}
	h^{1}(t|E=1)-h^{0}(t|E=1)\geq 0 \ ,
	\stepcounter{equation}\tag{\textit{monotone treatment response for eligible types}, \theequation}
\end{gather}
the left hand side of condition~\eqref{eq:decomp3} is greater or equal zero. If
medical review actually decreases the chances of the ineligible to get insurance
benefits, the weighted decrease in the hazard for the ineligible must be less in
absolute value than the weighted increase in the hazard for the eligible for the
condition to be fulfilled. In this case, any observed inflow reduction can be
interpreted as a net reduction in DI award errors.

This assumption is not directly testable with the available data. It relies on
the fact that medical review is an intervention to improve screening quality
and, unlike variations in screening stringency, does not involve a trade-off
between false positives and false negatives
\citep[][]{Parsons1991self,Kleven2011transfer,low2015}. Alternatively, the
condition in~\eqref{eq:decomp3} is trivially fulfilled if~(10) is violated and
medical review actually has the perverse effect of worsening the chances of the
truly eligible to get insurance, reducing their DI inflow hazard.

I will consider the consequences of violations of these assumptions and how they
can be relaxed in turn. Assumption (7) posits that medical review does not
change the composition of applications. This assumption could be weakened by
assuming that medical review decreases the propensity of ineligible types to
apply, i.e., $p^{1}(E=1)\geq p^{0}(E=1)$.%
\footnote{I am grateful to an anonymous referee for pointing out this
	possibility.}
This coincides with \citeapos{Parsons1991self} empirical result that
self-screening is non-perverse. The finding is also confirmed by
\citet{low2015}, who find that false applications decrease with program
stringency. Since medical review extracts information, those at the margin of
being discouraged from applying are those that are more likely to be found
undeserving. The lower bound interpretation can be retained with non-perverse
self-screening, because a larger share of the effect can be attributed to
medical review of the eligible (although the bound will be less informative).
Alternatively, perverse self-screening attributes a larger share of the effect
to ineligible types, the intended target population, complicating the
interpretation (but diminishing the importance of assumption 10). Assuming no
effect on self-screening is neutral with regard to composition and eases
interpretation.

Regarding the implications of assumption (10), there are four different
scenarios to consider (illustrated in the effect matrix in Appendix
\autoref{fig:effect-matrix}). For each eligibility type, medical review can
either be perverse (unintended) or non-perverse (intended). Consider first the
cases where screening is perverse for the ineligible, i.e., ineligible types are
accepted at higher rates under medical review. These cases can be dismissed.
First, if eligible types are accepted at higher rates as well (non-perverse), we
would not observe a reduction in inflow. Second, if eligible types would be
rejected at higher rates due to medical review, this perverse effect on the
eligible would have to exceed the perverse effect on the ineligible. This
scenario is implausible.

Consider the remaining two cases where medical review for the ineligible is
non-perverse, i.e., ineligible types are rejected more due to medical review as
intended. In this case, if (10) is violated and medical review is perverse and
reduces the chances of the truly eligible indiscriminately, the effect can still
be interpreted as an upper bound. Finally, if medical review has the intended
effect of being non-perverse and potentially also increases the chances of the
eligible to be allowed benefits, the effect can be interpreted as a lower bound.
This is the case implied by assumption (10).

The upper bound interpretation is the most likely case in which assumption (10)
would be violated. In this case, medical review has a perverse effect and
induces an even larger number of false negative errors. One scenario in which
this might occur is that if lower DI incidence is politically desired,
individual physicians might be pressured to be generally more critical when
reviewing new applications due to a fear of being laid off. However, the
additional staff at the DI offices were hired on permanent employment contracts
and could not have been easily laid off, irrespective of the development of the
insurance rolls. In addition, the scope of the federal government to influence
local public entities is limited due to the decentralized nature of the Swiss
political system. Hiring, medical review and the DI decision are made on the
local level, even though staff and DI benefits are paid out of federal funds.
The role of the federal government was limited to providing funding for the
program. It was generally recognized that the insurance offices' structure, last
revised 1973, needed to be overhauled and that they were insufficiently staffed
with physicians. The physicians responsible for medical review had the explicit
mandate to improve the accuracy of medical diagnoses of functional limitations.%
\footnote{The leading physician in one office was aware of the fact that more
	intense medical review could increase DI incidence. She stated that in her
	experience, rejection errors do occur and are sometimes encountered during
	revisions, but are much less frequent in relation to the amount of award errors
	uncovered ex post.}
As discussed, the institutional structure of the DI offices remains
unchanged---the final decision to grant benefits still lies with the DI
caseworker.

There is also some empirical support for assumption (10). If differential
changes in stringency where to occur due to medical review during the pilot,
these would most likely manifest in higher rates of legal claims regarding DI
entitlements. Comprehensive data on legal claims is sparse due to limited
reporting coverage, but I collected data on the amount of legal claims for each
canton from yearly reports of the cantonal courts. Both the number of total and
rejected lawsuits in treated and control regions evolve very similarly over time
(Appendix \autoref{fig:diclaims}). This suggests that it is unlikely that
differential changes in stringency occur during the pilot, supporting the lower
bound interpretation. Even if the upper bound interpretation applies, given the
institutional and empirical evidence it is likely that a non-negligible fraction
of the effect can be related to reductions in false positives.

In this section, I have outlined additional assumptions that are required to
bound the effect of medical review on award errors and have provided some
empirical and institutional support in favor of these assumptions. The
reduced-form estimate in sections~\ref{sec:did} does not permit inference about
targeting efficiency of the DI program. Given assumptions (7) and (10) in this
section, targeting efficiency improves if medical review reduces inflow.
However, this interpretation should be made with utmost care as it is highly
dependent on the assumptions that are made. Assuming no change in self-screening
as in (7) is neutral with regard to targeting, but directional self-screening
would not be. Importantly, constraining the direction of the effect for eligible
types as in (10) has direct implications for targeting efficiency. In case this
assumption is violated, and rejections are a mix of eligible and ineligible
types (the lower left quadrant in Appendix \autoref{fig:effect-matrix}), the
interpretation ceases to be valid. In this case, rejections of the ineligible
have to exceed rejections of the eligible for net targeting to improve, assuming
that the same weight is placed on false positive and false negative decision
errors, which might not be desirable.

%%%%%%%%%%%%%%%%%%%% RESULTS %%%%%%%%%%%%%%%%%%%%%%%%%%%%%%%%%%%%%%%%%%%%%%%%%%

\section{Results}
\label{sec:Results}

\subsection{Disability incidence and award errors}

The main results are presented in \autoref{tab:mainresults}, separately for the
unrestricted and the local sample. The first column for each sample considers
only spells which are censored or result in failure before the end of the pilot
period in 2005, the remaining columns use all recorded spells and control for
the post-treatment period in which the intervention was extended nationwide.
The last column adds individual control variables, including gender,
education, marital status, number of children and foreign citizenship. All
specifications stratify the baseline hazard by five year birth cohort intervals
to account for cohort specific differences in health environment. Survey
weights are applied in the full sample such that estimates are representative
of the Swiss population. Observations in the local sample are weighted for
pairwise nearest-neighbor estimation. All tables report hazard ratios, i.e.,
exponentiated coefficients and corresponding standard errors.

All estimates of the effect of the reform are negative (corresponding to a
hazard ratio less than one) and significant at conventional levels, indicating
that third-party medical review significantly reduced insurance inflow. The
estimate for the full sample implies a 14\% reduction. The magnitude for the
local sample is slightly higher and corresponds to a 23\% lower inflow rate.
Both estimates are stable in magnitude across specifications. The post
coefficient estimates are negative as well, reflecting the fact that the reform
was extended to the federal level after 2004 and funding increased even further.
However, the post estimates for the local sample are imprecise as the failure
density in the local sample is not dense enough in later years, when many
observations are censored at the sampling date.

The preferred specification for the remainder of the paper is given in column
(5), since adding covariates does not affect the results in a notable way. The
remaining analysis focuses on the local sample. Results for the main sample are
qualitatively similar.

\begin{table}
	\centering
	\caption{Disability incidence}
	\label{tab:mainresults}
	\begin{threeparttable}
		\scriptsize{
            \begin{tabular}{l*{3}{S[table-number-alignment=center,%
                                    table-figures-integer=1,%
                                    table-figures-decimal=3]}%
                                    p{3mm}%
                                    *{3}{S[table-number-alignment=center,%
                                    table-figures-integer=1,%
                                    table-figures-decimal=3]}}
\toprule \addlinespace[1em]
	 & \multicolumn{3}{c}{(a) Full sample} & & \multicolumn{3}{c}{(b) Local sample (within 20 km)} \\ \cmidrule(lr{.75em}){2-4} \cmidrule(lr{.75em}){6-8}
	 & \multicolumn{1}{c}{(1)}         & \multicolumn{1}{c}{(2)}                           &  \multicolumn{1}{c}{(3)}                           &  & \multicolumn{1}{c}{(4)}       & \multicolumn{1}{c}{(5)}       & \multicolumn{1}{c}{(6)}        \\
\midrule
Treat                           &       1.322***&       1.322***&       1.236***& &       1.150***&       1.151***&       1.148***\\
                                &     (0.041)   &     (0.041)   &     (0.039)   & &     (0.061)   &     (0.061)   &     (0.061)   \\
Pilot time                      &       1.083   &       1.088   &       1.110   & &       1.257*  &       1.267** &       1.298** \\
                                &     (0.089)   &     (0.089)   &     (0.090)   & &     (0.148)   &     (0.148)   &     (0.152)   \\
\rowcolor{gray!15}Treat x pilot &       0.856** &       0.856** &       0.860*  & &       0.770** &       0.771** &       0.766** \\
\rowcolor{gray!15}              &     (0.067)   &     (0.067)   &     (0.068)   & &     (0.087)   &     (0.087)   &     (0.086)   \\
Post time                       &               &       0.690***&       0.731***& &               &       0.867   &       0.918   \\
                                &               &     (0.068)   &     (0.072)   & &               &     (0.151)   &     (0.160)   \\
Treat x post                    &               &       0.971   &       0.970   & &               &       0.841   &       0.829   \\
                                &               &     (0.078)   &     (0.078)   & &               &     (0.105)   &     (0.104)   \\
\midrule
Other controls          & \multicolumn{1}{r}{-}         & \multicolumn{1}{r}{-}         & \multicolumn{1}{r}{\checkmark} &  & \multicolumn{1}{r}{-}         & \multicolumn{1}{r}{-}         & \multicolumn{1}{r}{\checkmark} \\
N municipalities       & \multicolumn{1}{r}{2,337}     & \multicolumn{1}{r}{2,338}     & \multicolumn{1}{r}{2,338}      &  & \multicolumn{1}{r}{1,086}     & \multicolumn{1}{r}{1,087}     & \multicolumn{1}{r}{1,087}      \\
N individuals           & \multicolumn{1}{r}{249,750}    & \multicolumn{1}{r}{259,323}    & \multicolumn{1}{r}{259,323}    & & \multicolumn{1}{r}{128,536}    & \multicolumn{1}{r}{133,549}    & \multicolumn{1}{r}{133,549}    \\
N failures              & \multicolumn{1}{r}{7,877}      & \multicolumn{1}{r}{9,204}      & \multicolumn{1}{r}{9,204}      & & \multicolumn{1}{r}{3,985}      & \multicolumn{1}{r}{4,693}      & \multicolumn{1}{r}{4,693}      \\
N failures during pilot & \multicolumn{1}{r}{1,713}      & \multicolumn{1}{r}{1,713}      & \multicolumn{1}{r}{1,713}      & & \multicolumn{1}{r}{885}       & \multicolumn{1}{r}{885}       & \multicolumn{1}{r}{885}       \\
\bottomrule
		\end{tabular}}
		\begin{tablenotes}[para,flushleft]
            \item Note: Cox Proportional Hazard estimates for individuals in
                treated and control regions based on SESAM individual-level
                survey and administrative data sampled during 1999--2011.
                Estimations separately for a complete representative sample of
                the Swiss population and only for individuals in the vicinity
                of the border between treated and non-treated regions.
                Baseline hazard for all regressions stratified by 5-year birth
                cohorts. Survey weights applied for the full sample.
                Observations in the local sample are weighted for
                nearest-neighbor pairwise differences. Results are reported in
                exponentiated form as hazard ratios. The hazard ratio for
                `Treat x pilot' corresponds to the relative average treatment
                effect on the treated as defined in \autoref{sec:Strategy}.
                Standard errors clustered at the individual level in
                parentheses, number of observations given below. *, ** and ***
                denote significance at the 10\%, 5\% and 1\% level
                respectively.
		\end{tablenotes}
	\end{threeparttable}
\end{table}

External medical review is also likely to affect the classification of the
severity of health impediments for new awards. I analyse whether medical review
changes the relative incidence of partial and full benefit awards. Results in
\autoref{tab:partialdi} show that incidence reductions occur only for full
benefit awards (columns 2 and 3) and those due to limitations classified as very
serious (disability degree of 70\% or larger, columns 4 and 5). Estimates for
partial benefit awards and those classified as less serious are too imprecisely
estimated to draw a clear conclusion, but may be unaffected. One possible
explanation is that incidence reductions occur mainly for full benefit
applicants. However, it is unlikely that only applicants claiming 100\% work
incapability constitute the affected marginal cases. A more likely scenario is
that DI incidence reductions occur at all latent health levels. After
introducing medical review, some individuals who would have received the full
benefit amount previously are now downgraded, resulting in a zero net effect for
partial DI benefits. This finding is also reflected by a moderate decrease in
the aggregate share of full benefit awards---in 2005, 58\% of new beneficiaries
are awarded full benefits compared to 68\% in 2002.

\begin{table}
    \centering
    \caption{Disability classification}
    \label{tab:partialdi}
    \begin{threeparttable}
        \scriptsize{
            \begin{tabular}{l*{5}{S[table-number-alignment=center,%
                                    table-figures-integer=1,%
                                    table-figures-decimal=3]}}
\toprule \addlinespace[1em]
& \multicolumn{1}{c}{All} & \multicolumn{1}{c}{Partial} & \multicolumn{1}{c}{Full} & \multicolumn{1}{c}{DD < 70} & \multicolumn{1}{c}{DD $\geq$ 70} \\
\cmidrule(lr{.75em}){2-2} \cmidrule(lr{.75em}){3-4} \cmidrule(lr{.75em}){5-6}
& \multicolumn{1}{c}{(1)} & \multicolumn{1}{c}{(2)}     & \multicolumn{1}{c}{(3)}  & \multicolumn{1}{c}{(4)}     & \multicolumn{1}{c}{(5)}          \\
\midrule
Treated region                  & 1.151*** & 1.071   & 1.169** & 1.043   & 1.219*** \\
                                & (0.061)  & (0.115) & (0.073) & (0.104) & (0.081)  \\
Pilot period                    & 1.267**  & 1.541** & 1.118   & 1.509** & 1.166    \\
                                & (0.148)  & (0.305) & (0.165) & (0.290) & (0.183)  \\
\rowcolor{gray!15}Treat x pilot & 0.771**  & 0.925   & 0.710** & 0.981   & 0.646*** \\
\rowcolor{gray!15}              & (0.087)  & (0.178) & (0.102) & (0.181) & (0.099)  \\
Post time                       & 0.867    & 1.446   & 0.584** & 1.423   & 0.633*   \\
                                & (0.151)  & (0.400) & (0.133) & (0.382) & (0.151)  \\
Treat x post                    & 0.841    & 0.722   & 1.003   & 0.717*  & 0.992    \\
                                & (0.105)  & (0.147) & (0.164) & (0.141) & (0.169)  \\
\midrule
N municipalities        & \multicolumn{1}{r}{1,087}   & \multicolumn{1}{r}{1,087}   & \multicolumn{1}{r}{1,087}   & \multicolumn{1}{r}{1,087}   & \multicolumn{1}{r}{1,087}   \\
N individuals           & \multicolumn{1}{r}{133,549} & \multicolumn{1}{r}{133,549} & \multicolumn{1}{r}{133,549} & \multicolumn{1}{r}{133,549} & \multicolumn{1}{r}{133,549} \\
N failures              & \multicolumn{1}{r}{4,693}   & \multicolumn{1}{r}{1,352}   & \multicolumn{1}{r}{3,283}   & \multicolumn{1}{r}{1,481}   & \multicolumn{1}{r}{2,879}   \\
N failures during pilot & \multicolumn{1}{r}{885}     & \multicolumn{1}{r}{338}     & \multicolumn{1}{r}{538}     & \multicolumn{1}{r}{357}     & \multicolumn{1}{r}{474}     \\
\bottomrule
        \end{tabular}}
        \begin{tablenotes}[para,flushleft]
            \item Note: Cox Proportional Hazard estimates for individuals in
                treated and control regions based on SESAM individual-level
                survey and administrative data sampled during 1999--2011.
                Sample is based on individuals living within 20 km of the
                border between treated and non-treated regions. Columns
                distinguish between partial/full DI benefit awards and awards
                due to less serious/serious health limitations (disability
                degree smaller/greater than 70).  Baseline hazard for all
                regressions stratified by 5-year birth cohorts.  Observations
                are weighted for nearest-neighbor pairwise differences. Results
                are reported in exponentiated form as hazard ratios. The hazard
                ratio for `Treat x pilot' corresponds to the relative average
                treatment effect on the treated as defined in
                \autoref{sec:Strategy}. Standard errors clustered at the
                individual level in parentheses, number of observations given
                below. *, ** and *** denote significance at the 10\%, 5\% and
                1\% level respectively.
        \end{tablenotes}
    \end{threeparttable}
\end{table}

\subsection{Incidence of difficult-to-diagnose conditions}

The main analysis indicates that DI awards declined substantially due to
external medical review, most likely due to a reduction in false positive
benefit awards. If the effect is driven by more accurate health and functional
capacity diagnoses, then incidence reductions are more likely to occur for
diseases which are difficult to diagnose and verify for treating physicians, the
first DI gatekeeper. The reduction will be most pronounced for illnesses which
are both difficult to diagnose and whose functional capacity implications are
more likely to be misjudged.

\begin{table}
    \centering
    \caption{Disability types}
    \label{tab:moreoutcomes}
    \begin{threeparttable}
        \scriptsize{
            \begin{tabular}{l*{7}{S[table-number-alignment=center,%
                                    table-figures-integer=1,%
                                    table-figures-decimal=3]}}
\toprule \addlinespace[1em]
                                 &                                &                                & \multicolumn{1}{c}{Illness:}   & \multicolumn{1}{c}{Illness:}  & \multicolumn{1}{c}{Illness:}   &                               & \multicolumn{1}{c}{Congenital/} \\ % & \multicolumn{1}{c}{Illness:}  & \multicolumn{1}{c}{}           & \multicolumn{1}{c}{Congenital/} \\
                                 & \multicolumn{1}{c}{All}        & \multicolumn{1}{c}{Illness}    & \multicolumn{1}{c}{Psych.}     & \multicolumn{1}{c}{Nerve}     & \multicolumn{1}{c}{MSC}        & \multicolumn{1}{c}{Accident}  & \multicolumn{1}{c}{Other}       \\ % & \multicolumn{1}{c}{Other}     & \multicolumn{1}{c}{Congenital} & \multicolumn{1}{c}{Other}       \\
\cmidrule(lr{.75em}){2-2} \cmidrule(lr{.75em}){3-6} \cmidrule(lr{.75em}){7-8}
                     & \multicolumn{1}{c}{(1)}        & \multicolumn{1}{c}{(2)}        & \multicolumn{1}{c}{(3)}        & \multicolumn{1}{c}{(4)}       & \multicolumn{1}{c}{(5)}        & \multicolumn{1}{c}{(6)}       & \multicolumn{1}{c}{(7)}         \\ % & \multicolumn{1}{c}{(6)}       & \multicolumn{1}{c}{(8)}        & \multicolumn{1}{c}{(8)}         \\
\midrule
Treatment region                & 1.151*** & 1.229***        & 1.185*        & 1.100        & 1.245**      & 0.843       & 1.293**             \\ % & 1.322*** & 1.326***       & 1.159***     & 0.982       & 1.430***    & 1.742***     & 1.213**    & 1.006         & 1.630***           & 1.379**       & 1.007
                                & (0.061)  & (0.072)         & (0.106)       & (0.216)      & (0.136)      & (0.148)     & (0.162)             \\ % & (0.041)  & (0.047)        & (0.062)      & (0.117)     & (0.090)     & (0.142)      & (0.116)    & (0.186)       & (0.122)            & (0.181)       & (0.331)
Pilot period                    & 1.267**  & 1.384**         & 1.450*        & 2.373*       & 1.412        & 0.900       & 0.795               \\ % & 1.088    & 1.159          & 1.096        & 1.020       & 1.241       & 1.228        & 0.658*     & 2.398*        & 1.308*             & 0.846         & 0.557
                                & (0.148)  & (0.178)         & (0.282)       & (1.185)      & (0.330)      & (0.362)     & (0.201)             \\ % & (0.089)  & (0.105)        & (0.115)      & (0.241)     & (0.247)     & (0.216)      & (0.155)    & (1.159)       & (0.213)            & (0.229)       & (0.365)
\rowcolor{gray!15}Treat x pilot & 0.771**  & 0.683***        & 0.699*        & 0.377**      & 0.633**      & 1.729       & 1.150               \\ % & 0.856**  & 0.862*         & 1.089        & 0.993       & 0.781       & 0.571***     & 0.721      & 0.443*        & 0.562***           & 1.060         & 1.748
\rowcolor{gray!15}              & (0.087)  & (0.084)         & (0.129)       & (0.167)      & (0.145)      & (0.656)     & (0.290)             \\ % & (0.067)  & (0.075)        & (0.117)      & (0.240)     & (0.147)     & (0.104)      & (0.173)    & (0.218)       & (0.095)            & (0.283)       & (1.183)
Post time                       & 0.867    & 0.974           & 0.667         & 1.737        & 1.285        & 0.175***    & 1.220               \\ % & 0.690*** & 0.731***       & 0.639***     & 0.613       & 0.687*      & 1.106        & 0.227***   & 1.127         & 1.109              & 1.172         & 2.094
                                & (0.151)  & (0.183)         & (0.188)       & (1.211)      & (0.460)      & (0.102)     & (0.441)             \\ % & (0.068)  & (0.076)        & (0.100)      & (0.199)     & (0.138)     & (0.243)      & (0.077)    & (0.843)       & (0.233)            & (0.448)       & (1.645)
Treat x post                    & 0.841    & 0.733**         & 0.897         & 0.607        & 0.596**      & 6.436***    & 0.748               \\ % & 0.971    & 0.990          & 1.165        & 1.271       & 0.901       & 0.748*       & 2.110**    & 0.773         & 0.776*             & 0.696         & 1.154
                                & (0.105)  & (0.097)         & (0.176)       & (0.272)      & (0.156)      & (2.942)     & (0.197)             \\ % & (0.078)  & (0.075)        & (0.142)      & (0.304)     & (0.136)     & (0.114)      & (0.655)    & (0.449)       & (0.114)            & (0.190)       & (0.916)
\midrule
N municipalities         & \multicolumn{1}{r}{1,087}     & \multicolumn{1}{r}{1,087}     & \multicolumn{1}{r}{1,087}     & \multicolumn{1}{r}{1,087}    & \multicolumn{1}{r}{1,087}     & \multicolumn{1}{r}{1,087}     & \multicolumn{1}{r}{1,087}    \\ %                            & \multicolumn{1}{r}{1,087}     & \multicolumn{1}{r}{1,087}     & \multicolumn{1}{r}{1,087}     \\
N individuals            & \multicolumn{1}{r}{133,549}   & \multicolumn{1}{r}{133,549}   & \multicolumn{1}{r}{133,549}   & \multicolumn{1}{r}{133,549}   & \multicolumn{1}{r}{133,549}   & \multicolumn{1}{r}{133,549}   & \multicolumn{1}{r}{133,549}   \\ % & \multicolumn{1}{r}{259,323}   & \multicolumn{1}{r}{259,323}   & \multicolumn{1}{r}{259,323}   & \multicolumn{1}{r}{259,323}  & \multicolumn{1}{r}{259,323}   & \multicolumn{1}{r}{259,323}   & \multicolumn{1}{r}{259,323}  & \multicolumn{1}{r}{259,323}  & \multicolumn{1}{r}{259,323}                 & \multicolumn{1}{r}{133,549}     & \multicolumn{1}{r}{133,549}
N failures               & \multicolumn{1}{r}{4,693}     & \multicolumn{1}{r}{3,827}     & \multicolumn{1}{r}{1,685}     & \multicolumn{1}{r}{339}       & \multicolumn{1}{r}{1,090}     & \multicolumn{1}{r}{409}       & \multicolumn{1}{r}{835}       \\ % & \multicolumn{1}{r}{9,204}     & \multicolumn{1}{r}{7,445}     & \multicolumn{1}{r}{3,141}     & \multicolumn{1}{r}{663}      & \multicolumn{1}{r}{2,165}     & \multicolumn{1}{r}{1,476}     & \multicolumn{1}{r}{847}      & \multicolumn{1}{r}{225}      & \multicolumn{1}{r}{1,701}                   & \multicolumn{1}{r}{713}         & \multicolumn{1}{r}{122}
N failures during pilot  & \multicolumn{1}{r}{885}       & \multicolumn{1}{r}{753}       & \multicolumn{1}{r}{352}       & \multicolumn{1}{r}{61}        & \multicolumn{1}{r}{210}       & \multicolumn{1}{r}{59}        & \multicolumn{1}{r}{149}       \\ % & \multicolumn{1}{r}{1,713}     & \multicolumn{1}{r}{1,468}     & \multicolumn{1}{r}{651}       & \multicolumn{1}{r}{125}      & \multicolumn{1}{r}{439}       & \multicolumn{1}{r}{253}       & \multicolumn{1}{r}{107}      & \multicolumn{1}{r}{32}       & \multicolumn{1}{r}{285}                     & \multicolumn{1}{r}{130}         & \multicolumn{1}{r}{19}
\bottomrule
        \end{tabular}}
		\begin{tablenotes}[para,flushleft]
            \item Note: Cox Proportional Hazard estimates for individuals in
                treated and control regions based on SESAM individual-level
                survey and administrative data sampled during 1999--2011.
                Sample is based on individuals living within 20 km of the
                border between treated and non-treated regions. Columns
                distinguish between DI awards due to different health
                impairments. Baseline hazard for all regressions stratified by
                5-year birth cohorts.  Observations are weighted for
                nearest-neighbor pairwise differences.  Results are reported in
                exponentiated form as hazard ratios. The hazard ratio for
                `Treat x pilot' corresponds to the relative average treatment
                effect on the treated as defined in \autoref{sec:Strategy}.
                Standard errors clustered at the municipality level in
                parentheses, number of observations given below. *, ** and ***
                denote significance at the 10\%, 5\% and 1\% level
                respectively.
        \end{tablenotes}
    \end{threeparttable}
\end{table}

\autoref{tab:moreoutcomes} investigates this by differentiating between health
impairments leading to benefit awards. The results confirm that reductions occur
most frequently for difficult-to-diagnose conditions, while conditions which can
typically be diagnosed unambiguously are not affected. Looking at column (3) and
(4), the effect is pronounced for psychological diseases and illnesses related
to nerve problems. Benefit awards due to mental health problems are reduced by
30\%. Nerve-related handicaps are reduced by over 60\%, but incidence in this
group is generally very low. Column (5) looks at the incidence of
musculoskeletal conditions (MSC). This category also includes a variety of
conditions which are difficult to verify (e.g.\ whiplash injuries, back pain).
The hazard ratio suggest a substantial reduction in incidence as well. The
specification in column (6) looks at disability benefit awards due to handicaps
incurred in accidents; the last column considers disabilities due to congenital
defects and other diseases. These conditions are unlikely to be subject to award
errors, as there is rarely any ambiguity and they are typically well-documented.
Indeed, there is no effect on conditions which are unaffected by intensified
medical review.

\subsection{Labor market responses to medical review}

This section investigates the labor market reaction in response to external
medical review. In case reductions in DI incidence are driven by rejections of
individuals capable of returning to the labor market, medical review should also
have a positive effect on labor market participation. Conversely, if the
reduction is largely driven by rejections of individuals incapable of working,
medical review should not have an effect on employment, but possibly on
the inflow into other social security programs
\citep[e.g.][]{Inderbitzin2016}. \autoref{tab:labormarket-mid} uses the
pooled cross-sectional administrative SESAM data to estimate a
differences-in-differences specification using a linear model.

The results in \autoref{tab:labormarket-mid} for the full sample show that the
share of individuals in registered employment increases. Similarly, the share of
individuals with positive (non-benefit) earnings increases as well. In addition,
the share of individuals registered with the employment office as job seekers
also decreases (columns 1--3). In columns (4) and (5), I consider other pathways
from unemployment and reasons for not being registered with the employment
office anymore. I find no effect on dismissal from the employment office (and
the associated return-to-work measures) due to exhausting unemployment the
maximum duration for unemployment benefits. Similarly, I find no effect on the
receipt of social assistance, the minimum social security provision. If rejected
DI applicants were incapable of working, we would expect to see an increase in
these measures. However, the results do not provide evidence for this channel.
The results for the local sample are comparable in sign and magnitude to the
estimates for the full sample. However, they are insignificant, most likely due
to a lack of power ($p = 0.17$ for the main employment estimate in the local
sample).

An explanation for these results is that DI applications are partly made by
people capable of gainful employment and driven by moral hazard. One possible
mechanism behind this result is the canonical substitution effect
interpretation---applicants seek benefits due to a distortion in the relative
price of leisure. This distortion is caused by an implicit tax on work due to DI
('cash cliffs'). An alternative explanation is that applications are (partly)
due to income effects, i.e., even if work is not implicitly taxed by the DI
program, given the transfer payments, beneficiaries may prefer leisure to labor
\citep[e.g.][]{Autor2007distinguishing,Eugster2017,gelber2017effect}. These
effects have different welfare implications. If DI reduces labor supply through
the substitution effect this implies a deadweight loss, which would be reduced
by medical review. Alternatively, medical review would not be welfare improving
if all of the labor supply increase is due to a reduced income effect. Since DI
is provided (partially) contingent on work, I am unable to separate these
effects. Taken together, the evidence from the analysis suggests that distorted
incentives are likely to matter in this context.

\begin{table}
	\centering
	\caption{Labor market responses to medical review}
  \label{tab:labormarket-mid}
	\scriptsize
	\begin{threeparttable}
		\begin{tabular}{l*{5}{S[table-number-alignment=center,%
                            table-figures-integer=1,%
                        table-figures-decimal=3]}}%
\toprule
\addlinespace[1em]
	 & \multicolumn{5}{c}{(a) Full sample}\\
\addlinespace[1em]
                      % & {Work}       & {Positive}      & \multicolumn{2}{c}{Employment office} & {Social}                           \\
                      % & {registered} & {labor income}  & {registration}                        & {dismissal}        & {assistance}  \\
 & {}        & {}                & \multicolumn{2}{c}{Employment office} & {}                   \\
 & {Working} & {Positive income} & {registration}                        & {dismissal} & {Social assistance} \\
\cmidrule(lr{.75em}){2-6}
                      & {(1)} & {(2)}  & {(3)} & {(4)}        & {(5)}  \\
\cmidrule(lr{.75em}){2-6}
Treat x pilot&       0.009***&       0.008** &      -0.007***&      -0.002   &      -0.000   \\
            &     (0.004)   &     (0.003)   &     (0.002)   &     (0.001)   &     (0.001)   \\
\cmidrule(lr{.75em}){2-6}
Individual covariates & {\checkmark} & {\checkmark}    & {\checkmark}                          & {\checkmark}       & {\checkmark}  \\
Canton FE             & {\checkmark} & {\checkmark}    & {\checkmark}                          & {\checkmark}       & {\checkmark}  \\
Year FE               & {\checkmark} & {\checkmark}    & {\checkmark}                          & {\checkmark}       & {\checkmark}  \\
N                     & {556,540}    & {557,270}       & {411,461}                             & {411,461}          & {411,461}     \\
\cmidrule(lr{.75em}){2-6}
\addlinespace[1em]
\addlinespace[1em]
                      & \multicolumn{5}{c}{(b) Local sample (within 20 km)}                                                         \\
\addlinespace[1em]
 & {}        & {}                & \multicolumn{2}{c}{Employment office} & {}                   \\
 & {Working} & {Positive income} & {registration}                        & {dismissal} & {Social assistance} \\
\cmidrule(lr{.75em}){2-6}
                      & {(1)} & {(2)}  & {(3)} & {(4)}        & {(5)}  \\
\cmidrule(lr{.75em}){2-6}
Treat x pilot&       0.007   &       0.006   &      -0.003   &       0.000   &       0.001   \\
            &     (0.005)   &     (0.005)   &     (0.003)   &     (0.002)   &     (0.002)   \\
\cmidrule(lr{.75em}){2-6}
Individual covariates & {\checkmark} & {\checkmark}    & {\checkmark}                          & {\checkmark}       & {\checkmark}  \\
Canton FE             & {\checkmark} & {\checkmark}    & {\checkmark}                          & {\checkmark}       & {\checkmark}  \\
Year FE               & {\checkmark} & {\checkmark}    & {\checkmark}                          & {\checkmark}       & {\checkmark}  \\
N                     & {282,858}    & {283,111}       & {208,340}                             & {208,340}          & {208,340}     \\
\bottomrule
		\end{tabular}
	\begin{tablenotes}[para,flushleft]
            \item Note: Linear model estimates for individuals in treated and
control regions based on SESAM individual-level survey and administrative data
sampled during 1999--2011. Estimations separately for a complete representative
sample of the Swiss population (panel a) and only for individuals in the
vicinity of the border between treated and non-treated regions (panel b). All
models include cantonal and year specific effects and control for gender, age
and native status. Standard errors clustered at the municipality level given in
parentheses. *, ** and *** denote significance at the 10\%, 5\% and 1\% level
respectively.
	\end{tablenotes}
	\end{threeparttable}
\end{table}

\subsection{Disability degree and benefit revisions in the recipient stock}

Although the primary task of the medical staff is to screen applicants, they
also aid with reviews of recipients' disability degree classification. While
scheduled by law to occur regularly, revisions seldom resulted in actual
disability degree or benefit cuts and typically involved DI caseworkers going
over beneficiaries files without personal contact. Revisions also commonly take
place if applicants have submitted new medical information, typically
documenting deteriorating health, and often result in benefit increases. With
the new regime in place, files that are scheduled for review are now also passed
to the DI physicians in charge of medical review.

% version w/ constant
\begin{table}
    \centering
    \caption{Stock reclassification and pension cuts}
    \label{tab:stockclass}
    \begin{threeparttable}
        \scriptsize{
            \begin{tabular}{l*{6}{S[table-number-alignment=center,%
                                    table-figures-integer=4,%
                                    table-figures-decimal=2,
                                    round-precision=2]}}
\toprule \addlinespace[1em]
                                    & \multicolumn{6}{c}{(a) Disability degree} \\ [1ex]
                                    & \multicolumn{1}{c}{All}                &  \multicolumn{1}{c}{All illnesses}        & \multicolumn{1}{c}{Psychological}  & \multicolumn{1}{c}{MSC}            & \multicolumn{1}{c}{Accident}       & \multicolumn{1}{c}{Congenital}     \\  %  & \multicolumn{1}{c}{Other}
    \cmidrule(lr{.75em}){2-7}
Treated region      & 3.043***                       & 3.408***                          & 3.661***                          & 2.775***                          & 1.666***                          & 1.186***                          \\  %   & 3.597***                          & 3.000***
                    & (0.079)                        & (0.092)                           & (0.129)                           & (0.175)                           & (0.258)                           & (0.177)                           \\  %   & (0.174)                           & (0.132)
Pilot               & 0.485***                       & 0.604***                          & 0.602***                          & 0.388***                          & 0.457***                          & 0.295**                           \\  %   & 0.342***                          & 0.316***
                    & (0.056)                        & (0.065)                           & (0.092)                           & (0.126)                           & (0.171)                           & (0.127)                           \\  %   & (0.120)                           & (0.092)
\rowcolor{gray!15}Treat x pilot       & -0.350***                      & -0.420***                         & -0.584***                         & -0.389*                           & -0.243                            & -0.102                            \\  %   & -0.405**                          & -0.305**
\rowcolor{gray!15}                    & (0.091)                        & (0.105)                           & (0.146)                           & (0.201)                           & (0.296)                           & (0.205)                           \\  %   & (0.202)                           & (0.153)
Post                & 1.629***                       & 1.803***                          & 1.442***                          & 0.866***                          & 0.894***                          & 1.390***                          \\  %   & 1.324***                          & 2.019***
                    & (0.053)                        & (0.062)                           & (0.086)                           & (0.120)                           & (0.161)                           & (0.118)                           \\  %   & (0.115)                           & (0.087)
Treat x post        & -0.521***                      & -0.607***                         & -0.827***                         & -0.466**                          & -0.393                            & -0.282                            \\  %   & -0.750***                         & -0.743***
                    & (0.086)                        & (0.100)                           & (0.137)                           & (0.191)                           & (0.278)                           & (0.191)                           \\  %   & (0.193)                           & (0.144)
Constant            & 78.544***                      & 77.977***                         & 82.749***                         & 72.306***                         & 74.526***                         & 86.074***                         \\  %   & 77.026***                         & 79.951***
                    & (0.049)                        & (0.057)                           & (0.081)                           & (0.110)                           & (0.149)                           & (0.110)                           \\  %   & (0.104)                           & (0.080)
    \cmidrule(lr{.75em}){2-7}
    \addlinespace[1em]
                                    & \multicolumn{6}{c}{(b) Pension amount} \\ [1ex]
                                    & \multicolumn{1}{c}{All}                &  \multicolumn{1}{c}{All illnesses}        & \multicolumn{1}{c}{Psychological}  & \multicolumn{1}{c}{MSC}            & \multicolumn{1}{c}{Accident}       & \multicolumn{1}{c}{Congenital}     \\  %  & \multicolumn{1}{c}{Other}
    \cmidrule(lr{.75em}){2-7}
Treated region & 124.681***                     & 141.873***                     & 101.224***                        & 133.262***                        & 88.413***                         & 8.355***                          \\  %   & 209.994***                        & 143.719***
               & (2.196)                        & (2.607)                        & (3.672)                           & (4.942)                           & (7.249)                           & (3.143)                           \\  %   & (5.322)                           & (3.620)
Pilot          & 37.036***                      & 39.922***                      & 33.124***                         & 39.023***                         & 33.292***                         & 28.431***                         \\  %   & 40.563***                         & 36.469***
               & (1.546)                        & (1.848)                        & (2.622)                           & (3.559)                           & (4.827)                           & (2.252)                           \\  %   & (3.667)                           & (2.528)
\rowcolor{gray!15}Treat x pilot  & -17.251***                     & -21.765***                     & -17.792***                        & -21.899***                        & -8.100                            & -0.454                            \\  %   & -20.853***                        & -13.787***
\rowcolor{gray!15}               & (2.518)                        & (2.983)                        & (4.174)                           & (5.664)                           & (8.330)                           & (3.637)                           \\  %   & (6.153)                           & (4.186)
Post           & 143.120***                     & 148.366***                     & 126.958***                        & 139.818***                        & 122.499***                        & 126.264***                        \\  %   & 154.250***                        & 155.089***
               & (1.458)                        & (1.746)                        & (2.462)                           & (3.375)                           & (4.535)                           & (2.097)                           \\  %   & (3.504)                           & (2.391)
Treat x post   & -41.252***                     & -48.739***                     & -33.495***                        & -50.512***                        & -19.171**                         & -1.606                            \\  %   & -49.914***                        & -45.034***
               & (2.375)                        & (2.819)                        & (3.923)                           & (5.375)                           & (7.836)                           & (3.385)                           \\  %   & (5.875)                           & (3.956)
Constant       & 1232.084***                    & 1221.014***                    & 1311.780***                       & 1134.607***                       & 1199.859***                       & 1343.853***                       \\  %   & 1182.156***                       & 1234.422***
               & (1.346)                        & (1.611)                        & (2.301)                           & (3.102)                           & (4.197)                           & (1.946)                           \\  %   & (3.170)                           & (2.185)
\midrule                                                                                                                                                                                                                           %
N            & \multicolumn{1}{r}{2,489,323}  & \multicolumn{1}{r}{1,884,876}  & \multicolumn{1}{r}{887,604}       & \multicolumn{1}{r}{537,191}       & \multicolumn{1}{r}{282,224}       & \multicolumn{1}{r}{274,918}       \\  %   & \multicolumn{1}{r}{460,081}       & \multicolumn{1}{r}{734,999}
\bottomrule
        \end{tabular}}
		\begin{tablenotes}[para,flushleft]
            \item Note: Estimates from a linear model. Outcomes are the
disability degree in percent (panel a) and the effective benefit amount paid to
recipients in panel (b). The reference group are individuals in the non-treated
regions in 2001. Based on administrative panel data provided by the Swiss
Federal Ministry of Social insurances which tracks the complete stock of Swiss
DI benefit recipients in 2001 until 2011. Standard errors in parentheses, number
of observations given below. *, ** and *** denote significance at the 10\%, 5\%
and 1\% level respectively.
		\end{tablenotes}
    \end{threeparttable}
\end{table}

To assess whether stock reclassifications occur, I estimate a linear
difference-in-difference model using data for the stock of all DI beneficiaries
in Switzerland in 2001. I condition on benefit receipt prior to treatment and
track the changes to the disability degree and the effective benefit payments of
existing beneficiaries over time. Results are given in \autoref{tab:stockclass}.
The sample is again stratified by disease groups. The outcome in Panel (a) is
the individual disability degree, Panel (b) looks at the benefit amount. On
average, recipients are classified less disabled by 0.35 percentage points and
lose about 17 CHF in monthly benefits. The effect magnitudes are small since
reclassification remains a rare event. Summary statistics indicate that only
9.3\% of individuals of the 2001 stock are reclassified during the three years
of the pilot period. Complete denial of benefits after a revision occurs only in
exceptional cases.%
\footnote{Complete benefit denial is legally difficult, unless fraud or
	malingering are proven beyond reasonable doubt. These cases also require high
	up-front investment from DI offices and are initiated only in extreme cases.}
Upward revisions are far more common, downward changes only account for 2.3
percentage points. Still, introducing mandatory medical review appears to cause
revisions of the disability status of beneficiaries whose documentation is
deemed insufficient, suspicious or whose health has improved. Both the
disability classification and payouts are again only adjusted for those
beneficiaries with illnesses which are more difficult to screen. Again, cuts are
most pronounced for those who receive DI due to mental health problems or
musculoskeletal conditions, while beneficiaries with congenital diseases or
handicaps incurred in accidents are unaffected. Unlike previously, nerve-related
diseases are not declared in this data.

\subsection{Robustness checks}%
\label{sec:robustness}

To assess the validity of the main identifying assumption, I test the effect of
a placebo reform prior to the treatment period and assume a pseudo-treatment to
be effective during 1999--2001. Hazard ratio estimates across all specifications
are close to one, precisely estimated and insignificant at conventional levels,
supporting the validity of the identification strategy (Appendix
\autoref{tab:robustnessplacebo}). Placebo results for employment also do not
indicate any violation of the common trend assumption
(\autoref{tab:placebo-work}, Online Appendix).

Another potential concern is that the results are sensitive to the choice of
distance window. \autoref{fig:distances} addresses this issue by plotting
treatment effect estimates across a large set of bandwidths, using both actual
travel distance and travel time as distance measures. The coefficient of
interest remains stable in size and significant across a large set of
distances. The estimates consistently suggest at least a 20\% reduction in
incidence in the treatment group during the pilot program. More detailed
estimates over selected distances are provided in Appendix \autoref{tab:robustnessdist}.

As discussed in section~\ref{sec:threats}, potential outflow effects of the reform
might confound the main result. Since previous benefit receipt is unobserved,
outflow effects would lead to inflow being measured with error in the sample. I
use the stock data to test for outflow effects. A duration model similar to
the main specification is estimated for those who are beneficiaries prior to
treatment in 2001. Exit from the DI rolls is considered failure, individuals are
censored at the sampling limit in 2011 or when they exit at the relevant pension
age. Variable measurements are less clean-cut in this case. Exit due to work or
expulsion cannot be separated. However, there is no explicit reason why trends
in work take-up by insurees (a similarly rare event) should differ between
regions. Results are given in Appendix \autoref{tab:outflow}, separately for all
individuals and those below age 50 in 2001, an age requirement which prohibits
early retirement within the analysis horizon and selects a younger and possibly
healthier group more likely to exit. All estimates are consistently
indistinguishable from zero and precisely estimated.

Another concern is that external medical review might simply prolong the
decision process and delay benefit approval. Note that the DI entry measurement
effectively precludes this possibility. Entry is observed for those who
effectively enter the insurance system at the time when they register with the
insurance office and file their application, not when they are finally granted
benefits. This is due to the fact that benefits are paid retroactively from the
filing date after approval.

As illustrated, the main results are robust to a series of checks and very
stable in magnitude. The results for both samples are also robust to model
changes. Replacing the main model using treatment group and pilot period
interactions with a model including canton and time fixed effects yields very
similar treatment effect estimates (\autoref{tab:mainresults-with-fe},
Online Appendix). Similarly, the results are not dependent on the application of
weights, stratification or the stratification level. Moreover, even though the
identifying assumptions are different and the analysis is underpowered, relative
effects obtained from a linear probability model for inflow in the full sample
are similar in magnitude to the relative average treatment effect estimates in
\autoref{tab:mainresults} (\autoref{tab:mainresults-lpm}, Online Appendix).

Equally persistent through variations is the approximately 7\% difference in
magnitude between estimates for the global and the local sample. It is
illustrative to trace where the difference in results stems from. To shed light
on the differences between the local and the full sample, I estimate a Probit
model for the probability to be included in the local sample, separately for
treated and control regions. Appendix \autoref{tab:probitrepresentative}
presents the results. The local treated sample closely resembles the rest of the
treated region. However, the local control sample differs from the rest of the
control population. It has a higher share of foreigners (about 10\% at the
mean), more women and more well-educated individuals---all factors which
contribute to a lower overall incidence and are likely to drive the difference
in results.

%%%%%%%%%%%%%%%%%%%% DISCUSSION & CONCLUSION %%%%%%%%%%%%%%%%%%%%%%%%%%%%%%%%%%

\section{Conclusion}
\label{sec:Discussion}

This paper provides a comprehensive evaluation of the introduction of
medical review for DI applications in a setting in which treating physician
testimony is decisive. The results indicate that external medical review can
reduce insurance inflow substantially. The main estimate suggests that medical
review reduces DI uptake by 23\%. Reductions are closely tied to
difficult-to-diagnose conditions, suggesting a more accurate assessment of
complex or multidisciplinary diseases. This is corroborated by the fact that
disability status and benefit revisions in the stock of recipients occur only
for individuals with the same types of conditions and the fact that medical
review also increases labor market participation. Under additional assumptions,
the results suggest that medical review is likely to reduce the amount of false
positive award errors and that these errors occur frequently in the absence of
medical review.

Results from the local approach (sample restricted to commuting distance around
borders) have the same sign and are comparable in magnitude to the global
approach (using the full sample). The distance variations in
\autoref{fig:distances} consistently suggest a reduction in the hazard of about
20\%. Considering the sizeable effect of medical review on the DI hazard, it is
illustrative to assess how large the absolute effects induced by introduction of
the medical reviews are. Looking at the main specification, without treatment,
the baseline DI hazard in the treated regions is about 0.38\%, i.e., on average
3.8 persons per thousand enter DI.\@ The medical review process reduces this by
about 23\% to 0.29\%, implying that approximately one person less per thousand
enters DI due to a second medical assessment.

Given the substantial present-discounted value of DI benefits, it is
interesting to examine whether external medical review is a cost-effective
policy. Simple back-of-the-envelope calculations indicate that outlays for
hiring physicians are more than offset by reductions in the beneficiary payload.
The calculations are based on the observed increase in the number of physicians,
a conservative effect estimate and the average benefit amount and remaining
spell duration until retirement, assuming rejections are permanent. Based on
these parameters, the yearly savings only in the treated regions during the
pilot are likely to be above 650 million Swiss Francs (approximately 650 million
US\$ in 2018). Extending medical review nationwide in 2005 may have saved in excess of
1.2 billion Swiss Francs in that year alone. Even if all rejected applicants
never reenter the labor market and immediately receive social assistance,
estimated yearly savings for 2005 are upwards of 500 million Swiss Francs. These
calculations disregard the fact that benefit decisions are tied to additional
occupational benefits and private pension schemes, which are substantially more
generous than the main state DI benefits and would result in further savings.
Nevertheless, the yearly savings far exceed potential outlays for the medical
personnel that was hired. Introducing external medical review is a highly
cost-effective tool to reduce insurance inflow.

Taken together, the results cast doubt on the practice to assign a large weight
to the treating physician's opinion in DI insurance decisions. Considering that
inflow reductions are restricted to difficult-to-diagnose conditions and the
results indicate that work take-up increases when medical review is done by
clinical specialists, treating physicians may not be well-suited to serve as the
main gatekeeper to DI.\@ This result corroborates medical studies which posit
that specialists may be better suited to judge social insurance eligibility than
personal physicians
\citep[e.g.][]{Novack1989physicians,Zinn1996physician,Freeman1999lying,%
	Wynia2000physician,Everett2011lie}. In addition, treating physicians have
often voiced discomfort with being both care-takers of patients and gatekeepers
to public insurance systems. In surveys, physicians are overwhelmingly in favor
of designating independent third-party physicians to determine disability status
to prevent damaging physician-patient relations
\citep[e.g.][]{Zinn1996physician}.

Since external medical review by DI physicians appears to be effective in the
Swiss setting, it might provide a viable policy option for other countries which
are burdened by high disability insurance costs and rely on treating physician
assessments for DI.\@ However, it is important to bear in mind that prior to the
reform, medical review was conducted almost exclusively by treating physicians
and DI physicians could not examine patients. Both the policy impact and the
size of award errors are likely to depend on the initial level of screening
intensity. Still, treating physician testimony is influential for DI
determinations in many OECD countries. The results suggest that subjecting
treating physcicians' opinions to medical review by a third party is a
cost-effective policy to regulate inflow and award errors. Since the policy also
lifted bans on personal medical examinations, the changes in Switzerland can
potentially also provide some insight about extending medical review in systems
which exclusively rely on file-based review.

It is important to note that screening during the pilot does not necessarily
come at the cost of increased program complexity \citep[e.g.\ as modeled
	by][]{Kleven2011transfer}. The additional administrative hassle is low, and
there are few visible additional up-front costs borne by the applicant. As such,
external medical review is unlikely to discourage take-up strongly in the
long-term. This situation might differ if medical review is announced publicly.
Since medical review extracts information, it may also discourage ineligible
applicants from applying for benefits, as they have higher chances to be
ultimately denied. This deterrence effect is found to be pronounced by
\citet[][]{low2015}.

The mechanisms behind the results in this paper merit further investigation. One
possible channel behind the incidence reductions are inaccurate diagnoses by
treating physicians, the first gatekeeper to the DI system. However, whether and
how much application behavior suffers from moral hazard remains ultimately
unclear. Applicants could be largely myopic or actively engage in malingering.
Still, the overall reduction in inflow provides a tentative suggestion that
award errors exceed rejection errors in award decisions. This result diverges
from previous analyses for the US.\@ However, given that benefits are
substantially more generous in Switzerland, this finding is in line with
\citeapos{low2015} result that false applications are strongly increasing with
benefit generosity. Hence, the result is also a first indication that the
relative prevalence of errors may be different in European DI systems which
offer higher replacement rates. Separating type-I and type-II classification
errors more cleanly and examining the mechanisms through which they occur
remains a promising pursuit for further research.

%%%%%%%%%%%%%%%%%%%% References %%%%%%%%%%%%%%%%%%%%%%%%%%%%%%%%%%%%%%%%%%%%%%%

\clearpage
\bibliographystyle{custom_natbib}
\bibliography{submission-JHE_2018_356}

\begin{thebibliography}{61}
\providecommand{\natexlab}[1]{#1}
\providecommand{\url}[1]{\texttt{#1}}
\providecommand{\urlprefix}{}
\providecommand{\eprint}[2][]{\url{#2}}

\bibitem[{Adam et~al.(2010)Adam, Bozio and Emmerson}]{Adam2010reforming}
Adam, S., Bozio, A. and Emmerson, C. (2010).
\newblock Reforming disability insurance in the uk: Evaluation of the pathways
  to work programme.
\newblock \emph{Working paper}, Insitute for Fiscal Studies, London.

\bibitem[{Akerlof(1978)}]{Akerlof1978economics}
Akerlof, G.~A. (1978).
\newblock The economics of ``tagging'' as applied to the optimal income tax,
  welfare programs, and manpower planning.
\newblock \emph{The American Economic Review} 68(1), 8--19.

\bibitem[{Autor and Duggan(2003)}]{Autor2003rise}
Autor, D. and Duggan, M. (2003).
\newblock The rise in the disability rolls and the decline in unemployment.
\newblock \emph{The Quarterly Journal of Economics} 118(1), 157--205.

\bibitem[{Autor and Duggan(2007)}]{Autor2007distinguishing}
Autor, D.~H. and Duggan, M.~G. (2007).
\newblock Distinguishing income from substitution effects in disability
  insurance.
\newblock \emph{American Economic Review} 97(2), 119--124.

\bibitem[{Benitez-Silva et~al.(2004)Benitez-Silva, Buchinsky and
  Rust}]{Benitez-Silva2004how}
Benitez-Silva, H., Buchinsky, M. and Rust, J. (2004).
\newblock {How Large are the Classification Errors in the Social Security
  Disability Award Process?}
\newblock \emph{NBER Working Papers 10219}, National Bureau of Economic
  Research, Inc.

\bibitem[{Bolduc et~al.(2002)Bolduc, Fortin, Labrecque and
  Lanoie}]{Bolduc2002workers}
Bolduc, D., Fortin, B., Labrecque, F. and Lanoie, P. (2002).
\newblock Workers' compensation, moral hazard and the composition of workplace
  injuries.
\newblock \emph{The Journal of Human Resources} 37(3), 623--652.

\bibitem[{Borghans et~al.(2014)Borghans, Gielen and Luttmer}]{Borghans2014}
Borghans, L., Gielen, A.~C. and Luttmer, E. F.~P. (2014).
\newblock {Social Support Substitution and the Earnings Rebound: Evidence from
  a Regression Discontinuity in Disability Insurance Reform}.
\newblock \emph{American Economic Journal: Economic Policy} 6(4), 34--70.

\bibitem[{Bound(1989)}]{Bound1989health}
Bound, J. (1989).
\newblock The health and earnings of rejected disability insurance applicants.
\newblock \emph{Working Paper 2816}, National Bureau of Economic Research.

\bibitem[{BSV(2012)}]{BSV2012statistiken}
BSV (2012).
\newblock \emph{Statistiken zur sozialen Sicherheit -- IV-Statistik 2011}.
\newblock Bundesamt für Sozialversicherungen.

\bibitem[{Butler et~al.(1987)Butler, Burkhauser, Mitchell and
  Pincus}]{Butler1987measurement}
Butler, J.~S., Burkhauser, R.~V., Mitchell, J.~M. and Pincus, T.~P. (1987).
\newblock Measurement error in self-reported health variables.
\newblock \emph{The Review of Economics and Statistics} 69(4), 644--650.

\bibitem[{B{\"u}tler et~al.(2015)B{\"u}tler, Deuchert, Lechner, Staubli and
  Thiemann}]{Buetler2015financial}
B{\"u}tler, M., Deuchert, E., Lechner, M., Staubli, S. and Thiemann, P. (2015).
\newblock Financial work incentives for disability benefit recipients: Lessons
  from a randomised field experiment.
\newblock \emph{IZA Journal of Labor Policy} 4(1), 1--18.

\bibitem[{Campolieti(2002)}]{Campolieti2002moral}
Campolieti, M. (2002).
\newblock Moral hazard and disability insurance: {On} the incidence of
  hard-to-diagnose medical conditions in the {Canada}/{Quebec} {Pension} {Plan}
  {Disability} {Program}.
\newblock \emph{Canadian Public Policy / Analyse de Politiques} 28(3),
  419--441.

\bibitem[{Campolieti(2006)}]{Campolieti2006disability}
Campolieti, M. (2006).
\newblock Disability insurance adjudication criteria and the incidence of
  hard-to-diagnose medical conditions.
\newblock \emph{Contributions to Economic Analysis \& Policy} 5(1), Article 15.

\bibitem[{Campolieti and Riddell(2012)}]{Campolieti2012disability}
Campolieti, M. and Riddell, C. (2012).
\newblock Disability policy and the labor market: Evidence from a natural
  experiment in {Canada}, 1998--2006.
\newblock \emph{Journal of Public Economics} 96(3--4), 306--316.

\bibitem[{Chen and van~der Klaauw(2008)}]{Chen2008work}
Chen, S. and van~der Klaauw, W. (2008).
\newblock The work disincentive effects of the disability insurance program in
  the 1990s.
\newblock \emph{Journal of Econometrics} 142(2), 757--784.

\bibitem[{Cox(1972)}]{Cox1972regression}
Cox, D.~R. (1972).
\newblock Regression models and life-tables.
\newblock \emph{Journal of the Royal Statistical Society. Series B
  (Methodological)} 34(2), 187--220.

\bibitem[{Englund et~al.(2000)Englund, Tibblin and
  Svärdsudd}]{Englund2000variations}
Englund, L., Tibblin, G. and Svärdsudd, K. (2000).
\newblock Variations in sick-listing practice among male and female physicians
  of different specialities based on case vignettes.
\newblock \emph{Scandinavian Journal of Primary Health Care} 18(1), 48--52.

\bibitem[{Eugster and Deuchert(2017)}]{Eugster2017}
Eugster, B. and Deuchert, E. (2017).
\newblock Income and substitution effects of a disability insurance reform.
\newblock \emph{Economics Working Paper Series 1709}, University of St. Gallen,
  School of Economics and Political Science.

\bibitem[{Eugster and Parchet(2018)}]{eugsterparchet2018}
Eugster, B. and Parchet, R. (2018).
\newblock Culture and taxes.
\newblock \emph{Journal of Political Economy} (forthcoming).

\bibitem[{Everett et~al.(2011)Everett, Walters, Stottlemyer, Knight, Oppenberg
  and Orr}]{Everett2011lie}
Everett, J.~P., Walters, C.~A., Stottlemyer, D.~L., Knight, C.~A., Oppenberg,
  A.~A. and Orr, R.~D. (2011).
\newblock To lie or not to lie: Resident physician attitudes about the use of
  deception in clinical practice.
\newblock \emph{Journal of Medical Ethics} 37(6), 333--338.

\bibitem[{Freeman et~al.(1999)Freeman, Rathore, Weinfurt, Schulman and
  Sulmasy}]{Freeman1999lying}
Freeman, V., Rathore, S., Weinfurt, K., Schulman, K. and Sulmasy, D. (1999).
\newblock Lying for patients: Physician deception of third-party payers.
\newblock \emph{Archives of Internal Medicine} 159(19), 2263--2270.

\bibitem[{French and Song(2014)}]{French2014}
French, E. and Song, J. (2014).
\newblock The effect of disability insurance receipt on labor supply.
\newblock \emph{American Economic Journal: Economic Policy} 6(2), 291--337.

\bibitem[{Fr\"olich and Lechner(2010)}]{Froelich2010exploiting}
Fr\"olich, M. and Lechner, M. (2010).
\newblock Exploiting regional treatment intensity for the evaluation of labor
  market policies.
\newblock \emph{Journal of the American Statistical Association} 105(491),
  1014--1029.

\bibitem[{Frueh et~al.(2003)Frueh, Elhai, Gold, Monnier, Magruder, Keane and
  Arana}]{Frueh2003disability}
Frueh, B.~C., Elhai, J.~D., Gold, P.~B., Monnier, J., Magruder, K.~M., Keane,
  T.~M. and Arana, G.~W. (2003).
\newblock Disability compensation seeking among veterans evaluated for
  posttraumatic stress disorder.
\newblock \emph{Psychiatric Services} 54(1), 84--91.

\bibitem[{{Garcia Mandico} et~al.(2018){Garcia Mandico}, Garcia-Gomez, Gielen
  and O'Donnell}]{GarciaMandico2018}
{Garcia Mandico}, S., Garcia-Gomez, P., Gielen, A. and O'Donnell, O. (2018).
\newblock Earnings responses to disability benefit cuts.
\newblock \emph{Working paper}, Tinbergen Institute.

\bibitem[{Gelber et~al.(2017)Gelber, Moore and Strand}]{gelber2017effect}
Gelber, A., Moore, T.~J. and Strand, A. (2017).
\newblock The effect of disability insurance payments on beneficiaries'
  earnings.
\newblock \emph{American Economic Journal: Economic Policy} 9(3), 229--61.

\bibitem[{Inderbitzin et~al.(2016)Inderbitzin, Staubli and
  Zweim{\"{u}}ller}]{Inderbitzin2016}
Inderbitzin, L., Staubli, S. and Zweim{\"{u}}ller, J. (2016).
\newblock Extended unemployment benefits and early retirement: Program
  complementarity and program substitution.
\newblock \emph{American Economic Journal: Economic Policy} 8(1), 253--288.

\bibitem[{de~Jong et~al.(2011)de~Jong, Lindeboom and van~der
  Klaauw}]{Jong2011screening}
de~Jong, P., Lindeboom, M. and van~der Klaauw, B. (2011).
\newblock Screening disability insurance applications.
\newblock \emph{Journal of the European Economic Association} 9(1), 106--129.

\bibitem[{Kankaanp{\"a}{\"a} et~al.(2012)Kankaanp{\"a}{\"a}, Franck and
  Tuominen}]{Kankaanpaeae2012variations}
Kankaanp{\"a}{\"a}, A.~T., Franck, J.~K. and Tuominen, R.~J. (2012).
\newblock Variations in primary care physicians{\textquoteright} sick leave
  prescribing practices.
\newblock \emph{The European Journal of Public Health} 22(1), 92--96.

\bibitem[{Karlstr\"om et~al.(2008)Karlstr\"om, Palme and
  Svensson}]{Karlstroem2008employment}
Karlstr\"om, A., Palme, M. and Svensson, I. (2008).
\newblock The employment effect of stricter rules for eligibility for {DI}:
  {E}vidence from a natural experiment in {S}weden.
\newblock \emph{Journal of Public Economics} 92(10--11), 2071--2082.

\bibitem[{Keele and Titiunik(2016)}]{keeletitiunik2016}
Keele, L. and Titiunik, R. (2016).
\newblock Natural experiments based on geography.
\newblock \emph{Political Science Research and Methods} 4(1), 65--95.

\bibitem[{Kleven and Kopczuk(2011)}]{Kleven2011transfer}
Kleven, H.~J. and Kopczuk, W. (2011).
\newblock Transfer program complexity and the take-up of social benefits.
\newblock \emph{American Economic Journal: Economic Policy} 3(1), 54--90.

\bibitem[{Kom et~al.(1997)Kom, Graubard and Midthune}]{Kom1997time}
Kom, E.~L., Graubard, B.~I. and Midthune, D. (1997).
\newblock Time-to-event analysis of longitudinal follow-up of a survey: Choice
  of the time-scale.
\newblock \emph{American Journal of Epidemiology} 145(1), 72--80.

\bibitem[{Kornfeld and Rupp(2000)}]{Kornfeld2000}
Kornfeld, R. and Rupp, K. (2000).
\newblock The net effects of the project network return-to-work case management
  experiment on participant earnings, benefit receipt, and other outcomes.
\newblock \emph{Social Security Bulletin} 63(1), 12--33.

\bibitem[{Kreider(1999)}]{Kreider1999latent}
Kreider, B. (1999).
\newblock Latent work disability and reporting bias.
\newblock \emph{Journal of Human Resources} 34(4), 734--769.

\bibitem[{Kreider and Pepper(2007)}]{Kreider2007disabilitya}
Kreider, B. and Pepper, J. (2007).
\newblock Disability and employment: Reevaluating the evidence in light of
  reporting errors.
\newblock \emph{Journal of the American Statistical Association} 102(478),
  432--441.

\bibitem[{Kreider and Pepper(2008)}]{Kreider2008inferring}
Kreider, B. and Pepper, J. (2008).
\newblock Inferring disability status from corrupt data.
\newblock \emph{Journal of Applied Econometrics} 23(3), 329--349.

\bibitem[{Lechner(2010)}]{Lechner2010estimation}
Lechner, M. (2010).
\newblock The estimation of causal effects by difference-in-difference methods.
\newblock \emph{Foundations and Trends in Econometrics} 4(3), 165--224.

\bibitem[{Low and Pistaferri(2015)}]{low2015}
Low, H. and Pistaferri, L. (2015).
\newblock Disability insurance and the dynamics of the incentive insurance
  trade-off.
\newblock \emph{American Economic Review} 105(10), 2986--3029.

\bibitem[{Maestas et~al.(2013)Maestas, Mullen and Strand}]{Maestas2013does}
Maestas, N., Mullen, K.~J. and Strand, A. (2013).
\newblock Does disability insurance receipt discourage work? {Using} examiner
  assignment to estimate causal effects of {SSDI} receipt.
\newblock \emph{American Economic Review} 103(5), 1797--1829.

\bibitem[{Mitra(2009)}]{Mitra2009disability}
Mitra, S. (2009).
\newblock Disability screening and labor supply: Evidence from {S}outh
  {A}frica.
\newblock \emph{American Economic Review} 99(2), 512--516.

\bibitem[{Moore(2015)}]{Moore2015}
Moore, T.~J. (2015).
\newblock {The employment effects of terminating disability benefits}.
\newblock \emph{Journal of Public Economics} 124, 30--43.

\bibitem[{Nagi(1969)}]{Nagi1969disability}
Nagi, S.~Z. (1969).
\newblock \emph{Disability and rehabilitation: Legal, clinical, and
  self-concepts and measurement.}
\newblock Columbus, Ohio State University Press.

\bibitem[{Novack et~al.(1989)Novack, Detering, Arnold, Forrow, Ladinsky and
  Pezzullo}]{Novack1989physicians}
Novack, D., Detering, B., Arnold, R., Forrow, L., Ladinsky, M. and Pezzullo, J.
  (1989).
\newblock Physicians' attitudes toward using deception to resolve difficult
  ethical problems.
\newblock \emph{JAMA} 261(20), 2980--2985.

\bibitem[{OECD(2003)}]{OECD2003transforming}
OECD (2003).
\newblock \emph{Transforming Disability into Ability}.
\newblock Paris, OECD Publishing.

\bibitem[{OECD(2006)}]{OECD2006}
OECD (2006).
\newblock \emph{Sickness, Disability and Work: Breaking the Barriers---Norway,
  Poland and Switzerland, Vol.\ 1}.
\newblock Paris, OECD Publishing.

\bibitem[{OECD(2009)}]{OECD2009sickness}
OECD (2009).
\newblock Sickness, disability and work: Keeping on track in the economic
  downturn.
\newblock \emph{Working paper}, High-Level Forum, Stockholm.

\bibitem[{OECD(2010)}]{OECD2010}
OECD (2010).
\newblock \emph{Sickness, Disability and Work: Breaking the Barriers---A
  Synthesis of Findings across OECD countries}.
\newblock Paris, OECD Publishing.

\bibitem[{Parsons(1991)}]{Parsons1991self}
Parsons, D.~O. (1991).
\newblock Self-screening in targeted public transfer programs.
\newblock \emph{Journal of Political Economy} 99(4), 859--876.

\bibitem[{Parsons(1996)}]{Parsons1996imperfect}
Parsons, D.~O. (1996).
\newblock Imperfect `tagging' in social insurance programs.
\newblock \emph{Journal of Public Economics} 62(1–2), 183--207.

\bibitem[{Rubin(1977)}]{Rubin1977assignment}
Rubin, D.~B. (1977).
\newblock Assignment to treatment group on the basis of a covariate.
\newblock \emph{Journal of Educational and Behavioral Statistics} 2(1), 1--26.

\bibitem[{Sheshinski(1978)}]{Sheshinski1978model}
Sheshinski, E. (1978).
\newblock A model of social security and retirement decisions.
\newblock \emph{Journal of Public Economics} 10(3), 337--360.

\bibitem[{Smith and Lilienfeld(1971)}]{Smith1971social}
Smith, R.~T. and Lilienfeld, A.~M. (1971).
\newblock \emph{The {Social Security Disability} program: An evaluation study}.
\newblock 39, US Social Security Administration, Office of Research and
  Statistics.

\bibitem[{Staten and Umbeck(1982)}]{Staten1982information}
Staten, M.~E. and Umbeck, J. (1982).
\newblock Information costs and incentives to shirk: Disability compensation of
  air traffic controllers.
\newblock \emph{The American Economic Review} 72(5), 1023--1037.

\bibitem[{Staubli(2011)}]{Staubli2011impact}
Staubli, S. (2011).
\newblock The impact of stricter criteria for disability insurance on labor
  force participation.
\newblock \emph{Journal of Public Economics} 95(9-10), 1223--1235.

\bibitem[{Thi\'{e}baut and B\'{e}nichou(2004)}]{Thiebaut2004choice}
Thi\'{e}baut, A. C.~M. and B\'{e}nichou, J. (2004).
\newblock Choice of time-scale in {Cox's} model analysis of epidemiologic
  cohort data: {A} simulation study.
\newblock \emph{Statistics in Medicine} 23(24), 3803--3820.

\bibitem[{Vittinghoff et~al.(2011)Vittinghoff, Glidden, Shiboski and
  McCulloch}]{vittinghoff2011regression}
Vittinghoff, E., Glidden, D.~V., Shiboski, S.~C. and McCulloch, C.~E. (2011).
\newblock \emph{Regression methods in biostatistics: linear, logistic,
  survival, and repeated measures models}.
\newblock Springer Science \& Business Media.

\bibitem[{von Wachter et~al.(2011)von Wachter, Song and
  Manchester}]{Wachter2011trends}
von Wachter, T., Song, J. and Manchester, J. (2011).
\newblock Trends in employment and earnings of allowed and rejected applicants
  to the social security disability insurance program.
\newblock \emph{American Economic Review} 101(7), 3308--29.

\bibitem[{Wapf and Peters(2007)}]{Wapf2007evaluation}
Wapf, B. and Peters, M. (2007).
\newblock Evaluation der regionalen ärztlichen {Dienste}.
\newblock \emph{Beiträge zur Sozialen Sicherheit,} Bericht im Rahmen des
  mehrjährigen Forschungsprogramms zu Invalidität und Behinderung,
  Forschungsbericht Nr. 13/07.

\bibitem[{Wynia et~al.(2000)Wynia, Cummins, VanGeest and
  Wilson}]{Wynia2000physician}
Wynia, M., Cummins, D., VanGeest, J. and Wilson, I. (2000).
\newblock Physician manipulation of reimbursement rules for patients: Between a
  rock and a hard place.
\newblock \emph{JAMA} 283(14), 1858--1865.

\bibitem[{Zinn and Furutani(1996)}]{Zinn1996physician}
Zinn, W. and Furutani, N. (1996).
\newblock Physician perspectives on the ethical aspects of disability
  determination.
\newblock \emph{Journal of General Internal Medicine} 11(9), 525--532.

\end{thebibliography}

%%%%%%%%%%%%%%%%%%%% Appendix %%%%%%%%%%%%%%%%%%%%%%%%%%%%%%%%%%%%%%%%%%%%%%%%%

\clearpage
\appendix
\renewcommand{\thesection}{A\arabic{section}}

\section*{Appendix A: Additional tables and figures}
\label{sec:tables}
\renewcommand{\thetable}{A\arabic{table}}
\renewcommand{\thefigure}{A\arabic{figure}}
\setcounter{table}{0}
\setcounter{figure}{0}

% TABLES
\begin{table}[h]
	\centering
  \caption{DI recipients before and after filing for benefits}
	\label{tab:beforeafter}
	\begin{threeparttable}
		\scriptsize
    \begin{tabular}{l*{5}{S}}
\toprule \addlinespace[1em]
                                 & \multicolumn{1}{c}{} & \multicolumn{1}{c}{} & \multicolumn{1}{c}{\textit{DI filing}} & \multicolumn{1}{c}{} & \multicolumn{1}{c}{} \\
                                 & \multicolumn{1}{c}{\textit{-2}} & \multicolumn{1}{c}{\textit{-1}} & \multicolumn{1}{c}{\textit{year}} & \multicolumn{1}{c}{\textit{+1}} & \multicolumn{1}{c}{\textit{+2}} \\
\midrule
Worked last week                 & 0.761                      & 0.622                      & 0.368                    & 0.320                      & 0.296                      \\
                                 & {(197)}                    & {(410)}                    & {(810)}                  & {(1268)}                   & {(1457)}                   \\
& & & & & \\
Looking for work last month      & 0.484                      & 0.333                      & 0.120                    & 0.096                      & 0.079                      \\
                                 & {(31)}                     & {(84)}                     & {(357)}                  & {(748)}                    & {(953)}                    \\
& & & & & \\
Work contract but absent at work last week & 0.326                      & 0.455                      & 0.294                    & 0.120                      & 0.057                      \\
                                 & {(46)}                     & {(154)}                    & {(506)}                  & {(845)}                    & {(1006)}                   \\
& & & & & \\
Yearly income (1k CHF)           & 53.113                     & 47.785                     & 33.262                   & 17.284                     & 11.574                     \\
                                 & {(197)}                    & {(410)}                    & {(810)}                  & {(1268)}                   & {(1457)}                   \\
& & & & & \\
Dismissed from unemployment office & 0.025                      & 0.053                      & 0.061                    & 0.056                      & 0.066                      \\
                                 & {(79)}                     & {(206)}                    & {(445)}                  & {(784)}                    & {(948)}                    \\
& & & & & \\
Social assistance                & 0.051                      & 0.058                      & 0.038                    & 0.051                      & 0.044                      \\
                                 & {(79)}                     & {(206)}                    & {(445)}                  & {(784)}                    & {(948)}                    \\
& & & & & \\
Age                              & 47.365                     & 48.480                     & 50.022                   & 50.445                     & 50.648                     \\
                                 & {(197)}                    & {(410)}                    & {(810)}                  & {(1268)}                   & {(1457)}                   \\
& & & & & \\
Mental or physical problem    & 0.234         & 0.393         & 0.688         & 0.844         & 0.836         \\
                        & {(124)}         & {(262)}         & {(523)}         & {{(841)}}         & {(980)}         \\
& & & & & \\
Accident within the last 12 months        & 0.208         & 0.176         & 0.118         & 0.057         & 0.082         \\
                        & {(48)}          & {(85)}          & {(136)}         & {(174)}         & {(184)}         \\
\bottomrule
		\end{tabular}
		\begin{tablenotes}[para,flushleft]
            \item Note: This table shows the mean values of selected variables
              for DI recipients from two years prior to filing the application
              until two years afterwards. The table utilizes the limited
              longitudinal information that is available in the SESAM data. The
              number of observations in a cell is given in parentheses. Note
              that sample sizes vary because not all recipients have the same
              historic coverage and not all survey modules are administered
              every year.
    \end{tablenotes}
	\end{threeparttable}
\end{table}

%%% Local Variables:
%%% mode: latex
%%% TeX-master: "../rad"
%%% End:

\begin{table}
    \centering
    \scriptsize{
    \caption{Descriptive statistics}
    \label{tab:desc}
    \begin{threeparttable}
		\begin{tabular}{l*{5}{r}}
\toprule
\addlinespace[1ex]
\multicolumn{6}{c}{(a) Full sample}\\
\midrule
  & \multicolumn{1}{c}{Mean} & \multicolumn{1}{c}{SD} & \multicolumn{1}{c}{Min} & \multicolumn{1}{c}{Max} & \multicolumn{1}{c}{N} \\ % & \multicolumn{1}{c}{Sum} & \multicolumn{1}{c}{CV}
\cmidrule{2-6}
All individuals                                 &           &         &         &          &         \\ % [1ex]
\hspace*{1em}Age                                 & 50.316    & 18.033  & 18.0    & 104.0    & 259,323 \\   %   & 13,048,180.0 & 0.358
\hspace*{1em}Female                              & 0.539     & 0.498   & 0.0     & 1.0      & 259,323 \\   %   & 139,841.0    & 0.924
\hspace*{1em}Married                             & 0.552     & 0.497   & 0.0     & 1.0      & 259,323 \\   %   & 143,180.0    & 0.901
\hspace*{1em}Foreign                             & 0.322     & 0.467   & 0.0     & 1.0      & 259,323 \\   %   & 83,424.0     & 1.452
\hspace*{1em}Nr.\ of children                    & 0.582     & 0.973   & 0.0     & 7.0      & 259,323 \\   %   & 151,035.0    & 1.671
\hspace*{1em}Education: Primary                  & 0.234     & 0.423   & 0.0     & 1.0      & 259,323 \\   %   & 60,726.0     & 1.808
\hspace*{1em}Education: Secondary                & 0.510     & 0.500   & 0.0     & 1.0      & 259,323 \\   %   & 132,378.0    & 0.979
\hspace*{1em}Education: Tertiary                 & 0.255     & 0.436   & 0.0     & 1.0      & 259,323 \\   %   & 66,219.0     & 1.708
\hspace*{1em}Gross annual earnings               & 41.450    & 107.251 & 0.0     & 42,317.4 & 259,323 \\   %   & 10,748,852.3 & 2.587
\hspace*{1em}Travel distance (km)                & 34.297    & 31.825  & 0.2     & 194.1    & 259,323 \\   %   & 8,894,111.3  & 0.928
\hspace*{1em}Travel time (min)                   & 31.411    & 23.167  & 0.6     & 169.5    & 259,323 \\   %   & 8,145,545.8  & 0.738
\hspace*{1em}Unemployed                          & 0.027     & 0.163   & 0.0     & 1.0      & 259,323 \\   %   & 7,092.0      & 5.964
\hspace*{1em}Receives DI                         & 0.035     & 0.185   & 0.0     & 1.0      & 259,323 \\   %   & 9,204.0      & 5.213
Region                                          &           &         &         &          &         \\ %[1ex]
\hspace*{1em}L\'eman                            & 0.191     & 0.393   & 0.0     & 1.0      & 259,323 \\   %   & 49,634.0     & 2.055
\hspace*{1em}Mittelland                         & 0.194     & 0.396   & 0.0     & 1.0      & 259,323 \\   %   & 50,413.0     & 2.036
\hspace*{1em}Nordwestschweiz                    & 0.136     & 0.343   & 0.0     & 1.0      & 259,323 \\   %   & 35,288.0     & 2.520
\hspace*{1em}Z\"urich                           & 0.166     & 0.372   & 0.0     & 1.0      & 259,323 \\   %   & 42,959.0     & 2.244
\hspace*{1em}Ostschweiz                         & 0.122     & 0.328   & 0.0     & 1.0      & 259,323 \\   %   & 31,730.0     & 2.678
\hspace*{1em}Zentralschweiz                     & 0.107     & 0.310   & 0.0     & 1.0      & 259,323 \\   %   & 27,861.0     & 2.882
\hspace*{1em}Tessin                             & 0.083     & 0.275   & 0.0     & 1.0      & 259,323 \\   %   & 21,438.0     & 3.331
\cmidrule{2-6}
DI recipients                                     &           &         &         &          &         \\ %[1ex]
\hspace*{1em}Years in DI                          & 9.415     & 6.847   & 0.0     & 48.0     & 9,204   \\   %  & 86,658.0     & 0.727
\hspace*{1em}Disability: Psych. problems          & 0.341     & 0.474   & 0.0     & 1.0      & 9,204   \\   %  & 3,141.0      & 1.389
\hspace*{1em}Disability: Nerve                    & 0.072     & 0.259   & 0.0     & 1.0      & 9,204   \\   %  & 663.0        & 3.589
\hspace*{1em}Disability: Muscoloskeletal cond.\   & 0.235     & 0.424   & 0.0     & 1.0      & 9,204   \\   %  & 2,165.0      & 1.803
\hspace*{1em}Disability: Accident                 & 0.092     & 0.289   & 0.0     & 1.0      & 9,204   \\   %  & 847.0        & 3.141
\hspace*{1em}Disability: Congenital disease/other & 0.185     & 0.388   & 0.0     & 1.0      & 9,204   \\   %  & 1,701.0      & 2.100
\midrule
\addlinespace[1.5ex]
\multicolumn{6}{c}{(b) Local sample (within 20 km)}\\
\midrule
  & \multicolumn{1}{c}{Mean} & \multicolumn{1}{c}{SD} & \multicolumn{1}{c}{Min} & \multicolumn{1}{c}{Max} & \multicolumn{1}{c}{N} \\ % & \multicolumn{1}{c}{Sum} & \multicolumn{1}{c}{CV}
\cmidrule{2-6}
All individuals                                       &           &         &         &          &         \\ % [1ex]
\hspace*{1em}Age                                      & 49.950    & 18.019  & 18.0    & 104.0    & 133,549 \\    %  & 6,670,755.0 & 0.361
\hspace*{1em}Female                                   & 0.538     & 0.499   & 0.0     & 1.0      & 133,549 \\    %  & 71,891.0    & 0.926
\hspace*{1em}Married                                  & 0.546     & 0.498   & 0.0     & 1.0      & 133,549 \\    %  & 72,939.0    & 0.912
\hspace*{1em}Foreign                                  & 0.329     & 0.470   & 0.0     & 1.0      & 133,549 \\    %  & 43,888.0    & 1.429
\hspace*{1em}Nr.\ of children                         & 0.580     & 0.972   & 0.0     & 7.0      & 133,549 \\    %  & 77,472.0    & 1.676
\hspace*{1em}Education: Primary                       & 0.226     & 0.418   & 0.0     & 1.0      & 133,549 \\    %  & 30,160.0    & 1.851
\hspace*{1em}Education: Secondary                     & 0.510     & 0.500   & 0.0     & 1.0      & 133,549 \\    %  & 68,065.0    & 0.981
\hspace*{1em}Education: Tertiary                      & 0.265     & 0.441   & 0.0     & 1.0      & 133,549 \\    %  & 35,324.0    & 1.668
\hspace*{1em}Gross annual earnings                    & 43.252    & 134.295 & 0.0     & 42,317.4 & 133,549 \\    %  & 5,776,299.0 & 3.105
\hspace*{1em}Travel distance (km)                     & 11.871    & 4.753   & 0.2     & 20.0     & 133,549 \\    %  & 1,585,326.5 & 0.400
\hspace*{1em}Travel time (min)                        & 14.981    & 5.170   & 0.6     & 30.1     & 133,549 \\    %  & 2,000,699.9 & 0.345
\hspace*{1em}Unemployed                               & 0.027     & 0.163   & 0.0     & 1.0      & 133,549 \\    %  & 3,665.0     & 5.953
\hspace*{1em}Receives DI                              & 0.035     & 0.184   & 0.0     & 1.0      & 133,549 \\    %  & 4,693.0     & 5.240
Region                                                &           &         &         &          &         \\ %[1ex]
\hspace*{1em}L\'eman                                  & 0.119     & 0.324   & 0.0     & 1.0      & 133,549 \\    %  & 15,940.0    & 2.716
\hspace*{1em}Mittelland                               & 0.156     & 0.363   & 0.0     & 1.0      & 133,549 \\    %  & 20,899.0    & 2.322
\hspace*{1em}Nordwestschweiz                          & 0.260     & 0.439   & 0.0     & 1.0      & 133,549 \\    %  & 34,729.0    & 1.687
\hspace*{1em}Z\"urich                                 & 0.256     & 0.436   & 0.0     & 1.0      & 133,549 \\    %  & 34,161.0    & 1.706
\hspace*{1em}Ostschweiz                               & 0.068     & 0.252   & 0.0     & 1.0      & 133,549 \\    %  & 9,070.0     & 3.705
\hspace*{1em}Zentralschweiz                           & 0.140     & 0.347   & 0.0     & 1.0      & 133,549 \\    %  & 18,749.0    & 2.474
\hspace*{1em}Tessin                                   & 0.000     & 0.003   & 0.0     & 1.0      & 133,549 \\    %  & 1.0         & 365.444
\cmidrule{2-6}
DI recipients                                         &           &         &         &          &         \\ %[1ex]
\hspace*{1em}Years in DI                              & 9.294     & 6.779   & 0.0     & 47.0     & 4,693   \\    %  & 43,617.0    & 0.729
\hspace*{1em}Disability: Psych. problems              & 0.359     & 0.480   & 0.0     & 1.0      & 4,693   \\    %  & 1,685.0     & 1.336
\hspace*{1em}Disability: Nerve                        & 0.072     & 0.259   & 0.0     & 1.0      & 4,693   \\    %  & 339.0       & 3.584
\hspace*{1em}Disability: Muscoloskeletal cond.\       & 0.232     & 0.422   & 0.0     & 1.0      & 4,693   \\    %  & 1,090.0     & 1.818
\hspace*{1em}Disability: Accident                     & 0.087     & 0.282   & 0.0     & 1.0      & 4,693   \\    %  & 409.0       & 3.237
\hspace*{1em}Disability: Congenital disease/other     & 0.178     & 0.382   & 0.0     & 1.0      & 4,693   \\    %  & 835.0       & 2.150
\bottomrule
        \end{tabular}
        \begin{tablenotes}[para,flushleft]
            \item Note: Descriptive statistics for the unrestricted and the
                local estimation sample.  Based on the 1999--2011 SESAM data.
        \end{tablenotes}
    \end{threeparttable}}
\end{table}

\newgeometry{left=0.5cm,right=0.5cm,top=2.0cm,bottom=2.0cm,nofoot}
\begin{table}
    \centering
    \scriptsize{
    \caption{Pre-treatment covariate balance}
    \label{tab:desc_ttest_combined}
    \begin{threeparttable}
        \begin{tabular}{l*{3}{S[table-number-alignment=center,%
                                table-figures-integer=2,%
                                table-figures-decimal=2,
                                round-precision=2]}%
                              S[table-number-alignment=center,%
                                table-figures-integer=1,%
                                table-figures-decimal=3,
                            table-space-text-post={\rlap{***}}]%}
                           p{2pt}%
                           *{3}{S[table-number-alignment=center,%
                                table-figures-integer=2,%
                                table-figures-decimal=2,%
                                round-precision=2]}%
                              S[table-number-alignment=center,%
                                table-figures-integer=1,%
                                table-figures-decimal=3,
                            table-space-text-post={\rlap{***}}]%
                            }
\toprule \addlinespace[1em]
                                  & \multicolumn{4}{c}{(a) Full sample} &                             & \multicolumn{4}{c}{(b) Local sample (within 20 km)}       \\
\cmidrule(){2-5} \cmidrule(){7-10}
\addlinespace[1ex]
\addlinespace[1em]
                                       & \multicolumn{1}{c}{Total}   & \multicolumn{1}{c}{Treated} & \multicolumn{1}{c}{Control} & \multicolumn{1}{c}{Difference} &  & \multicolumn{1}{c}{Total}  & \multicolumn{1}{c}{Treated} &  \multicolumn{1}{c}{Control} & \multicolumn{1}{c}{Difference} \\ % old columns local sample unweighted    %   & \multicolumn{1}{c}{Total}  & \multicolumn{1}{c}{Treated} &  \multicolumn{1}{c}{Control} & \multicolumn{1}{c}{Difference} \\ % old columns local sample unweighted
\midrule\addlinespace[1.5ex]
All individuals                        &           &          &           &                 &  &           &         &         &           \\ [1em]
\hspace*{1em} Age                      & 48.336    & 47.736   & 48.662   & -0.926\rlap{***} & & 48.554   & 48.530   & 48.683    & -0.153      \\
\hspace*{1em}                          & (18.281)  & (18.829) & (17.945) & (0.309)     & & (18.555) & (10.610) & (40.056)  & (0.605)     \\
\hspace*{1em} Female                   & 0.542     & 0.548    & 0.539    & 0.009       & & 0.548    & 0.549    & 0.540     & 0.009       \\
\hspace*{1em}                          & (0.498)   & (0.521)  & (0.486)  & (0.009)     & & (0.498)  & (0.285)  & (1.077)   & (0.016)     \\
\hspace*{1em} Married                  & 0.524     & 0.575    & 0.497    & 0.078\rlap{***} & & 0.524    & 0.527    & 0.506     & 0.021       \\
\hspace*{1em}                          & (0.499)   & (0.517)  & (0.487)  & (0.009)     & & (0.499)  & (0.285)  & (1.080)   & (0.016)     \\
\hspace*{1em} Foreign                  & 0.090     & 0.118    & 0.075    & 0.043\rlap{***} & & 0.132    & 0.136    & 0.109     & 0.027\rlap{***} \\
\hspace*{1em}                          & (0.286)   & (0.337)  & (0.257)  & (0.005)     & & (0.339)  & (0.196)  & (0.674)   & (0.010)     \\
\hspace*{1em} Nr.\ of children         & 0.564     & 0.656    & 0.513    & 0.142\rlap{***} & & 0.574    & 0.570    & 0.593     & -0.023      \\
\hspace*{1em}                          & (0.975)   & (1.083)  & (0.913)  & (0.018)     & & (0.976)  & (0.556)  & (2.150)   & (0.035)     \\
\hspace*{1em} Education: Primary       & 0.207     & 0.225    & 0.197    & 0.028\rlap{***} & & 0.239    & 0.243    & 0.219     & 0.024\rlap{*}  \\
\hspace*{1em}                          & (0.405)   & (0.437)  & (0.387)  & (0.007)     & & (0.427)  & (0.245)  & (0.893)   & (0.014)     \\
\hspace*{1em} Education: Secondary     & 0.592     & 0.586    & 0.595    & -0.010      & & 0.579    & 0.576    & 0.597     & -0.021      \\
\hspace*{1em}                          & (0.491)   & (0.515)  & (0.478)  & (0.009)     & & (0.494)  & (0.283)  & (1.060)   & (0.016)     \\
\hspace*{1em} Education: Tertiary      & 0.202     & 0.190    & 0.208    & -0.019\rlap{***} & & 0.181    & 0.181    & 0.184     & -0.004      \\
\hspace*{1em}                          & (0.401)   & (0.410)  & (0.395)  & (0.007)     & & (0.385)  & (0.220)  & (0.838)   & (0.012)     \\
\hspace*{1em} Gross annual earnings    & 36.094    & 35.359   & 36.494   & -1.135      & & 34.188   & 33.931   & 35.589    & -1.658      \\
\hspace*{1em}                          & (48.351)  & (50.571) & (47.095) & (0.877)     & & (45.807) & (26.259) & (97.475)  & (1.444)     \\
\hspace*{1em} Travel distance (km)     & 28.688    & 43.022   & 20.898   & 22.125\rlap{***} & & 10.281   & 10.256   & 10.415    & -0.158      \\
\hspace*{1em}                          & (27.217)  & (37.620) & (15.953) & (0.506)     & & (4.795)  & (2.742)  & (10.345)  & (0.150)     \\
\hspace*{1em} Travel time (min)        & 27.798    & 37.150   & 22.716   & 14.434\rlap{***} & & 13.252   & 13.215   & 13.455    & -0.240      \\
\hspace*{1em}                          & (20.270)  & (27.785) & (12.976) & (0.378)     & & (5.243)  & (3.002)  & (11.234)  & (0.165)     \\
\hspace*{1em} Unemployed               & 0.015     & 0.018    & 0.013    & 0.005       & & 0.020    & 0.021    & 0.013     & 0.008\rlap{**} \\
\hspace*{1em}                          & (0.120)   & (0.139)  & (0.109)  & (0.002)     & & (0.138)  & (0.081)  & (0.246)   & (0.004)     \\
\hspace*{1em} Receives DI in 2001      & 0.042     & 0.038    & 0.043    & -0.005      & & 0.038    & 0.037    & 0.042     & -0.004      \\
\hspace*{1em}                          & (0.200)   & (0.200)  & (0.199)  & (0.004)     & & (0.191)  & (0.109)  & (0.430)   & (0.008)     \\
\midrule\addlinespace[1.5ex]
DI recipients                          &           &          &           &                   &  &           &         &         &           \\ [1em]
\hspace*{1em} Years in DI              & 7.902     & 7.643    & 8.034    & -0.391      & & 7.407    & 7.671    & 6.090     & 1.582       \\
\hspace*{1em}                          & (6.936)   & (7.476)  & (6.616)  & (0.646)     & & (6.679)  & (3.713)  & (12.701)  & (0.967)     \\
\hspace*{1em} Entry age                & 43.105    & 44.202   & 42.548   & 1.654       & & 45.046   & 45.261   & 43.991    & 1.270       \\
\hspace*{1em}                          & (11.689)  & (13.249) & (10.835) & (1.142)     & & (11.507) & (6.233)  & (24.991)  & (2.271)     \\
\hspace*{1em} DI: Psych.\ problems     & 0.292     & 0.273    & 0.301    & -0.028      & & 0.285    & 0.271    & 0.352     & -0.081      \\
\hspace*{1em}                          & (0.455)   & (0.491)  & (0.434)  & (0.043)     & & (0.452)  & (0.244)  & (1.014)   & (0.091)     \\
\hspace*{1em} DI: Nerve                & 0.107     & 0.085    & 0.118    & -0.033      & & 0.114    & 0.114    & 0.110     & 0.004       \\
\hspace*{1em}                          & (0.309)   & (0.307)  & (0.305)  & (0.029)     & & (0.318)  & (0.175)  & (0.664)   & (0.051)     \\
\hspace*{1em} DI: MSK                  & 0.214     & 0.273    & 0.183    & 0.089\rlap{**}  & & 0.234    & 0.257    & 0.121     & 0.136\rlap{**} \\
\hspace*{1em}                          & (0.410)   & (0.491)  & (0.366)  & (0.041)     & & (0.424)  & (0.240)  & (0.693)   & (0.064)     \\
\hspace*{1em} DI: Other illness        & 0.207     & 0.206    & 0.208    & -0.002      & & 0.194    & 0.200    & 0.166     & 0.034       \\
\hspace*{1em}                          & (0.406)   & (0.446)  & (0.384)  & (0.039)     & & (0.396)  & (0.220)  & (0.789)   & (0.063)     \\
\hspace*{1em} DI: Accident             & 0.103     & 0.086    & 0.111    & -0.025      & & 0.079    & 0.057    & 0.186     & -0.129      \\
\hspace*{1em}                          & (0.304)   & (0.309)  & (0.297)  & (0.029)     & & (0.270)  & (0.127)  & (0.826)   & (0.090)     \\
\midrule\addlinespace[1.5ex]
All individuals & {15,522} & {5,983} & {9,539} &  &  & {8,570} & {2,367} & {6,203} &  \\
DI recipients   & {506}    & {207}   & {299}   &  &  & {280}   & {70}    & {210}    &  \\
\bottomrule
        \end{tabular}
        \begin{tablenotes}[para,flushleft]
            \item Note: Means of selected covariates for individuals in treated
                and control regions sampled between 1999--2001, prior to the
                pilot period. Separate statistics for all individuals and those
                within a distance of 20 kilometers in border regions. Standard
                deviation in parentheses. The last column in each block shows
                the difference between treated and control individuals for each
                variable, standard error in parentheses. Survey weights applied
                for the full sample. Observations weighted for pairwise
                differences in the local sample.  *, ** and *** denote
                significance at the 10\%, 5\% and 1\% level respectively.
        \end{tablenotes}
    \end{threeparttable}}
\end{table}
\restoregeometry

\begin{table}
    \centering
    \caption{Placebo reform}
    \label{tab:robustnessplacebo}
    \begin{threeparttable}
        \scriptsize{
            \begin{tabular}{l*{4}{S[table-number-alignment=center,%
                                    table-figures-integer=1,%
                                    table-figures-decimal=3]}%
                                    p{3mm}%
                                    *{4}{S[table-number-alignment=center,%
                                    table-figures-integer=1,%
                                    table-figures-decimal=3]}}
\toprule \addlinespace[1em]
     & \multicolumn{4}{c}{(a) Full sample} & & \multicolumn{4}{c}{(b) Local sample (within 20 km)} \\ \cmidrule(lr{.75em}){2-5} \cmidrule(lr{.75em}){7-10}
     & \multicolumn{1}{c}{(1)}         & \multicolumn{1}{c}{(2)}                           &  \multicolumn{1}{c}{(3)}                           & \multicolumn{1}{c}{(4)}     &    & \multicolumn{1}{c}{(5)}       & \multicolumn{1}{c}{(6)}  & \multicolumn{1}{c}{(7)}& \multicolumn{1}{c}{(8)}      \\
\midrule
Treatment region                   & 1.337*** & 1.337*** & 1.337*** & 1.248*** & & 1.150** & 1.150**  & 1.150**  & 1.148**  \\
                                   & (0.051)  & (0.051)  & (0.051)  & (0.048)  & & (0.076) & (0.076)  & (0.076)  & (0.076)  \\
Pre-pilot time                     & 1.235*** & 1.241*** & 1.241*** & 1.274*** & & 1.204   & 1.213    & 1.213    & 1.253*   \\
                                   & (0.082)  & (0.082)  & (0.082)  & (0.084)  & & (0.146) & (0.146)  & (0.146)  & (0.150)  \\
\rowcolor{gray!15}Treat x pre      & 0.970    & 0.970    & 0.970    & 0.975    & & 0.999   & 0.999    & 0.999    & 0.996    \\
\rowcolor{gray!15}                 & (0.064)  & (0.064)  & (0.064)  & (0.064)  & & (0.111) & (0.111)  & (0.111)  & (0.111)  \\
Pilot time                         &          & 1.320*** & 1.326*** & 1.390*** & &         & 1.514*** & 1.525*** & 1.612*** \\
                                   &          & (0.129)  & (0.129)  & (0.135)  & &         & (0.228)  & (0.229)  & (0.241)  \\
\rowcolor{gray!15}Treat x pilot    &          & 0.847**  & 0.846**  & 0.852**  & &         & 0.770**  & 0.771**  & 0.765**  \\
\rowcolor{gray!15}                 &          & (0.069)  & (0.069)  & (0.069)  & &         & (0.092)  & (0.092)  & (0.091)  \\
Post time                          &          &          & 0.842    & 0.917    & &         &          & 1.046    & 1.142    \\
                                   &          &          & (0.094)  & (0.103)  & &         &          & (0.207)  & (0.226)  \\
Treat x post                       &          &          & 0.960    & 0.961    & &         &          & 0.841    & 0.829    \\
                                   &          &          & (0.080)  & (0.080)  & &         &          & (0.110)  & (0.109)  \\
\midrule
Other controls         & \multicolumn{1}{r}{-}         & \multicolumn{1}{r}{-}         & \multicolumn{1}{r}{-}         & \multicolumn{1}{r}{\checkmark} &  & \multicolumn{1}{r}{-}         & \multicolumn{1}{r}{-}         & \multicolumn{1}{r}{-}         & \multicolumn{1}{r}{\checkmark} \\
N municipalities       & \multicolumn{1}{r}{2,336}     & \multicolumn{1}{r}{2,337}     & \multicolumn{1}{r}{2,338}     & \multicolumn{1}{r}{2,338}      &  & \multicolumn{1}{r}{1,086}     & \multicolumn{1}{r}{1,086}     & \multicolumn{1}{r}{1,087}     & \multicolumn{1}{r}{1,087}      \\
N individuals          & \multicolumn{1}{r}{242,531}   & \multicolumn{1}{r}{249,750}   & \multicolumn{1}{r}{259,323}   & \multicolumn{1}{r}{259,323}   & & \multicolumn{1}{r}{124,747}   & \multicolumn{1}{r}{128,633}   & \multicolumn{1}{r}{133,648}   & \multicolumn{1}{r}{133,648}   \\
N failures             & \multicolumn{1}{r}{6,164}     & \multicolumn{1}{r}{7,877}     & \multicolumn{1}{r}{9,204}     & \multicolumn{1}{r}{9,204}     & & \multicolumn{1}{r}{3,100}     & \multicolumn{1}{r}{3,985}     & \multicolumn{1}{r}{4,693}     & \multicolumn{1}{r}{4,693}     \\
N fail during pilot    & \multicolumn{1}{r}{0}         & \multicolumn{1}{r}{1,713}     & \multicolumn{1}{r}{1,713}     & \multicolumn{1}{r}{1,713}     & & \multicolumn{1}{r}{0}         & \multicolumn{1}{r}{885}       & \multicolumn{1}{r}{885}       & \multicolumn{1}{r}{885}       \\
N fail during prepilot & \multicolumn{1}{r}{1,950}     & \multicolumn{1}{r}{1,950}     & \multicolumn{1}{r}{1,950}     & \multicolumn{1}{r}{1,950}     & & \multicolumn{1}{r}{989}       & \multicolumn{1}{r}{989}       & \multicolumn{1}{r}{989}       & \multicolumn{1}{r}{989}       \\
N                      & \multicolumn{1}{r}{439,761}   & \multicolumn{1}{r}{631,782}   & \multicolumn{1}{r}{787,954}   & \multicolumn{1}{r}{787,954}   & & \multicolumn{1}{r}{226,345}   & \multicolumn{1}{r}{325,321}   & \multicolumn{1}{r}{406,221}   & \multicolumn{1}{r}{406,221}   \\
\bottomrule
        \end{tabular}}
        \begin{tablenotes}[para,flushleft]
            \item Note: Cox Proportional Hazard estimates for individuals in
                treated and control regions based on SESAM individual-level
                survey and administrative data sampled during 1999--2011.
                Baseline hazard for all regressions stratified by 5-year birth
                cohorts. Survey weights applied for the full sample.
                Observations in the local sample are weighted for pairwise
                estimation. Results are reported in exponentiated form as
                hazard ratios. The hazard ratio for `Treat x pilot' corresponds
                to the relative average treatment effect on the treated as
                defined in \autoref{sec:Strategy}. Standard errors clustered at
                the individual level in parentheses, number of observations
                given below. *, ** and *** denote significance at the 10\%, 5\%
                and 1\% level respectively.
        \end{tablenotes}
    \end{threeparttable}
\end{table}

\begin{table}
    \centering
    \caption{Distance windows}
    \label{tab:robustnessdist}
\makebox[\textwidth][c]{
        \scriptsize{
    \begin{threeparttable}
            \begin{tabular}{l*{6}{S[table-number-alignment=center,%
                                    table-figures-integer=1,%
                                    table-figures-decimal=2,%
                                    round-precision=2]}%
                                    p{1mm}%
                                    *{6}{S[table-number-alignment=center,%
                                    table-figures-integer=1,%
                                    table-figures-decimal=2,%
                                    round-precision=2]}}%
\toprule \addlinespace[1em]
                                & \multicolumn{5}{c}{(a) Travel distance (km)} &                               & \multicolumn{5}{c}{(b) Travel time (min)} \\ \cmidrule(lr{.75em}){2-6} \cmidrule(lr{.75em}){8-12} \addlinespace[1ex]
                                & \multicolumn{1}{c}{10 km}                & \multicolumn{1}{c}{15 km}     & \multicolumn{1}{c}{20 km}             &  \multicolumn{1}{c}{25 km}                                               & \multicolumn{1}{c}{30 km}     &                         & \multicolumn{1}{c}{10 min}   & \multicolumn{1}{c}{15 min}    & \multicolumn{1}{c}{20 min}    & \multicolumn{1}{c}{25 min}    &  \multicolumn{1}{c}{30 min}    \\ %                        & \multicolumn{1}{c}{5 min} &  \multicolumn{1}{c}{5 km}
\midrule
Treatment region                &       1.132   &       1.199***&       1.151***&       1.161***&       1.195***&  &      1.040   &       1.125   &       1.176***&       1.157***&       1.088   \\
                                &     (0.099)   &     (0.079)   &     (0.061)   &     (0.058)   &     (0.057)   &  &    (0.115)   &     (0.094)   &     (0.071)   &     (0.064)   &     (0.059)   \\
Pilot time                      &       1.291   &       1.380** &       1.267** &       1.247** &       1.251** &  &      1.469*  &       1.434** &       1.299** &       1.318** &       1.196   \\
                                &     (0.229)   &     (0.192)   &     (0.148)   &     (0.138)   &     (0.134)   &  &    (0.333)   &     (0.249)   &     (0.172)   &     (0.161)   &     (0.143)   \\
\rowcolor{gray!15}Treat x pilot &       0.745*  &       0.711** &       0.771** &       0.775** &       0.779** &  &      0.740   &       0.658** &       0.728** &       0.756** &       0.812*  \\
\rowcolor{gray!15}              &     (0.128)   &     (0.095)   &     (0.087)   &     (0.082)   &     (0.079)   &  &    (0.164)   &     (0.110)   &     (0.092)   &     (0.088)   &     (0.092)   \\
Post time                       &       0.916   &       0.909   &       0.867   &       0.815   &       0.838   &  &      1.086   &       0.873   &       0.784   &       0.801   &       0.801   \\
                                &     (0.236)   &     (0.184)   &     (0.151)   &     (0.135)   &     (0.132)   &  &    (0.337)   &     (0.211)   &     (0.145)   &     (0.138)   &     (0.134)   \\
Treat x post                    &       0.792   &       0.827   &       0.841   &       0.850   &       0.849   &  &      0.995   &       0.853   &       0.859   &       0.902   &       0.940   \\
                                &     (0.157)   &     (0.126)   &     (0.105)   &     (0.100)   &     (0.096)   &  &    (0.241)   &     (0.162)   &     (0.119)   &     (0.115)   &     (0.116)   \\
\midrule
N municipalities        & \multicolumn{1}{r}{549}       & \multicolumn{1}{r}{825}       & \multicolumn{1}{r}{1,087}     & \multicolumn{1}{r}{1,286}     & \multicolumn{1}{r}{1,414}     & & \multicolumn{1}{r}{372}      & \multicolumn{1}{r}{649}       & \multicolumn{1}{r}{922}       & \multicolumn{1}{r}{1,159}     & \multicolumn{1}{r}{1,371}     \\
N individuals           & \multicolumn{1}{r}{47,403}    & \multicolumn{1}{r}{88,990}    & \multicolumn{1}{r}{133,549}   & \multicolumn{1}{r}{151,215}   & \multicolumn{1}{r}{163,852}   & & \multicolumn{1}{r}{26,956}   & \multicolumn{1}{r}{56,609}    & \multicolumn{1}{r}{119,572}   & \multicolumn{1}{r}{143,504}   & \multicolumn{1}{r}{166,486}   \\
N failures              & \multicolumn{1}{r}{1,626}     & \multicolumn{1}{r}{3,230}     & \multicolumn{1}{r}{4,693}     & \multicolumn{1}{r}{5,223}     & \multicolumn{1}{r}{5,690}     & & \multicolumn{1}{r}{942}      & \multicolumn{1}{r}{1,948}     & \multicolumn{1}{r}{4,253}     & \multicolumn{1}{r}{5,031}     & \multicolumn{1}{r}{5,752}     \\
N failures during pilot & \multicolumn{1}{r}{332}       & \multicolumn{1}{r}{612}       & \multicolumn{1}{r}{885}       & \multicolumn{1}{r}{980}       & \multicolumn{1}{r}{1,063}     & & \multicolumn{1}{r}{180}      & \multicolumn{1}{r}{379}       & \multicolumn{1}{r}{811}       & \multicolumn{1}{r}{961}       & \multicolumn{1}{r}{1,087}     \\
N                       & \multicolumn{1}{r}{107,479}   & \multicolumn{1}{r}{200,431}   & \multicolumn{1}{r}{300,432}   & \multicolumn{1}{r}{340,370}   & \multicolumn{1}{r}{369,235}   & & \multicolumn{1}{r}{61,269}   & \multicolumn{1}{r}{128,479}   & \multicolumn{1}{r}{269,155}   & \multicolumn{1}{r}{323,290}   & \multicolumn{1}{r}{375,210}   \\
\bottomrule
        \end{tabular}
        \begin{tablenotes}[para,flushleft]
            \item Note: Cox Proportional Hazard estimates for individuals in
                treated and control regions across various distance windows
                from the border. Based on SESAM individual-level survey and
                administrative data sampled during 1999--2011.  Observations
                are weighted for pairwise estimation. Results are reported in
                exponentiated form as hazard ratios. The hazard ratio for
                `Treat x pilot' corresponds to the relative average treatment
                effect on the treated as defined in \autoref{sec:Strategy}.
                Standard errors clustered at the individual level in
                parentheses, number of observations given below. *, ** and ***
                denote significance at the 10\%, 5\% and 1\% level
                respectively.
        \end{tablenotes}
    \end{threeparttable}
        }
    }
\end{table}

\begin{table}
	\centering
	\caption{Stock outflow}
	\label{tab:outflow}
	\begin{threeparttable}
		%\footnotesize{
		\scriptsize{
            \begin{tabular}{l*{3}{S[table-number-alignment=center,%
                                    table-figures-integer=1,%
                                    table-figures-decimal=3]}%
                                    p{3mm}%
                                    *{3}{S[table-number-alignment=center,%
                                    table-figures-integer=1,%
                                    table-figures-decimal=3]}}
\toprule \addlinespace[1em]
	 & \multicolumn{3}{c}{(a) All individuals} & & \multicolumn{3}{c}{(b) Age $\leq$ 50 in 2001} \\ \cmidrule(lr{.75em}){2-4} \cmidrule(lr{.75em}){6-8}
	 & \multicolumn{1}{c}{(1)}         & \multicolumn{1}{c}{(2)}                           &  \multicolumn{1}{c}{(3)}                           &  & \multicolumn{1}{c}{(4)}       & \multicolumn{1}{c}{(5)}       & \multicolumn{1}{c}{(6)}        \\
\midrule
Treat                           & 0.925*** & 0.923*** & 0.911*** & & 0.871*** & 0.871*** & 0.872*** \\
                                & (0.027)  & (0.027)  & (0.027)  & & (0.041)  & (0.041)  & (0.041)  \\
Pilot time                      & 7.698*** & 7.677*** & 7.825*** & & 7.515*** & 7.479*** & 7.652*** \\
                                & (0.157)  & (0.156)  & (0.160)  & & (0.243)  & (0.240)  & (0.247)  \\
\rowcolor{gray!15}Treat x pilot & 0.985    & 0.986    & 0.992    & & 0.995    & 0.997    & 0.997    \\
\rowcolor{gray!15}              & (0.033)  & (0.033)  & (0.033)  & & (0.053)  & (0.053)  & (0.053)  \\
Post time                       &          & 7.518*** & 7.728*** & &          & 7.676*** & 7.931*** \\
                                &          & (0.152)  & (0.157)  & &          & (0.236)  & (0.246)  \\
Treat x post                    &          & 1.008    & 1.014    & &          & 1.036    & 1.035    \\
                                &          & (0.032)  & (0.033)  & &          & (0.052)  & (0.051)  \\
\midrule
Other controls          & \multicolumn{1}{r}{-}           & \multicolumn{1}{r}{-}           & \multicolumn{1}{r}{\checkmark}  &                                & \multicolumn{1}{r}{-}           & \multicolumn{1}{r}{-}           &  \multicolumn{1}{r}{\checkmark} \\
N individuals           & \multicolumn{1}{r}{314,249}     & \multicolumn{1}{r}{327,580}     & \multicolumn{1}{r}{327,580}     & & \multicolumn{1}{r}{145,018}    & \multicolumn{1}{r}{154,020}     & \multicolumn{1}{r}{154,020}     \\
N failures              & \multicolumn{1}{r}{20,481}      & \multicolumn{1}{r}{44,529}      & \multicolumn{1}{r}{44,529}      & & \multicolumn{1}{r}{8,904}      & \multicolumn{1}{r}{23,547}      & \multicolumn{1}{r}{23,547}      \\
N failures during pilot & \multicolumn{1}{r}{15,389}      & \multicolumn{1}{r}{15,389}      & \multicolumn{1}{r}{15,389}      & & \multicolumn{1}{r}{6,957}      & \multicolumn{1}{r}{6,957}       & \multicolumn{1}{r}{6,957}       \\
N                       & \multicolumn{1}{r}{1,032,666}   & \multicolumn{1}{r}{2,489,323}   & \multicolumn{1}{r}{2,489,323}   & & \multicolumn{1}{r}{504,801}    & \multicolumn{1}{r}{1,470,137}   & \multicolumn{1}{r}{1,470,137}   \\
\bottomrule
		\end{tabular}}
		\begin{tablenotes}[para,flushleft]
			\item Note: Cox Proportional Hazard estimates for individuals in
				treated and control regions based on SESAM individual-level
				survey and administrative data sampled during 1999--2011.
				Baseline hazard for all regressions stratified by 5-year birth
				cohorts. Survey weights applied for the full sample.
				Observations in the local sample are weighted for
				nearest-neighbor pairwise differences. Results are reported in
				exponentiated form as hazard ratios. Standard errors clustered
				at the municipality level in parentheses, number of
				observations given below. *, ** and *** denote significance at
				the 10\%, 5\% and 1\% level respectively.
		\end{tablenotes}
	\end{threeparttable}
\end{table}

\begin{table}
	\centering
	\caption{Determinants of local sample}
	\label{tab:probitrepresentative}
	\begin{threeparttable}
		\scriptsize{
			\begin{tabular}{l*{3}{S[table-number-alignment=center,%
									table-figures-integer=3,%
									table-figures-decimal=4,%
                                    round-precision=4,
									tight-spacing=false]}}
\toprule \addlinespace[1em]
                     & \multicolumn{1}{c}{Full sample} & \multicolumn{1}{c}{Treated} & \multicolumn{1}{c}{Control} \\ \cmidrule(lr{.75em}){2-4}
                     & \multicolumn{1}{c}{(1)}         & \multicolumn{1}{c}{(2)}     & \multicolumn{1}{c}{(3)}     \\
\midrule
Age                  & -0.0004031\rlap{*}  & -0.0008198\rlap{***} & 0.0002472           \\
                     & (0.0002282)         & (0.0002091)          & (0.0002508)         \\
Female               & 0.0039648           & -0.0079749           & 0.0159401\rlap{***} \\
                     & (0.0054449)         & (0.0063448)          & (0.0056666)         \\
Married              & -0.0115389          & 0.0093218            & -0.0320088\rlap{*}  \\
                     & (0.0165437)         & (0.0181248)          & (0.0180795)         \\
Foreign              & 0.0174959           & -0.0360146           & 0.1117031\rlap{***} \\
                     & (0.0258203)         & (0.0270384)          & (0.0209790)         \\
Nr.\ of children     & -0.0030342          & 0.0036742            & -0.0049889          \\
                     & (0.0041433)         & (0.0033275)          & (0.0049158)         \\
Education: Secondary & 0.0194663\rlap{***} & 0.0040979            & 0.0189137\rlap{**}  \\
                     & (0.0068359)         & (0.0066390)          & (0.0083376)         \\
Education: Tertiary  & 0.0373294           & 0.0008139            & 0.0489619\rlap{**}  \\
                     & (0.0228079)         & (0.0239767)          & (0.0238884)         \\
\midrule
N                    & {259,323}           & {117,701}            & {141,622}           \\
\bottomrule
		\end{tabular}}
		\begin{tablenotes}[para,flushleft]
            \item Note: Probit estimates for the probability to be included in
the local sample separately for treated and control regions. Marginal effects at
the mean reported. Standard errors clustered at the municipality level in
parentheses, number of observations given below. *, ** and *** denote
significance at the 10\%, 5\% and 1\% level respectively.
		\end{tablenotes}
	\end{threeparttable}
\end{table}

% FIGURES
\begin{figure}
    \centering
	\caption{Sample composition}
    \label{fig:map2}
\makebox[\textwidth][c]{
            \includegraphics[width=1.3\textwidth]{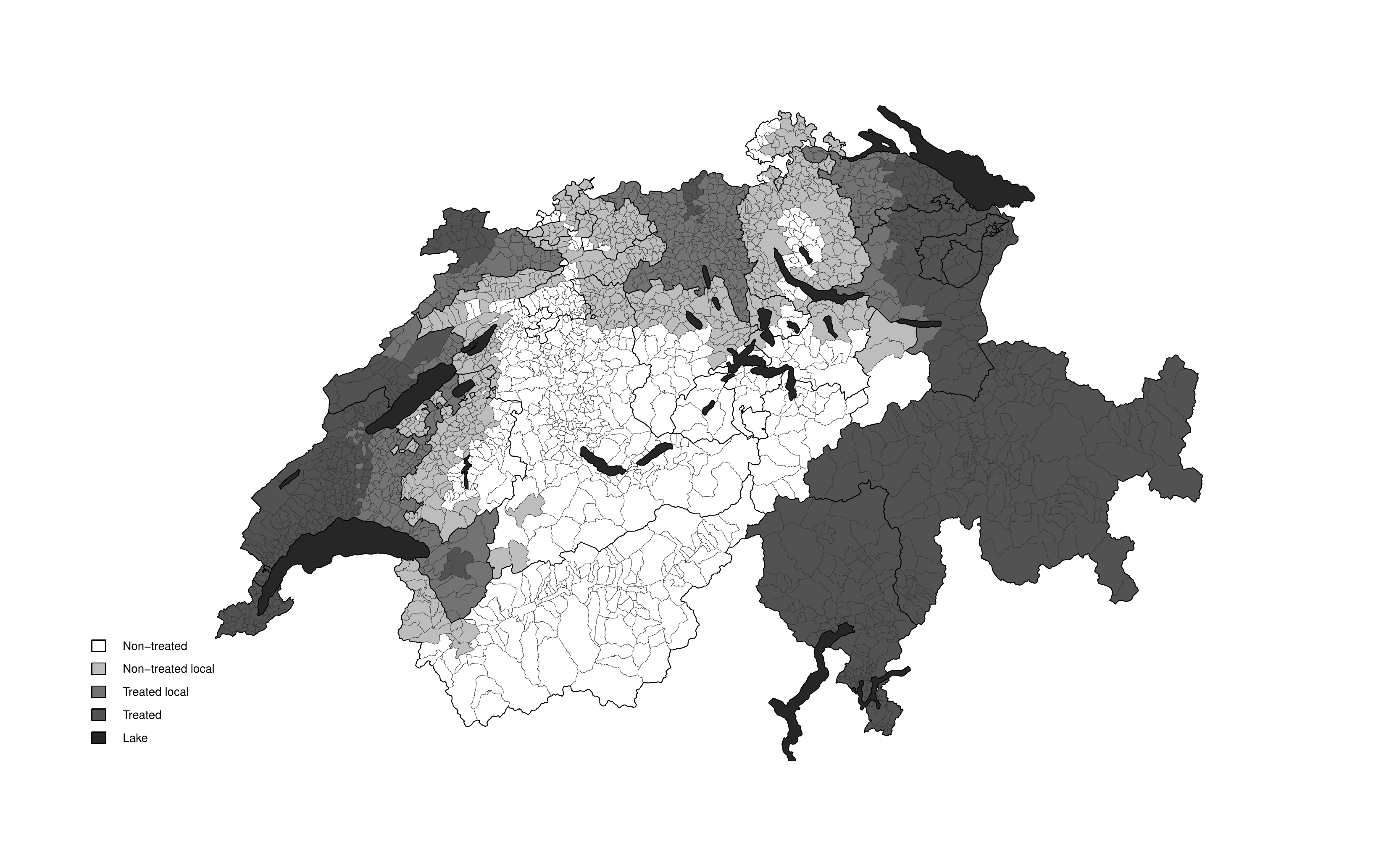}
        }
            \floatfoot{\scriptsize
            Note: Pilot cantons in shaded in dark and medium grey, control cantons shaded
            in light grey and white. Intermediate shades
            indicate the municipalities that are included in the local sample.
            Lakes shown in black.%
          }
\end{figure}

%%% Local Variables:
%%% mode: latex
%%% TeX-master: "../submission-JHE_2018_356"
%%% End:

\begin{figure}
    \centering
    \caption{Trends in DI physicians and caseload}
    \footnotesize
    \label{fig:diphys}
            \includegraphics[width=0.70\textwidth]{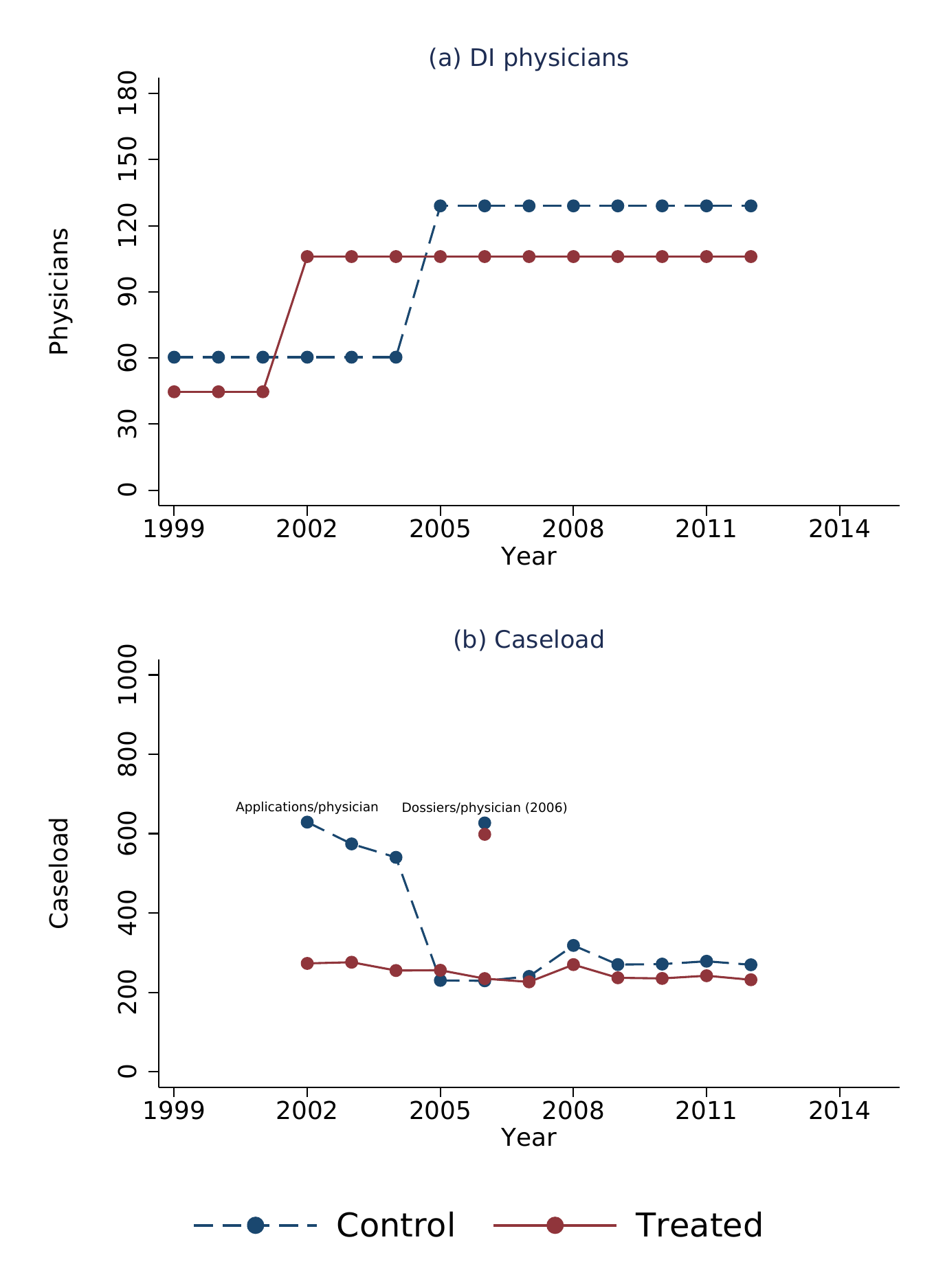}
            \floatfoot*{%
                Note: Panel (a) shows the number of full-time equivalent medical
staff positions before and after the reform changes, panel (b) approximates the
application caseload per physician. Cantons in western Switzerland for which the
electronic reporting system is known to have been faulty are omitted from the
sample for the statistics in panel (b) (Fribourg, Gen\`eve, Jura, Neuch\^atel, Vaud). Applications are only available from 2002.}
\end{figure}

%%% Local Variables:
%%% mode: latex
%%% TeX-master: "../rad"
%%% End:

\begin{figure}
    \centering
    \caption{Trends for aggregate disability insurance stock}
    \footnotesize
    \label{fig:aggstock}
            \includegraphics[width=0.70\textwidth]{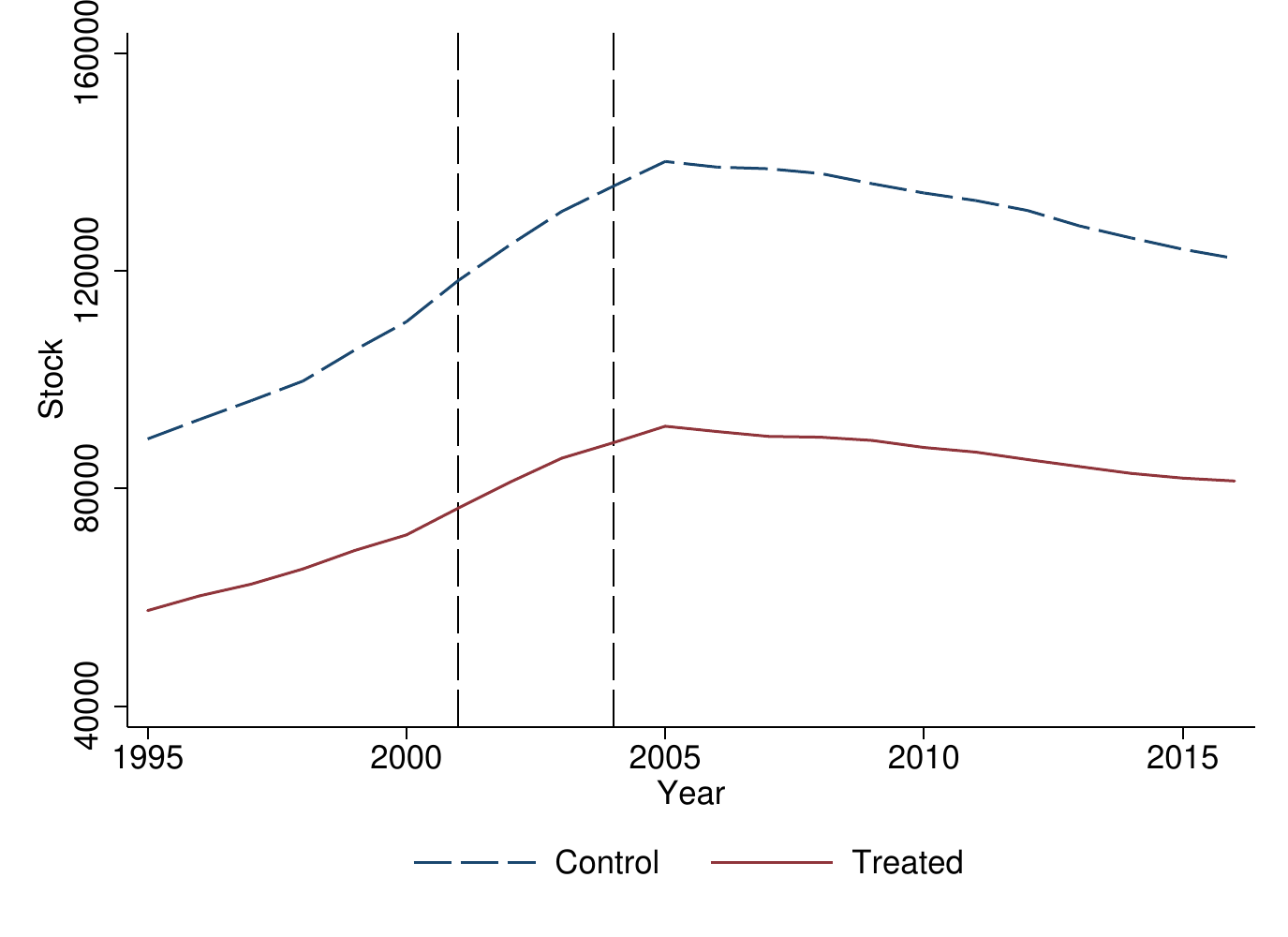}
            \floatfoot*{Note: Disability insurance stock for treated and
            control regions.}
\end{figure}

%%% Local Variables:
%%% mode: latex
%%% TeX-master: "../rad"
%%% End:

\begin{figure}%[t]
    \centering
    \caption{Trends for disability insurance applications}
    \footnotesize
    \label{fig:diapp}
            \includegraphics[width=0.75\textwidth]{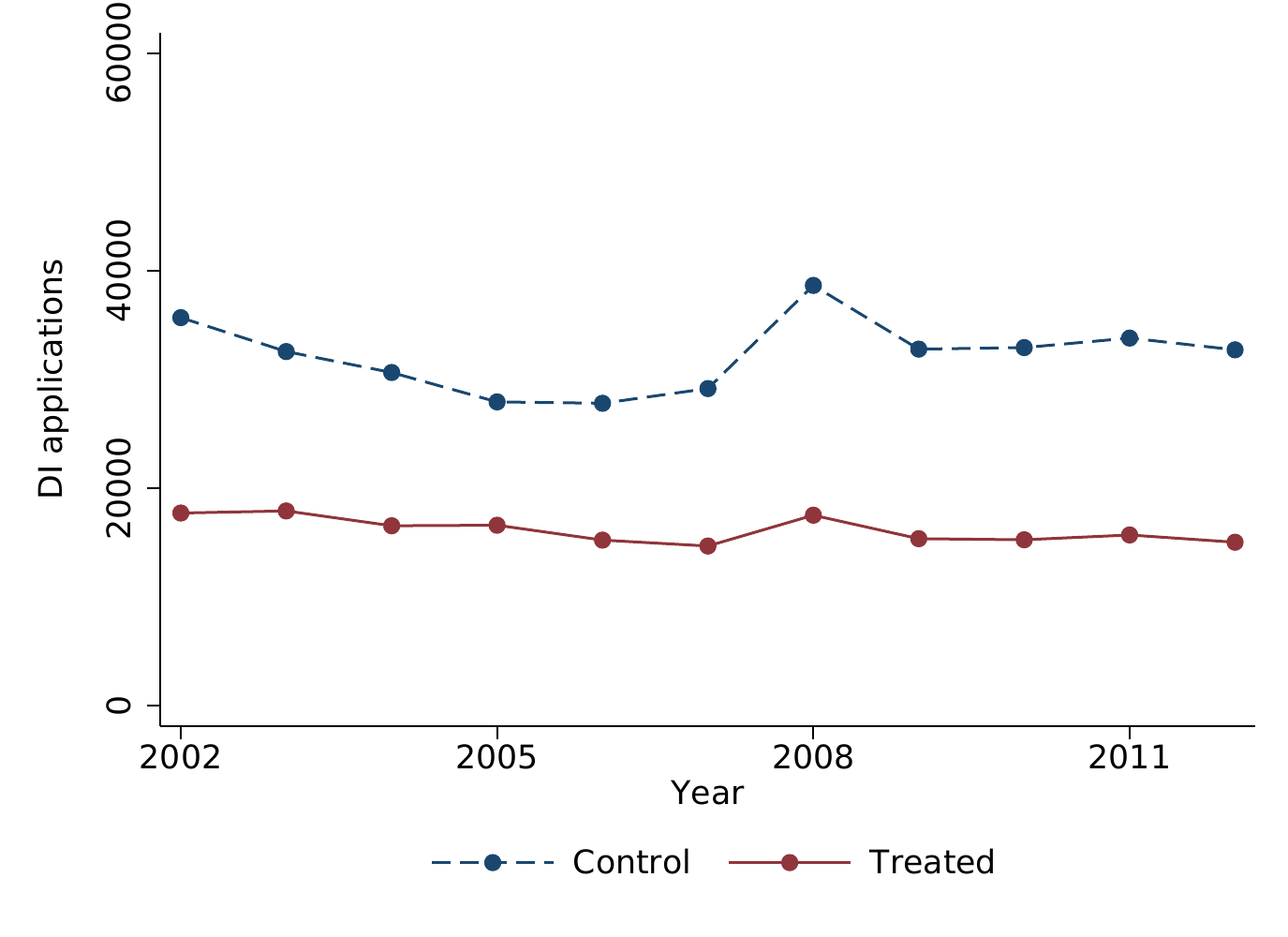}
            \floatfoot*{%
                Note: Disability insurance applications for the years
2002--2012. Cantons in western Switzerland for which the electronic reporting
system is known to have been faulty are omitted from the sample (Fribourg,
Gen\`eve, Jura, Neuch\^atel, Vaud).}
\end{figure}

%%% Local Variables:
%%% mode: latex
%%% TeX-master: "../rad"
%%% End:

\begin{figure}%[ht]
	\centering
	\caption{Possible effects of medical review on latent types}
	\label{fig:effect-matrix}
  \includegraphics[width=\textwidth]{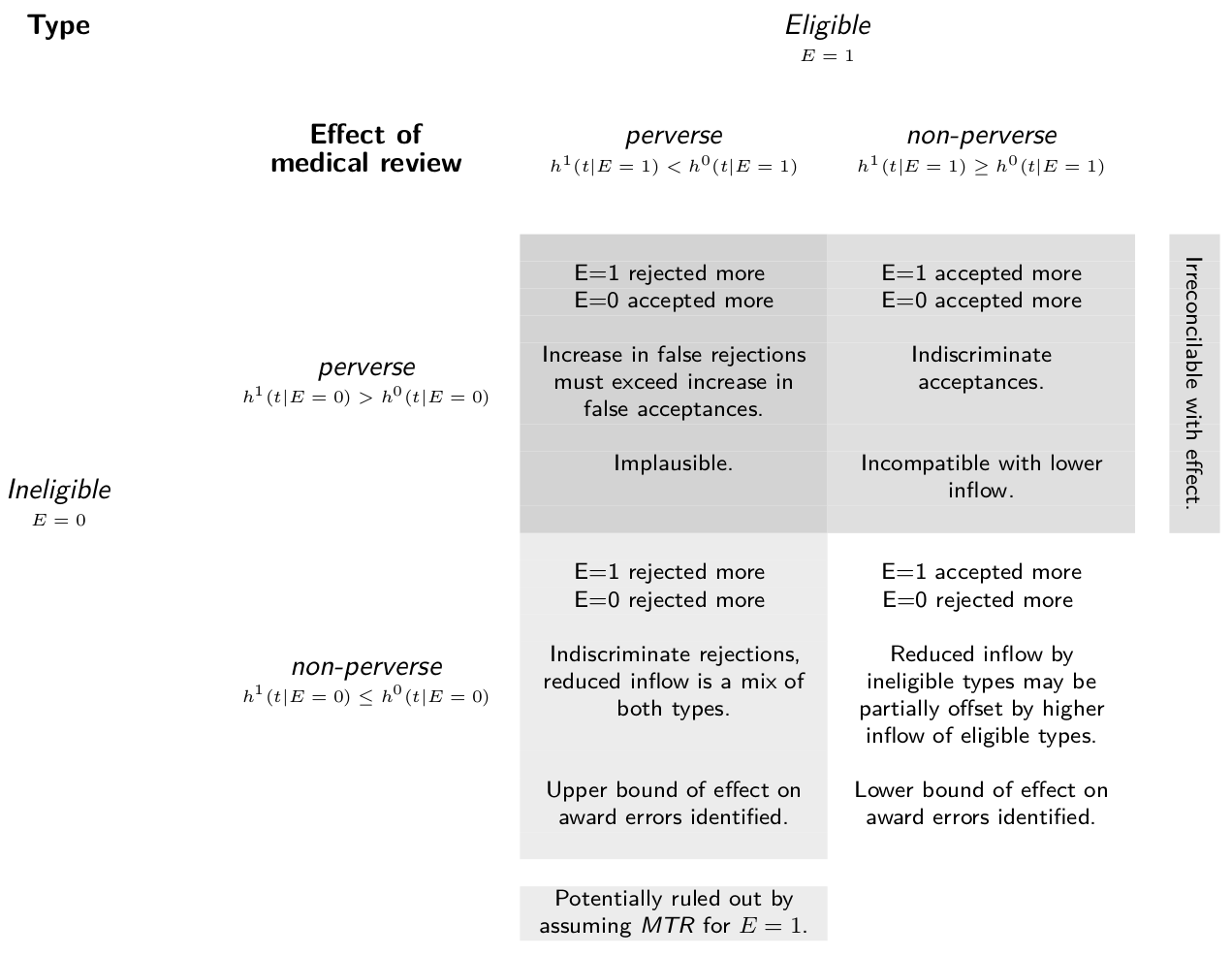}
\end{figure}
 % this is now a figure, was previously a table
\begin{figure}
    \centering
    \caption{Disability insurance court cases}
    \footnotesize
    \label{fig:diclaims}
            \includegraphics[width=\textwidth]{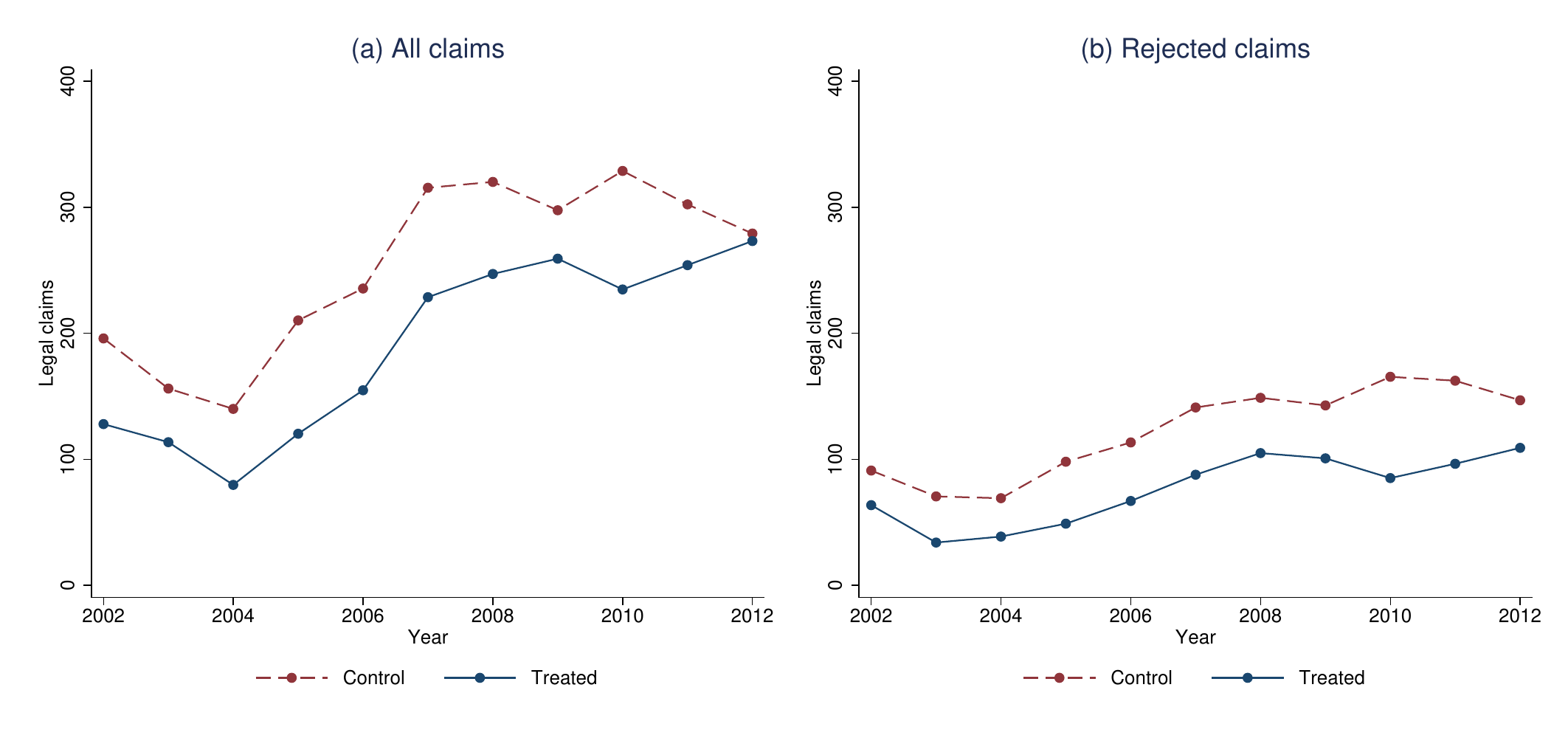}
            \floatfoot*{%
                Note: Mean cantonal total and rejected disability insurance legal claims
                for the years 2002--2012.}
\end{figure}

\begin{figure}
    \centering
    \caption{Distance windows}
    \footnotesize
    \label{fig:distances}
            \vspace{1.5ex}
            \begin{subfloatrow}
                \subfloat[Travel distance (km)]{
                    \begin{minipage}{.47\textwidth}
                        \includegraphics[width=1.05\textwidth]{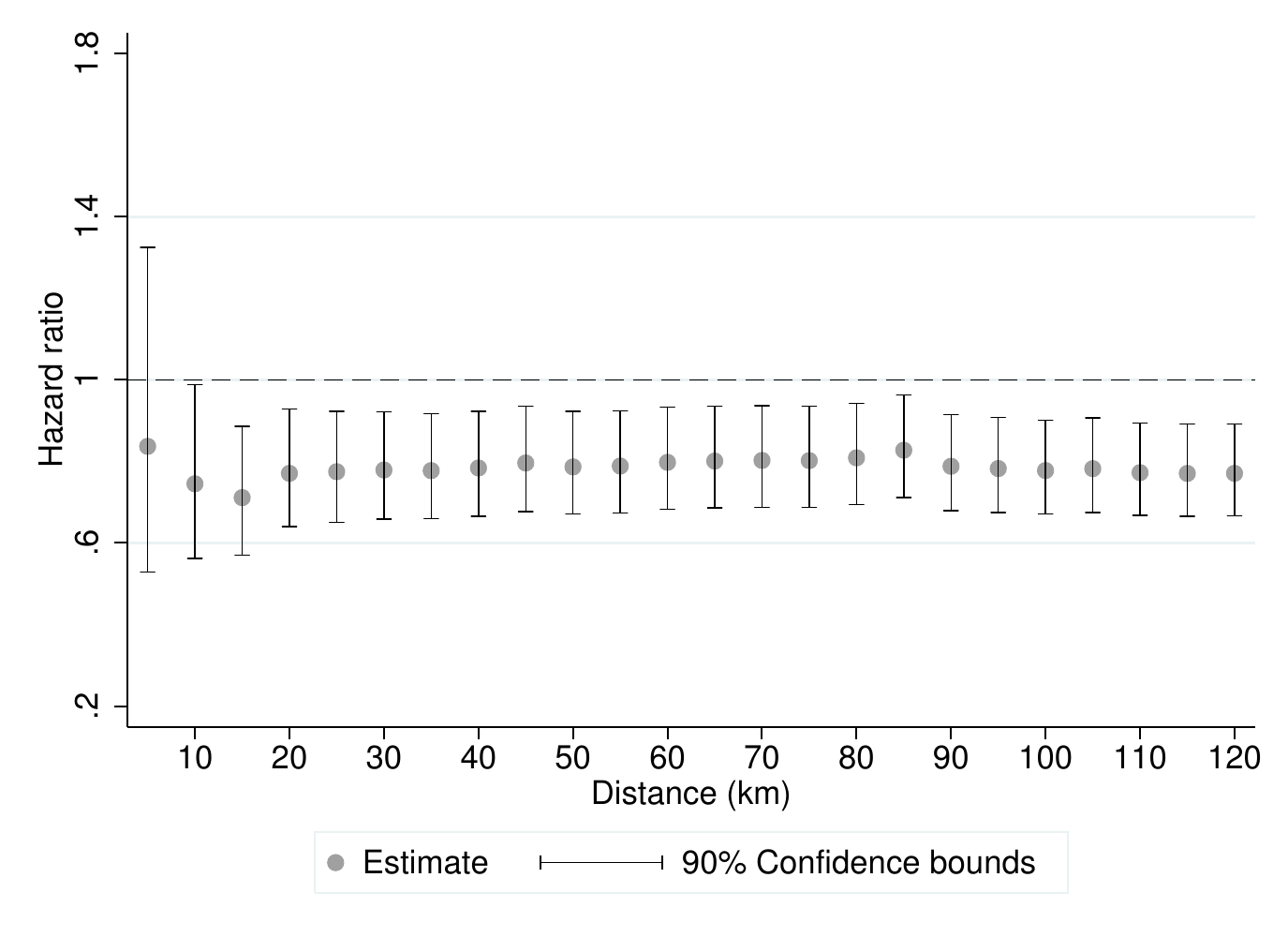}
                    \end{minipage}}
                \subfloat[Travel time (min)]{
                    \begin{minipage}{.47\textwidth}
                        \includegraphics[width=1.05\textwidth]{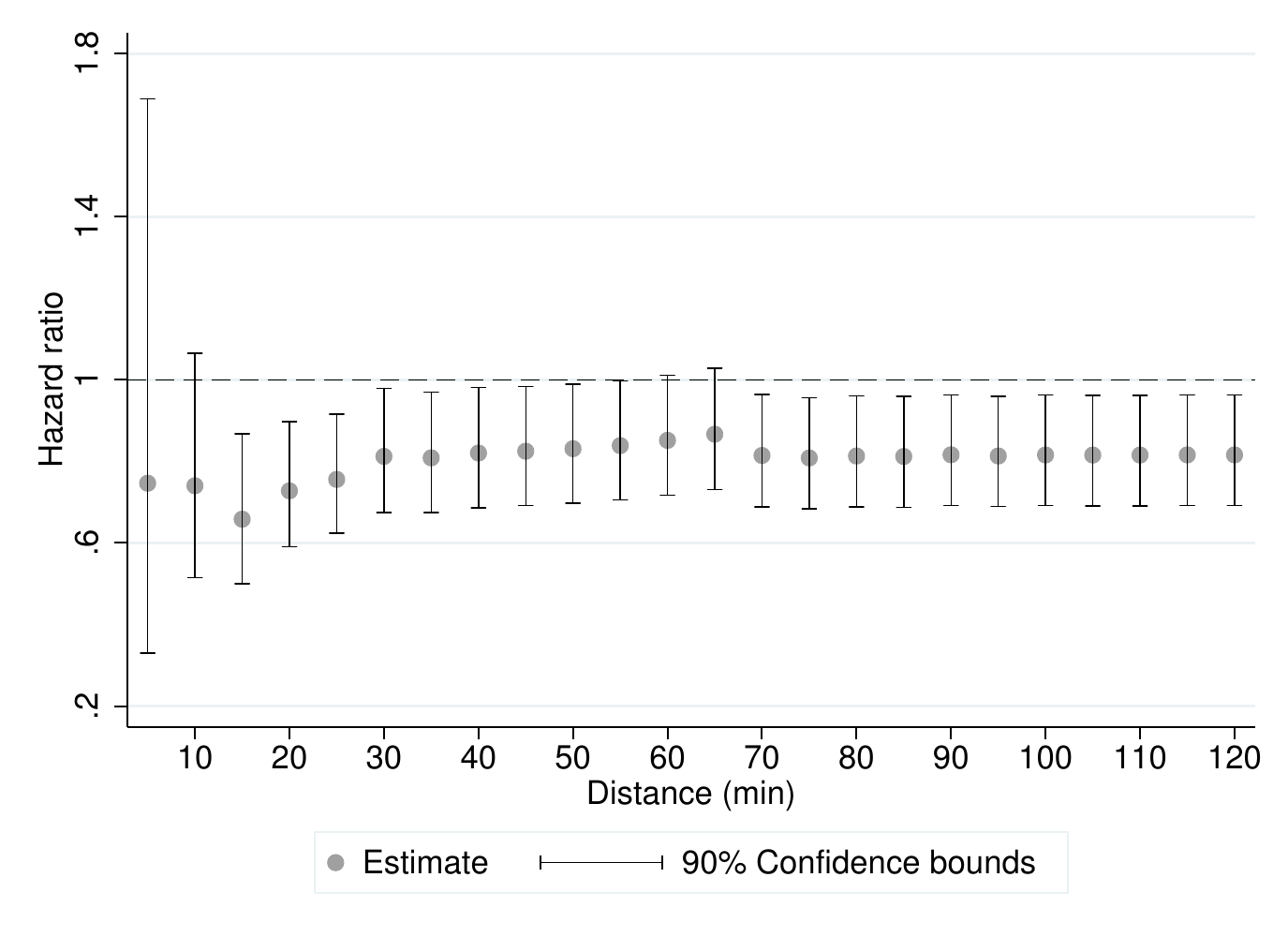}
                    \end{minipage}}
            \end{subfloatrow}
            \floatfoot*{%
                Note: Treatment effect estimates and 90\% confidence bounds
                from the main specification for different distance windows
                measured using actual travel distance and travel time.
            }
\end{figure}
 % this is now in the appendix, was previously in the main paper

\clearpage
\section*{Supplementary material for online publication only}
\section*{Appendix B: Further tables and figures}
\label{sec:online}
\renewcommand{\thetable}{B\arabic{table}}
\renewcommand{\thefigure}{B\arabic{figure}}
\setcounter{table}{0}
\setcounter{figure}{0}

\textit{Title}: Does external medical review reduce disability insurance inflow?\\
\textit{Author}: Helge Liebert\\ \bigskip

\begin{table}[ht]
	\centering
	\caption{Main disability incidence results, linear probability model}
	\label{tab:mainresults-lpm}
	\begin{threeparttable}
		\scriptsize
            \begin{tabular}{l*{2}{S[group-digits=false,round-mode=places,round-precision=6,table-number-alignment=center,%
                                    table-figures-integer=1,%
                                    table-figures-decimal=6]}%
                                    p{3mm}%
                                    *{2}{S[group-digits=false,round-mode=places,round-precision=6,table-number-alignment=center,%
                                    table-figures-integer=1,%
                                    table-figures-decimal=6]}}
\toprule \addlinespace[1em]
                                                           & \multicolumn{2}{c}{(a) Full sample} &                                & \multicolumn{2}{c}{(b) Local sample (within 20 km)} \\
\cmidrule(lr{.75em}){2-3} \cmidrule(lr{.75em}){5-6}
                                                           & \multicolumn{1}{c}{(1)}
                                                           & \multicolumn{1}{c}{(2)}
                                                           &                                     & \multicolumn{1}{c}{(3)}        & \multicolumn{1}{c}{(4)}                             \\
\midrule \addlinespace[1em]
Treat x pilot                                              & -0.000265                           & -0.000272                      &                                                     &  -0.000106                      &  -0.000107                      \\
                                                           & (0.000519)                          & (0.000518)                     &                                                     &  (0.001146)                     &  (0.001144)                     \\
\cmidrule(lr{.75em}){2-6}
\rowcolor{gray!15} relative ATT (implied)                  & {-0.1698}                             & {-0.1742}                        &                                                     &  {-0.0696}                        &  {-0.0701}                        \\
$\bar{y}$                                                  & {0.001559}                            & {0.001559}                       &                                                     &  {0.001530}                       &  {0.001530}                       \\
Other controls                                             & \multicolumn{1}{c}{-}               & \multicolumn{1}{c}{\checkmark} &                                                     &  \multicolumn{1}{c}{-}          &  \multicolumn{1}{c}{\checkmark} \\
Canton fixed effects                                       & \multicolumn{1}{c}{\checkmark}      & \multicolumn{1}{c}{\checkmark} &                                                     &  \multicolumn{1}{c}{\checkmark} &  \multicolumn{1}{c}{\checkmark} \\
Time fixed effects                                         & \multicolumn{1}{c}{\checkmark}      & \multicolumn{1}{c}{\checkmark} &                                                     &  \multicolumn{1}{c}{\checkmark} &  \multicolumn{1}{c}{\checkmark} \\
N                                                          & {592491}                              & {592491}                         &                                                     &  {299545}                         &  {299545}                         \\
\bottomrule
		\end{tabular}
		\begin{tablenotes}[para,flushleft]
            \item Note: Linear probability model estimates of DI receipt for
              individuals in treated and control regions based on SESAM
              individual-level survey and administrative data sampled during
              1999--2011. Estimations separately for a complete representative
              sample of the Swiss population and only for individuals in the
              vicinity of the border between treated and non-treated regions.
              Standard errors clustered at the cantonal level in parentheses.
		\end{tablenotes}
	\end{threeparttable}
\end{table}

%%% Local Variables:
%%% mode: latex
%%% TeX-master: "../rad"
%%% End:

\begin{table}[ht]
	\centering
	\caption{Main disability incidence results with canton fixed effects}
	\label{tab:mainresults-with-fe}
	\begin{threeparttable}
		\scriptsize{
            \begin{tabular}{l*{3}{S[table-number-alignment=center,%
                                    table-figures-integer=1,%
                                    table-figures-decimal=3]}%
                                    p{3mm}%
                                    *{3}{S[table-number-alignment=center,%
                                    table-figures-integer=1,%
                                    table-figures-decimal=3]}}
\toprule \addlinespace[1em]
 & \multicolumn{3}{c}{(a) Full sample} & & \multicolumn{3}{c}{(b) Local sample (within 20 km)} \\ \cmidrule(lr{.75em}){2-4} \cmidrule(lr{.75em}){6-8}
 & \multicolumn{1}{c}{(1)}         & \multicolumn{1}{c}{(2)}                           &  \multicolumn{1}{c}{(3)}                           &  & \multicolumn{1}{c}{(4)}       & \multicolumn{1}{c}{(5)}       & \multicolumn{1}{c}{(6)}        \\
\midrule
Treat x pilot &       0.857** &       0.855** &       0.859*  & &       0.765** &       0.765** &       0.760** \\
            &     (0.067)   &     (0.067)   &     (0.068)   & &     (0.086)   &     (0.087)   &     (0.086)   \\
\midrule
Canton fixed effects          & \multicolumn{1}{r}{\checkmark}         & \multicolumn{1}{r}{\checkmark}         & \multicolumn{1}{r}{\checkmark} &  & \multicolumn{1}{r}{\checkmark}         & \multicolumn{1}{r}{\checkmark}         & \multicolumn{1}{r}{\checkmark} \\
Time fixed effects          & \multicolumn{1}{r}{\checkmark}         & \multicolumn{1}{r}{\checkmark}         & \multicolumn{1}{r}{\checkmark} &  & \multicolumn{1}{r}{\checkmark}         & \multicolumn{1}{r}{\checkmark}         & \multicolumn{1}{r}{\checkmark} \\
Other controls          & \multicolumn{1}{r}{-}         & \multicolumn{1}{r}{-}         & \multicolumn{1}{r}{\checkmark} &  & \multicolumn{1}{r}{-}         & \multicolumn{1}{r}{-}         & \multicolumn{1}{r}{\checkmark} \\
N municipalities       & \multicolumn{1}{r}{2,337}     & \multicolumn{1}{r}{2,338}     & \multicolumn{1}{r}{2,338}      &  & \multicolumn{1}{r}{1,086}     & \multicolumn{1}{r}{1,087}     & \multicolumn{1}{r}{1,087}      \\
N individuals   & \multicolumn{1}{r}{249,750}    & \multicolumn{1}{r}{259,323}    & \multicolumn{1}{r}{259,323}    & & \multicolumn{1}{r}{128,536}    & \multicolumn{1}{r}{133,549}    & \multicolumn{1}{r}{133,549}    \\
N failures              & \multicolumn{1}{r}{7,877}      & \multicolumn{1}{r}{9,204}      & \multicolumn{1}{r}{9,204}      & & \multicolumn{1}{r}{3,985}      & \multicolumn{1}{r}{4,693}      & \multicolumn{1}{r}{4,693}      \\
N failures during pilot & \multicolumn{1}{r}{1,713}      & \multicolumn{1}{r}{1,713}      & \multicolumn{1}{r}{1,713}      & & \multicolumn{1}{r}{885}       & \multicolumn{1}{r}{885}       & \multicolumn{1}{r}{885}       \\
\bottomrule
		\end{tabular}}
		\begin{tablenotes}[para,flushleft]
            \item Note: Cox Proportional Hazard estimates for individuals in
                treated and control regions based on SESAM individual-level
                survey and administrative data sampled during 1999--2011.
                Estimations separately for a complete representative sample of
                the Swiss population and only for individuals in the vicinity
                of the border between treated and non-treated regions.
                Baseline hazard for all regressions stratified by 5-year birth
                cohorts. Survey weights applied for the full sample.
                Observations in the local sample are weighted for
                nearest-neighbor pairwise differences. Results are reported in
                exponentiated form as hazard ratios. The hazard ratio for
                `Treat x pilot' corresponds to the relative average treatment
                effect on the treated as defined in \autoref{sec:Strategy}.
                Standard errors clustered at the individual level in
                parentheses, number of observations given below. *, ** and ***
                denote significance at the 10\%, 5\% and 1\% level
                respectively.
		\end{tablenotes}
	\end{threeparttable}
\end{table}

\begin{table}[ht]
	\centering
	\caption{Robustness: Placebo test labor market participation}
	\label{tab:placebo-work}
\makebox[\textwidth][c]{%
	\begin{threeparttable}
		\scriptsize
            \begin{tabular}{l*{4}{S[group-digits=false,round-mode=places,round-precision=4,table-number-alignment=center,%
                                    table-figures-integer=1,%
                                    table-figures-decimal=4]}%
                             p{0.1mm}%
                             *{4}{S[group-digits=false,round-mode=places,round-precision=4,table-number-alignment=center,%
                                    table-figures-integer=1,%
                                    table-figures-decimal=4]}}
\toprule \addlinespace[1em]
 & \multicolumn{4}{c}{(a) Full sample} & & \multicolumn{4}{c}{(b) Local sample (within 20 km)} \\ \cmidrule(lr{.75em}){2-5} \cmidrule(lr{.75em}){6-10}
 & \multicolumn{1}{c}{(1)} & \multicolumn{1}{c}{(2)} & \multicolumn{1}{c}{(3)} & \multicolumn{1}{c}{(4)} & & \multicolumn{1}{c}{(5)} & \multicolumn{1}{c}{(6)} & \multicolumn{1}{c}{(7)} & \multicolumn{1}{c}{(8)} \\[1em]
\cmidrule(lr{.75em}){2-10}
Treat x pre                & 0.01070      &              & 0.00043      &              &  & 0.01266      &              & 0.00044      &              \\
 (2001)                    & (0.00758)    &              & (0.00629)    &              &  & (0.01085)    &              & (0.00897)    &              \\
Treat x pre                &              & 0.00449      &              & -0.00554     &  &              & 0.00545      &              & -0.00638     \\
 (2000, 2001)              &              & (0.00773)    &              & (0.00461)    &  &              & (0.01117)    &              & (0.00663)    \\
Treat x pilot              &              &              & 0.00913***   & 0.00866***   &  &              &              & 0.00684*     & 0.00627*     \\
                           &              &              & (0.00255)    & (0.00257)    &  &              &              & (0.00378)    & (0.00380)    \\
\cmidrule(lr{.75em}){2-10}
Individual covariates      & {\checkmark} & {\checkmark} & {\checkmark} & {\checkmark} &  & {\checkmark} & {\checkmark} & {\checkmark} & {\checkmark} \\
Canton FE                  & {\checkmark} & {\checkmark} & {\checkmark} & {\checkmark} &  & {\checkmark} & {\checkmark} & {\checkmark} & {\checkmark} \\
Year FE                    & {\checkmark} & {\checkmark} & {\checkmark} & {\checkmark} &  & {\checkmark} & {\checkmark} & {\checkmark} & {\checkmark} \\
Only years before 2002     & {\checkmark} & {\checkmark} &              &              &  & {\checkmark} & {\checkmark} &              &              \\
All  years                 &              &              & {\checkmark} & {\checkmark} &  &              &              & {\checkmark} & {\checkmark} \\
N                          & {52,016}     & {52,016}     & {556,540}    & {556,540}    &  & {27,887}     & {27,887}     & {282,858}    & {282,858}    \\
\bottomrule
		\end{tabular}
	\begin{tablenotes}[para,flushleft]
            \item Note: Linear model estimates for individuals in treated and
              control regions based on SESAM individual-level survey and
              administrative data sampled during 1999--2011. Estimations
              separately for a complete representative sample of the Swiss
              population (panel a) and only for individuals in the vicinity of
              the border between treated and non-treated regions (panel b). All
              models include cantonal and year specific effects and control for
              gender, age and native status. Robust standard errors given in
              parentheses. *, ** and *** denote significance at the 10\%, 5\%
              and 1\% level respectively.
	\end{tablenotes}
	\end{threeparttable}
}
\end{table}

\begin{figure}
    \centering
    \caption{Log cumulative hazard by age, treatment region and birth cohort strata}
    \footnotesize
    \label{fig:lncumhazstrata}
            \includegraphics[width=0.75\textwidth]{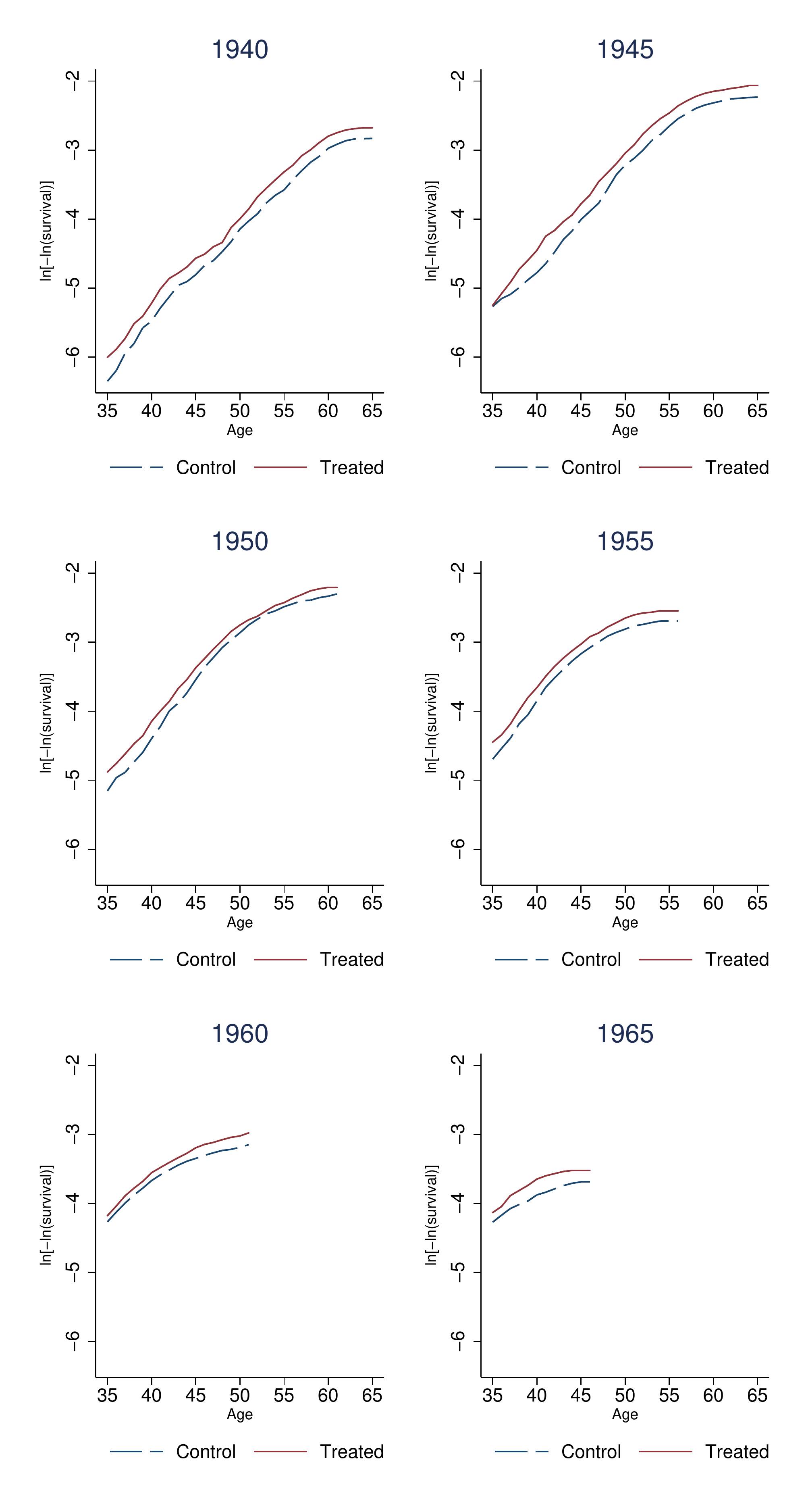}
            \floatfoot*{\centering
                Note: Log-log plot showing log cumulative hazard estimates by
age and birthcohort for individuals in treated and control regions, separately
for major birth cohort strata.}
\end{figure}

\restoregeometry

\end{document}